\documentclass[article,onecolumn, draftclsnofoot]{IEEEtran}  

\usepackage[utf8]{inputenc}
\usepackage[T1]{fontenc}
\usepackage[hidelinks]{hyperref}
\usepackage{url}
\usepackage{ifthen}
\usepackage{cite}
\usepackage[cmex10]{amsmath} 
\usepackage{amsfonts,amssymb}
\usepackage{color}
\usepackage{ifthen}
\usepackage[arxiv]{optional} 
\usepackage{bbm}
\usepackage{cite}
\usepackage{amsmath,amssymb,amsfonts}
\usepackage{algorithmic}
\usepackage{graphicx}
\usepackage{textcomp}
\usepackage{xcolor}
\usepackage{caption}
\usepackage{subcaption}
\usepackage{graphics} 
\usepackage{epsfig} 
\usepackage{times} 
\usepackage{amsmath} 
\usepackage{amssymb}  
\usepackage{caption}
\usepackage{hyperref}
\usepackage{bbold}

\newcommand{\eq}[1]{\begin{align}#1\end{align}}
\newcommand{\seq}[1]{\begin{subequations}#1\end{subequations}}

\newcommand{\lb}[1]{\Bigg\{ \begin{array}{ll} #1 \end{array} }

\newcommand{\bit}[1]{\begin{itemize}#1\end{itemize}}
\newcommand{\E}{\mathbb{E}}
\newcommand{\p}{\mathbb{P}}

\newcommand{\cT}{[T]}

\newcommand{\cP}{\mathcal{P}}

\newcommand{\cX}{\mathcal{X}}

\newcommand{\cN}{\mathcal{N}}

\newcommand{\cA}{\mathcal{A}}

\newcommand{\cZ}{\mathcal{Z}}

\newcommand{\tsigma}{\tilde{\sigma}}
\newcommand{\tgamma}{\tilde{\gamma}}

\newcommand{\defeq}{\buildrel\triangle\over =}

\newcommand{\nn}{\nonumber}

\newcommand{\cS}{\mathcal{S}}

\DeclareMathAlphabet{\mathcal}{OMS}{cmsy}{m}{n}

\usepackage{lipsum}
\usepackage{amsfonts}
\usepackage{graphicx}
\usepackage{epstopdf}
\usepackage{algorithmic}
\ifpdf
  \DeclareGraphicsExtensions{.eps,.pdf,.png,.jpg}
\else
  \DeclareGraphicsExtensions{.eps}
\fi

\newtheorem{lemma}{Lemma}

\newtheorem{theorem}{Theorem}

\newtheorem{definition}{Definition}
\newtheorem{claim}{Claim}
\newtheorem{corollary}{Corollary}

\title{Master equation of discrete-time Stackelberg mean field games with multiple leaders}

\author{Deepanshu Vasal\thanks{The author is with the Department of Electrical and Computer Engineering, Northwestern University,
  ({dvasal@umich.edu}).
  }
  }

\usepackage{amsopn}

\ifpdf
\fi

\begin{document}

\maketitle

\begin{abstract}
 In this paper, we consider a discrete-time Stackelberg mean field game with a finite number of leaders, a finite number of major followers and an infinite number of minor followers. The leaders and the followers each observe types privately that evolve as conditionally independent controlled Markov processes. The leaders are of ``Stackelberg" kind which means they \emph{commit} to a dynamic policy. We consider two types of followers: major and minor, each with a private type. All the followers best respond to the policies of the Stackelberg leaders and each other. Knowing that the followers would play a mean field game (with major players) based on their policy, each (Stackelberg) leader chooses a policy that maximizes her reward. We refer to the resulting outcome as a Stackelberg mean field equilibrium with multiple leaders (SMFE-ML). In this paper, we provide a master equation of this game that allows one to compute all SMFE-ML. We further extend this notion to the case when there are infinite number of leaders. 
    
\end{abstract}

\section{Introduction}

%
%

There are many scenarios in the real world where firms compete with each other (and many times with a government alternative such as United States Postal Service (USPS)) while engaging with the market such Lyft and Uber for ride sharing markets, or Amazon, Walmart and EBay online market platforms and so on. Thus it is an important question to investigate as to how to find the best strategies for such firms (or the government) while engaging with the market and while competing with each other.

With increasing amount of integration of technology in our society and with recent advancements in computation and algorithmic technologies, there is an unprecedented scale of interaction among people and devices. As an example, smartphones have penetrated our society in the last decade and more than 82\% of the world is connected through the internet.  
Such deep inter-connectedness of the society demands a  need to design and understand the behavior of the resulting \emph{large scale} interactions and a need to design policies by the government and private players to better govern and optimally respond. In this paper, we present a new methodology to analyze such interactions through Stackelberg mean-field dynamic games.

The theory of dynamic games is a powerful tool to model such sequential strategic interaction among selfish players, introduced by \cite{Sh53}. Discrete-time dynamic games with Markovian structure have been studied extensively to model many practical applications in both engineering and economics, such as as dynamic auctions~\cite{IyJoSu14,BeSa10}, security~\cite{EtBa19}, markets~\cite{Vill99,BoPaWu18}, traffic routing~\cite{JoScSt04,MePaOzAc20}, wireless systems~\cite{AdJoGo07}, social learning~\cite{VaAn16allerton_extended, LeSuBe14}, oligopolies-- i.e. competition among firms (e.g.~\cite{BaOl98, FiVr12}), and more. 

In dynamic games with perfect and symmetric information, subgame perfect equilibrium (SPE) is an appropriate equilibrium concept.
Markov Perfect Equilibria (MPE), introduced in \cite{MaTi01}, is a refinement of SPE that is also used, where players' strategies depend on a coarser Markovian state of the systems, instead of the whole history of the game which grows exponentially with time and thus becomes unwieldy.  An analogous notion to SPE for incomplete information games is perfect Bayesian equilibrium (PBE).
However, when the number of players is large, computing MPE/PBE becomes intractable.
To model the behavior of large population strategic interactions, mean-field games were introduced independently by \cite{HuMaCa06}, and \cite{LaLi07}. In such games, there are large number of homogeneous strategic players, where each player has infinitesimal effect on system dynamics and is affected by other players through a mean-field population state. There have been a number of applications such as economic growth, security in networks, oil production, volatility formation, population dynamics (see ~\cite{La08,GuLaLi11,SuMa19,HuMa16,HUMa17,HuMa17cdc, AdJoWe15} and references therein). 

An engineering side of game theory is the theory of Mechanism design that deals with the \emph{design} of games such that when acted upon by the strategic players, the equilibrium(s) of the game coincide with the outcome desired by the designer. It could be social welfare say desired by the government or profit maximization desired by private entities. 
Stackelberg equilibrium (SE) is a notion of equilibrium related to mechanism design.
A Bayesian Stackelberg game is played between two players: a leader and a follower.
The follower has a private type that only she observes, however, the leader knows the prior distribution
on that state. The leader commits to a strategy that is observable to the follower. The
follower then plays a best response to leader's strategy to maximize its utility. Knowing
that the follower will play a best response, the leader commits to and plays a strategy
that maximizes his utility. Such pair of strategies of the leader and the follower is called a Stackelberg equilibrium. It is
known that such strategies can provide higher utility to the leader than that obtained in a Nash
equilibrium of the game.

In this paper, we consider discrete-time Stackelberg mean field games where there are finitely many leaders (who have commitment power), finite major followers (who play Nash equilibrium i.e. do not commit) and infinitely many minor followers (who also play Nash equilibrium). All the leaders and followers sequentially make strategic decisions, simultaneously in each period $t$ and are affected by other players through a mean-field population state of the minor followers' actions, major followers actions and the leaders' actions. 
Each follower (both major and minor) and all the leaders individually have a private type that evolves through a controlled Markov process that only she observes and leaders and all the followers observe the current population state which is the distribution of all the minor followers' types. The leaders commit to a dynamic policy and all the followers, both major and minor, best respond to it while being in equilibrium with each other such that for each time $t$ and given the leaders' policies, a follower's policy (symmetric across all minor followers, non necessarily symmetric for the major followers) maximizes her reward to go so that she doesn't gain by unilaterally deviating while all the other followers play the equilibrium policy. Similarly each leader $i$ plays a strategy such that when all the followers' best respond to all the leaders' strategy and are in equilibrium with each other, and the other leaders best respond to leader $i$'s strategy then the leader's $i$'s strategy maximizes her reward to go. Note that leaders are in Nash equilibrium with each other and together they play Stackelberg equilibrium to the followers (as in~\cite{DeXu09}). We note that any of the leaders can be social welfare maximizing and her instantaneous reward is then the sum of expected reward of the followers, where expectation over the infinite minor followers is defined through the mean field state.

In such games, a Stackelberg Mean Field Equilibrium (SMFE) is defined through a coupled fixed-point equation as follows: the mean-field state evolves through a Fokker-Planck \emph{forward} equation given an SMFE policy profile of the leader and the  followers. The followers' (symmetric) policy is a best response to the leader's equilibrium policy, given the mean-field state evolution process. Finally, the leader's policy is optimum given that followers play the best response. 
As a result, in order to compute an SMFE, one needs to solve a coupled fixed-point equation in the space of mean-field states and the equilibrium policies of the leader and the followers.
In principle, one can solve this fixed-point equation across time for the whole game; however, the resulting complexity will increase double exponentially with time. In this paper, we present an algorithm which can equivalently solve for smaller fixed-point equations for each time $t$.
This algorithm is equivalent to the {\it master equation} of continuous-time mean field games~\cite{CaDeLaLi15} that allows one to compute all mean field equilibria (MFE) of the game sequentially.
  
     Our algorithm is motivated by the developments in the theory of dynamic games with asymmetric information in~\cite{VaSiAn16arxiv,VaAn16allerton, VaAn16cdc, Va20acc ,VaMiVi21, Va20St, VaBe22}, where authors in these works have considered different models of such games and provided a sequential decomposition framework to compute Markovian perfect Bayesian equilibria and Markov perfect Stackelberg equilibria of such games. A special case with a single leader was studied in~\cite{VaBe22}.


The paper is structured as follows. In Section~\ref{sec:Model}, we present the model, our notation and background. In Section~\ref{sec:Prelim}, we present the notion of a Stackelberg Mean Field Equilibrium (SMFE) and the common information approach.
In Section~\ref{sec:Result}, we present our main results, where we present an algorithm to compute a SMFE for the finite horizon game. We conclude in Section~\ref{sec:Conclusion}. All proofs are presented in Appendix.

\subsection{Notation}
We use uppercase letters for random variables and lowercase for their realizations. For any variable, subscripts represent time indices and superscripts represent player identities. We use notation $ -i$ to represent all players other than player $i$ i.e. $ -i = \{1,2, \ldots i-1, i+1, \ldots, N \}$. We use notation $a_{t:t'}$ to represent the vector $(a_t, a_{t+1}, \ldots a_{t'})$ when $t'\geq t$ or an empty vector if $t'< t$. We use $a_t^{-i}$ to mean $(a^1_t, a^2_{t}, \ldots, a_t^{i-1}, a_t^{i+1} \ldots, a^m_{t})$ . We remove superscripts or subscripts if we want to represent the whole vector, for example $a_t$  represents $(a_t^1, \ldots, a_t^m) $. We denote the indicator function of any set $A$ by $\mathbbm{1}\{A\}$. 
For any finite set $\mathcal{S}$, $\mathcal{P}(\mathcal{S})$ represents space of probability measures on $\mathcal{S}$ and $|\mathcal{S}|$ represents its cardinality. Given a set $\mathcal A$, we denote its $n$-fold Cartesian product by $({\mathcal A})^m$.   We denote the set of real numbers by $\mathbb{R}$. For a probabilistic strategy profile of players $(\sigma_t^i)_{i\in [N]}$ where probability of action $a_t^i$ conditioned on $z_{1:t},x_{1:t}^i$ is given by $\sigma_t^i(a_t^i|z_{1:t},x_{1:t}^i)$, we use the short hand notation $\sigma_t^{-i}(a_t^{-i}|z_{1:t},x_{1:t}^{-i})$ to represent $\prod_{j\neq i} \sigma_t^j(a_t^j|z_{1:t},x_{1:t}^j)$.   We denote by $P^{\sigma}$ (or $E^{\sigma}$) the probability measure generated by (or expectation with respect to) strategy profile $\sigma$.
All equalities and inequalities involving random variables are to be interpreted in the \emph{a.s.} sense. For any variable $a$, we define $\cS_a$ as the space of all possible $a$.

\section{Model}
\label{sec:Model}
We consider a stochastic Stackelberg mean field game over a time horizon $[T]\defeq$ $\{1, 2, \ldots T\}$ with simultaneous moves and perfect recall as follows. Suppose there are three kinds of players: $K$ Stackelberg leaders, $M$ (Nash) major followers, and an infinite number of homogeneous (Nash) minor followers. All leaders and followers have private types, $x_t^{l,i} \in \cX^{l}$ for $i$th leader, $x_t^{m,j} \in \cX^m$, for $j$th major follower and $x_t^{f} \in \cX^f$ for every minor follower at time $t$, where $x_t^{f,i},x_t^{m,j}$ evolve as a conditionally independent controlled Markov processes in the following way, where for any finite $K$, Stackelberg leaders, $M$ (Nash) major followers and $N$ (Nash) minor followers. Let $x_t^l = x_t^{l[1:K]}, x_t^m = x_t^{m[1:M]}$, $a_t^l = a_t^{l[1:K]}, a_t^m = a_t^{m[1:M]}$

\eq{
&P(x_t^{l[1:N]},x_t^{m[1:M]},x_t^{f[1:N]}|z_{1:t-1},a^{l,m}_{1:t-1},x^{l,m}_{1:t-1})\nn\\ 
&=  \prod_{i=1}^N Q^{l,i}(x_t^{l,i}|z_{t-1},a_{t-1}^{l,m},x_{t-1}^{l,m}) \prod_{j=1}^M Q^{m,j}(x_t^{m,j}|z_{t-1},a_{t-1}^{l,m},x_{t-1}^{l,m})\prod_{k=1}^K Q^{f}(x_t^{f,k}|z_{t-1},a_{t-1}^{l,m},x_{t-1}^{l,m},x_{t-1}^{f,k} ),
}
where $Q^{l,i},Q^{m,j}$ are known kernels, and $z_t$ is defined below. A Stackelberg leader $i$ takes action $a_t^{l,i}\in \cA^{l}$ at time $t$ on observing $z_{1:t},x_{1:t}^{l,i}$, a major follower $j$ takes action based on $z_{1:t},x_{1:t}^{m,j}$, and a minor follower $k$ takes action $a_t^{f,k}\in \cA^f$ at time $t$ on observing $z_{1:t}$ and $x_{1:t}^{f,k}$,
where $z_t$ is the mean field population state of the minor followers at time $t$, i.e.,
\eq{
z_t(x) \defeq \lim_{N\to\infty} \sum_{k=1}^m \frac{1}{N}1(x_t^{f,k} = x).
}
Here, $z_{1:t}$ is common information among players, and $x_{1:t}^{l,i}, x_{1:t}^{m,j},x_{1:t}^{f,k}$ are private information of the Stackelberg leader $i$, major follower $j$ and minor follower $k$, respectively. We denote the set of possible values of the mean-field state $z_t$ by $\cZ = \cP(X^f)$.

 At the end of interval $t$, Stackelberg leader $i$ receives an instantaneous reward $R_t^{l,i}(z_t,x_t^{l,m},a_t^{l,m})$, major follower $j$ receives an instantaneous reward $R_t^{m,j}(z_t,x_t^{l,m},a_t^{l,m})$ and the minor follower $k$ receives an instantaneous reward $R_t^f(z_t,x_t^{l,m},x_t^{f,k},a_t^{f,k},a_t^{l,m})$. Note that each leader's reward only depends on the followers' state and actions through the mean-field state. Likewise for each follower's reward function is the same for all followers (homogeneous followers) and it depends on the states and actions of the other followers through the mean-field state, but does depend directly on the states and actions of all the leaders and the follower's own action.

The sets $\cA^{l,i},\cA^{m,j},\cA^f, \cX^{l,i}, \cX^{m,j},\cX^f $ are assumed to be finite. Let $\sigma^{l,i} = ( \sigma^{l,i}_t)_{t \in [T]}$ be a probabilistic strategy of leader $i=1\ldots K$, $\sigma^{m,j} = ( \sigma^{m,j}_t)_{t \in [T]}$ be a probabilistic strategy of major follower $j=1\ldots K$, $\sigma^{f,k} = ( \sigma^{f,k}_t)_{t \in [T]}$ be a probabilistic strategy of the minor follower $k$ where $\sigma^{l,i}_t : (\mathcal{Z})^{t}\times(\cA^{l}\times\cA^m)^{t-1}\times(\cX^{l,i})^t \to \mathcal{P}(\cA^{l,i})$, $\sigma^{m,j}_t : (\mathcal{Z})^{t}\times(\cA^{l}\times\cA^m)^{t-1}\times(\cX^{m,j})^t \to \mathcal{P}(\cA^{m,j})$, and $\sigma^f_t : (\mathcal{Z})^{t}\times (\cA^{l}\times\cA^m)^{t-1}\times(\cX^f)^t \to \mathcal{P}(\cA^f)$ such that the leader plays action $A_t^{l,i}$ according to $ A_t^{l,i} \sim \sigma^{l,i}_t(\cdot|z_{1:t},a_{1:t-1}^{l,m},x_{1:t}^{l,i})$, and the follower plays action $A_t^f$ according to $ A_t^f \sim \sigma^f_t(\cdot|z_{1:t},a_{1:t-1}^{l,m},x_{1:t}^f)$, Let $ \sigma \defeq(\sigma^i)_{i\in \{l,f\}}$ be a strategy profile of all players. Suppose players discount their rewards by a discount factor $\delta\leq 1$.

\section{Preliminaries}
\label{sec:Prelim}
In this section, we first present the definition of a Stackelberg Mean Field Equilibrium with Multiple Leaders (SMFE-ML) which we will use in this paper. We then discuss the common agent approach that we will utilize in deriving an algorithm for finding an SMFE-ML.

\subsection{Stackelberg mean field equilibrium with multiple leaders}
\label{sec:PBSE}
In this paper, we will consider followers' Markovian equilibrium policy that only depends on her current states $x_t^f$, current mean field state $z_t$ and common marginal beliefs $\pi_t^{l,i},\pi_t^{m,j}, i=1\ldots K, j=1\ldots M$, where $\pi_t^{l,i}(x_t^{l,i}) = P^{\sigma^f,\sigma^{l},\sigma^{m}}(x_t^{l,i}|z_{1:t},a_{1:t-1}^{l,m}), \pi_t^{m,j}(x_t^{m,j}) = P^{\sigma^f,\sigma^{l},\sigma^{m}}(x_t^{m,j}|z_{1:t},a_{1:t-1}^{l,m})$ i.e. $\pi_t^{l,i}$ is the common belief on the Stackelberg leader $i$'s state given the common information $(z_{1:t},a_{1:t-1}^{l,m})$. Let ${\underline{\pi_t}} = \{ \pi_t^{l,i}, \pi_t^{m,j}, \}_{i=1\ldots N, j=1\ldots K }$ Thus, at equilibrium $a_t^{f,k}\sim \tsigma^{f,k}_t(\cdot|\underline{\pi_t},z_t,x_t^{f,k})$ and the Stackelberg leader $i$'s strategy $a_t^{l}\sim \tsigma^{l}_t(\cdot|\underline{\pi_t},z_t,x_t^{l,i})$ and major follower $j$'s strategy $a_t^{m,j}\sim \tsigma^{m,j}_t(\cdot|\underline{\pi_t},z_t,x_t^{m,j})$.\footnote{Note, however, that for the purpose of equilibrium, we allow for deviations in the space of all possible strategies that may depend on the entire observation history.} 

For the game considered, we first define best response mapping of the follower as follows, which we will use in turn to define a Stackelberg mean field equilibrium.

%
Let $BR_t^f$ is given by
\eq{
BR_t^f(\underline{\pi_t},z_{1:t},a_{1:t-1}^{l,m},x_{1:t}^{f},\sigma_{t:T}^{l},\sigma_{t:T}^{m}) &:=\arg\max_{\sigma^f} \E^{\sigma_{t:T}^{l,m},{\sigma}_{t:T}^f,\underline{\pi_t}}[\sum_{n=t}^T \delta^{n-t} R^f(Z_n,X_n^{l,m,f},A_n^{l,m,f})|\underline{\pi_t},z_{1:t},a_{1:t-1}^{l,m},x_{1:t}^{f} ].
}
 Let  $BR^f: \cZ^T\times \cS_{\sigma^{l}}\times \cS_{\sigma^{m}}\to{\cS_{\sigma^f}}$ be given by
\eq{
BR^f(z_{1:T},\sigma^{l},\sigma^{m}) &:=\bigcap_{t}\bigcap_{a_{1:t-1}^{l,m}}\bigcap_{x_{1:t}^{f}}BR_t^f(\underline{\pi_t},z_{1:t},a_{1:t-1}^{l,m},x_{1:t}^{f},\sigma_{t:T}^{l},\sigma_{t:T}^{m})
}
 This specifies a minor follower's best response at time $t$ given the history of the mean-field state and its private type up to time $t$ and the leader's strategy from time $t$ on-wards.   
 Note that this mapping specifies a complete policy for the follower for all time $t$. 
 
Similarly we define the best response of major follower $j$ as
\eq{
&BR_t^{m,j}(\underline{\pi_t},z_{1:t},a_{1:t-1}^{l,m},x_{1:t}^{m,j},\sigma_{t:T}^{l},\sigma_{t:T}^{m,-j},\sigma_{t:T}^f) \nn\\
&:=\arg\max_{ \sigma^{m,j}} \E^{\sigma_{t:T}^{l}\sigma_{t:T}^{m,j}\sigma_{t:T}^{m,-j},\sigma_{t:T}^f,\underline{\pi_t}} \big\{ \sum_{n=t}^T \delta^{n-t}R_n^{m,j}(Z_n,X_n^{l,m},A_n^{l,m}) |\underline{\pi_t},z_{1:t},a_{1:t-1}^{l,m},x_{1:t}^{m,j}\big\}
}
\eq{ 
&BR^{m,j}(z,\sigma^l,\sigma^{m,-j},\sigma^f) :=\bigcap_t \bigcap_{a_{1:t-1}^{m,j}}\bigcap_{x_{1:t}^{m,j}}  BR_t^{m,j}(\underline{\pi_t},z_{1:t},a_{1:t-1}^{l,m},x_{1:t}^{m,j},\sigma_{t:T}^{l},\sigma_{t:T}^{m,-j},\sigma_{t:T}^f).
}
With some abuse of notation, we will also say $\sigma_t^f\in BR_t^f(z_{1:T},\sigma_{1:T}^{l,m})$ if there exists $\hat{\sigma}^f \in BR^f(z_{1:T},\sigma_{1:T}^{l,m})$ such that $\sigma_t^f = \hat{\sigma}_t^f$. Similarly we say $\sigma_t^{m,j}\in BR_t^{m,j}(z_{1:T},\sigma_{1:T}^{l},\sigma_{1:T}^{m,-j},\sigma_{1:T}^f)$ if there exists $\hat{\sigma}^{m,j} \in BR^{m,j}(z_{1:T},\sigma_{1:T}^{l},\sigma_{1:T}^{m,-j},\sigma_{1:T}^f)$ such that $\sigma_t^{m,j} = \hat{\sigma}_t^{m,j}$
 

%

We now define best response of Stackelberg leader $i$ as follows
\eq{
BR^{l,i}(z_{1:T},\sigma^{l,-i}) &:=\bigcap_t \bigcap_{a_{1:t-1}^{l,i}}\bigcap_{x_{1:t}^{l,i}}  \arg\max_{ \sigma^{l,i}} \E^{\sigma^{l},\sigma^{m},{\sigma}^f,\underline{\pi_t}} \big\{ \sum_{n=t}^T \delta^{n-t}R_n^{l,i}(Z_nX_n^{l,m},A_n^{l,m}) |\underline{\pi_t},z_{1:t},a_{1:t-1}^{l,m},x_{1:t}^{l,i}\big\},\\
&\text{ where, } \hat{\sigma}^f \in BR^f(z,\sigma^{l},\hat{\sigma}^m),\hat{\sigma}^m\in BR^f(z,\sigma^{l},\hat{\sigma}^m,\hat{\sigma}^f)
}
Conversely, define a mapping $\Lambda: \cS_{\sigma^f}\times \cS_{\sigma^{l}}\times\cS_{\sigma^{m}} \to \cZ^T $ as follows: given $\sigma^f \in S_{\sigma^f},\sigma^{l} \in S_{\sigma^{l}},\sigma^m \in S_{\sigma^m}, z = \Lambda(\sigma^f,\sigma^{l},\sigma^{m})$, is constructed recursively as
$\forall t,z_{1:t},a_{1:t-1}^{l,m}$ 
\eq{z_{t+1}(\cdot) = \sum_{x_t,a_t}z_{1:t}(x_{1:t}^f)P(x^{l,m}_{1:t}|z_{1:t},a_{1:t-1}) Q^f( \cdot|z_t,x_t^f, a_t^f,a^{l,m}_t)\sigma^f_t(a^f_t|z_{1:t},a_{1:t-1}^{l,m},x_{1:t}^f)\sigma_t^{l,m}(a^{l,m}_t|z_{1:t},a_{1:t-1}^{l,m},x_{1:t}^{l,m}). 
}
This mapping determines the mean-field trajectory as a function of the leaders' and the followers' policies.



\begin{definition}
\label{def:MFE}
 $(\tsigma^l,\tsigma^{m},\tsigma^f,z)$ is a Stackelberg Mean-Field Equilibrium with multiple leaders (SMFE-ML) if

[(a)]: $\tsigma^f\in BR^f(z,\tsigma^{l},\tsigma^m)$,

[(b)]: $z=\Lambda(\tsigma^{l},\tsigma^m,\tsigma^f)$, and

[(c)]: $\tsigma^{l} \in BR^{l}(z,\tsigma^l,\tsigma^m,\tsigma^f)$

[(d)]:
$\tsigma^{m} \in BR^{n}(z,\tsigma^l,\tsigma^m,\tsigma^f)$

\end{definition}



\subsection{Common agent approach}
We recall that in general, the leaders and the followers generate their actions at time $t$ as follows, $a_t^{l,i}\sim \sigma_t^{l,i}(\cdot|z_{1:t},a_{1:t-1}^{l,m},x_{1:t}^{l,i}),a_t^{m,j}\sim \sigma_t^{m,j}(\cdot|z_{1:t},a_{1:t-1}^{l,m},x_{1:t}^{m,j})$ and $a_t^{f,k}\sim \sigma_t^f(\cdot|z_{1:t},a_{1:t-1}^{l,m},x_{1:t}^{f,k})$.
An alternative way to view the problem is as follows. As is done in the common information approach~\cite{NaMaTe13}, at time $t$, a fictitious common agent observes the common information $z_{1:t},a_{1:t-1}^{l,m}$ and generates prescription functions $\gamma_t = (\gamma_t^{l},\gamma_t^m,\gamma_t^f) = \psi_t[z_{1:t},a_{1:t-1}^{l,m}]$, where $\gamma_t^l = \{\gamma_t^{l,i}\}_{i=1\ldots N}, \gamma^{m,j} = \{\gamma_t^m\}_{j=1\ldots M}$. Follower $k$ uses its prescription function $\gamma_t^{f,k}$ to operate on its private information $x_{1:t}^{f,k}$ to produce her action $a_t^{f,k}$, i.e. $\gamma_t^{f,k}:(\cX^{f,k})^t\to \cP(\cA^{f,k})$ and $a_t^{f,k}\sim\gamma_t^{f,k}(\cdot|x_{1:t}^{f,k})$. Similarly, Stackelberg leader $i$ uses her prescription function $\gamma_t^{l,i}$ to operate on her private information $x_{1:t}^{l,i}$ to produce her action $a_t^{l,i}$, i.e. $\gamma_t^{l,i}:(\cX^{l,i})^t\to \cP(\cA^{l,i})$ and $a_t^{l,i}\sim\gamma_t^{l,i}(\cdot|x_{1:t}^{l,i})$. And major follower $j$ uses her prescription function $\gamma_t^{m,j}$ to operate on her private information $x_{1:t}^{m,j}$ to produce her action $a_t^{m,j}$, i.e. $\gamma_t^{m,j}:(\cX^{m,j})^t\to \cP(\cA^{m,j})$ and $a_t^{m,j}\sim\gamma_t^{m,j}(\cdot|x_{1:t}^{m,j})$. It is easy to see that for any $\sigma^{l,m,f}$ policy profile of the players, there exists an equivalent $\psi$ profile of the common agent (and vice versa) that generates the same control actions for every realization of the information of the players.

Here, we will consider Markovian common agent's policy as follows. We call a common agent's policy be of ``type $\theta$" if the common agent observes the mean field population state $z_t$ and common belief $\underline{\pi_t}$, and generates prescription functions $\gamma_t := (\gamma_t^{l},\gamma_t^m,\gamma_t^f) = \theta_t[\underline{\pi_t},z_t]$. 
The follower $k$ uses prescription function $\gamma_t^{f,k}$ to operate on her current private type $x_t^{f,k}$ to produce her action $a_t^{f,k}$, i.e. $\gamma_t^{f,k} : \cX^{f,k}\to \cP(\cA^{f,k})$ and $a_t^{f,k} \sim\gamma_t^{f,k}(\cdot|x_t^{f,k})$. Similarly, Stackelberg leader $i$ uses her prescription function $\gamma_t^{l,i}$ to operate on her private information $x_t^{l,i}$ to produce her action $a_t^{l,i}$, i.e. $\gamma_t^{l,i}:\cX^{l,i}\to \cP(\cA^{l,i})$ and $a_t^{l,i}\sim\gamma_t^{l,i}(\cdot|x_{t}^{l,i})$. Similarly, the major follower $j$ uses her prescription function $\gamma_t^{m,j}$ to operate on her private information $x_t^{m,j}$ to produce her action $a_t^{m,j}$, i.e. $\gamma_t^{m,j}:\cX^{m,j}\to \cP(\cA^{m,j})$ and $a_t^{m,j}\sim\gamma_t^{m,j}(\cdot|x_{t}^{m,j})$.

Then the mean field is updated as
\eq{z_{t+1}(\cdot) = \sum_{x_t,a_t}z_t(x_t^f)\underline{\pi_t}(x^{l,m}_t) Q^f( \cdot|z_t,x_t^{l,m,f}, a_t^{l,m,f})\gamma^f_t(a^f_t|x_t^f),\gamma_t^{l,m}(a_t^{l,m}|x^{l,m}_t) \label{eq:phi_def}
}
We also call the above equation as $z_{t+1} = \phi(\underline{\pi_t},z_t,\gamma_t)$

Recall that we defined common marginal beliefs ${\underline{\pi_t}} = \{ \pi_t^{l,i}, \pi_t^{m,j}, \}_{i=1\ldots N, j=1\ldots M }$ , where $\pi_t^{l,i},\pi_t^{m,j}, i=1\ldots N, j=1\ldots M$, where $\pi_t^{l,i}(x_t^{l,i}) = P^{\sigma^f,\sigma^{l},\sigma^{m}}(x_t^{l,i}|z_{1:t},a_{1:t-1}^{l,m}), \pi_t^{m,j}(x_t^{m,j}) = P^{\sigma^f,\sigma^{l},\sigma^{m}}(x_t^{m,j}|z_{1:t},a_{1:t-1}^{l,m})$ i.e. $\pi_t^{l,i}$ is the common belief on the Stackelberg leader $i$'s state $x_t^{l,i}$ given the common information $(z_{1:t},a_{1:t-1}^{l,m})$ and similarly $\pi_t^{m,j}$ is the common belief on the major follower $j$'s state $x_t^{m,j}$ given the common information $(z_{1:t},a_{1:t-1}^{l,m})$. 
In the following lemma, we show that the belief $\underline{\pi_t}$ can be updated using Bayes' rule.
\begin{lemma}
There exists  functions $F^{l,i}, F^{m,j}$ for $i=1\ldots N, j=1\ldots M$, independent of the strategy $\theta$ such that 
\eq{
\pi^{l,i}_{t+1}&=F^{l,i}(\underline{\pi}_t^{l,m},z_{t},\gamma^{l,m}_t,A^{l,m}_t)\\
\pi^{m,j}_{t+1}&=F^{m,j}(\underline{\pi}_t^{l,m},z_{t},\gamma^{l,m}_t,A^{l,m}_t)
}
Combining the above two we also say
\eq{
\pi^{l,m}_{t+1}=\underline{F}(\underline{\pi}_t^{l,m},z_{t},\gamma^{l,m}_t,a_t^{l,m})
}
\end{lemma}
\begin{IEEEproof}
We only consider the proof of the update of $\pi_t^{l,i}$ and the proof of the update of $\pi_t^{m,j}$ is similar which is skipped.
\eq{
&\pi_{t+1}^{l,i}(x_{t+1}^{l,i}) =  P^{\theta}(x_{1+t}^{l,i}|z_{1:t+1},a_{1:t}^{l,m})\\ 
&=\frac{\displaystyle \sum_{x_t^{l,m},a^{l,m}_t}\underline{\pi}_t^{l,m}(x^{l,m}_t)1(z_{t+1}=\phi(\underline{\pi_t},z_t,\gamma_t))\gamma_t^{l,m}(a^{l,m}_t|x^{l,m}_t)Q^{l,i}(x_{t+1}^{l,i}|z_t,x^{l,m}_t,a^{l,m}_t)}{\displaystyle \sum_{x_t^{l,m}}\underline{\pi}_t^{l,m}(x_t^{l,m})1(z_{t+1}=\phi(\underline{\pi_t},z_t,\gamma_t))\gamma_t^{l,m}(a^{l,m}_t|x^{l,m}_t)}\\
&=\frac{\displaystyle \sum_{x^{l,m}_t,a^{l,m}_t}\underline{\pi}_t(x^{l,m}_t)\gamma_t^{l,m}(a^{l,m}_t|x^{l,m}_t)Q^{l,i}(x_{t+1}^{l,i}|z_t,x^{l,m}_t,a^{l,m}_t)}{\displaystyle \sum_{x^{l,m}_t}\underline{\pi_t}(x^{l,m}_t)\gamma_t^{l,m}(a^{l,m}_t|x^{l,m}_t)}
\label{eq:piupdate}
}
\end{IEEEproof}


\section{Algorithm for SMFE-ML computation}
\label{sec:Result}
In the next section, we design an algorithm to compute SMFE-ML of the game.

\subsection{Backward Recursion}

In this section, we define an equilibrium generating function $\theta=(\theta^{l,i}_t,\theta_t^{m,j},\theta^{f}_t)_{i\in\{1\ldots N\}, j\in\{1\ldots M, \},t\in[T]}$, where $\theta^{l,i}_t :  (\prod_{i=1}^N\mathcal{P}(\cX^{l,i}))\times(\prod_{j=1}^M\cP(\cX^{m,j}))\times \cP(\cX^f) \to \big\{\cX^{{l,i}} \to \mathcal{P}(\cA^{l,i}) \big\}$, $\theta^{m,j}_t :  (\prod_{i=1}^K\mathcal{P}(\cX^{l,i}))\times(\prod_{j=1}^M\cP(\cX^{m,j}))\times \cP(\cX^f) \to \big\{\cX^{{m,j}} \to \mathcal{P}(\cA^{m,j}) \big\}$, $\theta^{f}_t :  (\prod_{i=1}^N\mathcal{P}(\cX^{l,i}))\times(\prod_{j=1}^M\cP(\cX^{m,j}))\times  \mathcal{P}(\cX^f) \to \big\{\cX^{f} \to \mathcal{P}(\cA^{f}) \big\}$
and a sequence of functions $(V_t^{l,i}, V_t^{m,j},V_t^f)_{t\in \{ 1,2, \ldots T+1\}}$, where $V^{l,i}_t :  (\prod_{i=1}^N\mathcal{P}(\cX^{l,i}))\times(\prod_{j=1}^M\cP(\cX^{m,j}))\times \cP(\cX^{f})\times \cX^{l,i} \to \mathbb{R}$, $V^{m,j}_t :  (\prod_{i=1}^N\mathcal{P}(\cX^{l,i}))\times(\prod_{j=1}^M\cP(\cX^{m,j}))\times\cP(\cX^f)\times  \cX^{m,j} \to \mathbb{R}$,$V^{f}_t :  (\prod_{i=1}^N\mathcal{P}(\cX^{l,i}))\times(\prod_{j=1}^M\cP(\cX^{m,j}))\times  \mathcal{P}(\cX^f)\times \cX^f \to \mathbb{R}$, in a backward recursive way, as follows. 
\begin{itemize}
\item[1.] Initialize $\forall \underline{\pi}_{T+1}\in\prod_{i=1}^N\mathcal{P}(\mathcal{X}^{l,i})\times\prod_{j=1}^M\mathcal{P}(\mathcal{X}^{m,j}), z_{T+1}\in \mathcal{P}(\cX^f), i=1\ldots N, j=1\ldots M, x_{T+1}^{l,i}\in \cX^{l,i}, x_{T+1}^{m,j}\in \cX^{m,j}, x_{T+1}^{f}\in \cX^f,$
\eq{
V^{l,i}_{T+1}(\underline{\pi}_{T+1},z_{T+1},x_{T+1}^{l,i}) &\defeq 0.   \label{eq:VT+1}\\
V^{m,j}_{T+1}(\underline{\pi}_{T+1},z_{T+1},x_{T+1}^{m,j}) &\defeq 0.   \label{eq:VT+1}\\
V^f_{T+1}(\underline{\pi}_{T+1},z_{T+1},x_{T+1}^f) &\defeq 0
}
\item[2.] For $t = T,T-1, \ldots 1, \ \forall \underline{\pi}_{t}\in\prod_{i=1}^N\mathcal{P}(\mathcal{X}^{l,i})\times\prod_{j=1}^M\mathcal{P}(\mathcal{X}^{m,j}), z_{t}\in \mathcal{P}(\cX^f)$, 
For a given $\underline{\pi_t},z_t,\gamma_t^{l},\gamma_t^{m}$, define $\bar{BR}_t^f(\underline{\pi_t},z_t,\gamma_t^{l},\gamma_t^{m})$ as follows, 
\seq{
\label{eq:SMFE_FP}
  \eq{
 &\bar{BR}_t^f(\underline{\pi_t},z_t,\gamma_t^{l,m}) :=\big\{ \tgamma_t^f: \forall x_t^f\in \cX^f, \tgamma_t^f(\cdot|x_t^f)\in  \arg\max_{\gamma^{m,j}_t(\cdot|x_t^{m,j})}\E^{\gamma^{m,j}_t(\cdot|x_t^{m,j}) {\gamma}^{l,m}_t,\,z_t,\underline{\pi_t}}\nn\\
 &  
\big\{ R_t^f(z_t, X_t,A_t) +\delta V_{t+1}^f(\underline{F}(\underline{\pi_t},z_t,\gamma_t^{l,m},A^{l,m}_t),\phi(\underline{\pi_t}, z_t,\gamma_t^{l,m},\tilde{\gamma}^f_t), X^f_{t+1}) \big\lvert \underline{\pi_t}, z_t,x_t^f \big\}  \big\}, \label{eq:m_FP}
}
where expectation in (\ref{eq:m_FP}) is with respect to random variables $(X^{l,m}_t,A_t,X^f_{t+1})$ through the measure\\
$\underline{\pi_t}(x^{l,m}_t)\gamma^f_t(a^f_t|x^f_t) {\gamma}^{l,m}_t(a^{l,m}_t|x^{l,m}_t)$ $Q^f(x^f_{t+1}|z_t,x^{l,m}_t,x^f_t,a^{l,m}_t,a_t^f)$ and $\phi$ is defined in \eqref{eq:phi_def}.

Similarly define $\forall \underline{\pi}_{t}\in\prod_{i=1}^N\mathcal{P}(\mathcal{X}^{l,i})\times\prod_{j=1}^M\mathcal{P}(\mathcal{X}^{m,j}), z_{t}\in \mathcal{P}(\cX^f)$, 
$\gamma_t^{l},\gamma_t^{m,-j}\gamma_t^{f}$, define\\ $\bar{BR}_t^{m,j}(\underline{\pi_t},z_t,\gamma_t^{l},\gamma_t^{m,-j},\gamma_t^f)$ as follows, 
\eq{
 &\bar{BR}_t^{m,j}(\underline{\pi_t},z_t,\gamma_t^{l},\gamma_t^{m,-j},\gamma_t^f) :=\big\{ \tgamma_t^{m,j}: \forall x_t^{m,j}\in \cX^{m,j}, \tgamma_t^{m,j}(\cdot|x_t^{m,j})\in  \arg\max_{\gamma^{m,j}_t(\cdot|x_t^{m,j})}\E^{\gamma^{m,j}_t(\cdot|x_t^{m,j}) {\gamma}^{l}_t,\gamma_t^{m,-j},\gamma_t^f,\,z_t,\underline{\pi_t}} \nn\\
 & 
\big\{ R_t^f(z_t, X^{l,m,f}_t,A_t) +\delta V_{t+1}^f(\underline{F}(\underline{\pi_t},z_t,\gamma_t^{l},\tgamma_t^{m,j},\gamma_t^{m,-j},A^{l,m}_t),\phi(\underline{\pi_t}, z_t,\gamma_t^{l},\tgamma_t^{m,j},\gamma_t^{m,-j},{\gamma}^f_t), X^f_{t+1}) \big\lvert \underline{\pi_t}, z_t,x_t^f \big\}  \big\}, \label{eq:m_FP_mj}
}
where the expectation in (\ref{eq:m_FP}) is with respect to random variables $(x^{l,m}_t,A_t,X^f_{t+1})$ through the measure
$\underline{\pi_t}(x^{l}_t,x^{m}_t)\gamma^f_t(a^f_t|x^f_t) {\gamma}^{l,m}_t(a^{l,m}_t|x^{l,m}_t)$ $Q^f(x^f_{t+1}|z_t,x^{l,m,f}_t,a^{l,m,f}_t)$ and $\phi$ is defined in \eqref{eq:phi_def}.
}
Then let for all $\underline{\pi_t},z_t$, $\theta[\underline{\pi_t},z_t] =(\tgamma_t^{l},\tgamma_t^{m},\tgamma_t^f)$ is a solution of the following fixed-point equation (if it exists),
\seq{
\eq{
\tgamma_t^f &\in \bar{BR}_t^f(\underline{\pi_t},z_t,\tgamma_t^{l,m})  \label{eq:FP1}\\
\tgamma_t^{m} &\in \bar{BR}_t^{m}(\underline{\pi_t},z_t,\tgamma_t^{l},\tgamma_t^{m},\tgamma_t^f) \label{eq:FP1b}
}
and for all $i=1\ldots N$
\eq{
\tgamma_t^{l,i} &\in \arg\max_{\gamma_t^{l,i}}\E^{ {\gamma}^{l,i}_t,\tgamma_t^{l,-j},\hat{\gamma}_t^m,\hat{\gamma}_t^f\,z_t} \big\{ R_t^{l,i}(z_t,X^{l,m}_t,A^{l,m}_t) +\nn\\
&\hspace{3cm}\delta V_{t+1}^{l,i}(\underline{F}(\underline{\pi_t},z_t,\gamma_t^{l,i},\tgamma_t^{l,-i},\hat{\gamma}^{m}_t,A^{l,m}_t),\phi(\underline{\pi_t},z_t,\gamma_t^{l,i},\tgamma_t^{l,-i},\hat{\gamma}^{m}_t,\hat{\gamma}_t^f),X_{t+1}^{l,i})|\underline{\pi_t},z_t,x^{l,i}_t\big\},  \label{eq:FP2}\\
&\text{where } \hat{\gamma}_t^f\in \bar{BR}_t^f(\underline{\pi_t}, z_t,\gamma_t^{l,i},\tgamma_t^{l,-i},\hat{\gamma}_t^{m}), \hat{\gamma}_t^m \in \bar{BR}_t^{m}(\underline{\pi_t}, z_t,\gamma_t^{l,i},\tgamma_t^{l,-i},\hat{\gamma}_t^{m},\hat{\gamma}_t^f)
}
where the above expectation is defined with respect to random variables $(X_t^{l,m},A^{l,m}_t)$ through the measure\\ $\underline{\pi_t}(x^{l,m}_t)z_t(x^f_t)\hat{\gamma}^f_t(a^f_t|x_t^f) {\gamma}^{l,i}_t(a^{l,i}_t|x_t^{l,i}){\tgamma}^{l,-i}_t(a^{l,-i}_t|x_t^{l,-i})\hat{\gamma}^{m}_t(a^{m}_t|x_t^{m})Q^{l,m}(x_{t+1}^{l,m}|z_t,x_t^{l,m},a_t^{l,m})$.

}
Let $(\tgamma_t^{l},\tgamma_t^{m},\tgamma_t^f)$ be a tuple of solution of the above operation. Then set $\forall i,j, x_t^f,x_{t}^{l,i},x_t^{m,j}$,
 \seq{
 \label{eq:Vdef}
  \eq{
  V^f_{t}(\underline{\pi_t},z_t,x_t^f) \defeq  &\;\E^{\tilde{\gamma}^{f}_t(\cdot|x_t) \tilde{\gamma}^{l,m}_t}\big\{ {R}_t^f (z_t,X^{l,m,f}_t,A_t) + \nn\\
  &\delta V_{t+1}^f (\underline{F}(\underline{\pi_t},z_t,\tgamma^{l,m}_t,A^{l,m}_t),\phi(\underline{\pi_t},z_t,\tgamma_t), X_{t+1}^f)\big\lvert \underline{\pi_t},z_t, x_t^f \big\}. 
  \\
   V^{l,i}_{t}(\underline{\pi_t},z_t,x_t^{l,i}) \defeq  &\;\E^{\tilde{\gamma}^{f}_t \tilde{\gamma}^{l,m}_t}\big\{ {R}_t^{l,i} (z_t,X^{l,m}_T,A_t^{l,m}) + \nn\\
   &\delta V_{t+1}^{l,i} (\underline{F}(\underline{\pi_t},z_t,\tgamma^{l,m}_t,A^{l,m}_t),\phi(\underline{\pi_t},z_t,\tgamma_t),X_{t+1}^{l,i})\big\lvert \underline{\pi_t}, z_t, x_t^{l,i} \big\}\\
   V^{m,j}_{t}(\underline{\pi_t},z_t,x_t^{m,j}) \defeq  &\;\E^{\tilde{\gamma}^{f}_t \tilde{\gamma}^{l,m}_t}\big\{ {R}_t^{m,j} (z_t,X^{l,m}_T,A_t^{l,m}) + \nn\\
   &\delta V_{t+1}^{m,j} (\underline{F}(\underline{\pi_t},z_t,\tgamma^{l,m}_t,A^{l,m}_t),\phi(\underline{\pi_t},z_t,\tgamma_t),X_{t+1}^{m,j})\big\lvert \underline{\pi_t}, z_t, x_t^{m,j} \big\}
   }
   }
   \end{itemize}

Based on $\theta$ defined in the backward recursion above, we now construct a set of strategies $\tsigma$ through forward induction as follows. 

For $t =1,2 \ldots T,i,j \underline{\pi_t}, z_{t}, x_{1:t}^f \in(\cX^f)^t,x_{1:t}^{l} \in(\cX^{l})^t,x_{1:t}^{m} \in(\cX^{m})^t,a_{1:t-1}^{l}\in(\cA^{l})^{t-1},a_{1:t-1}^{m}\in(\cA^{m})^{t-1}$
\eq{
\pi_1(x_1^{l,m}) &:= Q^{l,m}(x^{l}_1,x^{m}_1)\nn\\
\tsigma_{t}^{f}(a_{t}^{f}|z_{1:t},a_{1:t-1}^{l,m}, x_{1:t}^{f}) &:= \theta_{t}^{f}[\underline{\pi_t},z_t](a^{f}_{t}| x_{t}^{f})\\
\tsigma_{t}^{l,i}(a_{t}^{l,i}|z_{1:t},a_{1:t-1}^{l,i},x_{1:t}^{l,i}) &:= \theta_{t}^{l,i}[\underline{\pi_t},z_t](a^{l,i}_{t}|x_t^{l,i})  \\
\tsigma_{t}^{m,j}(a_{t}^{m,j}|z_{1:t},a_{1:t-1}^{m,j},x_{1:t}^{m,j}) &:= \theta_{t}^{m,j}[\underline{\pi_t},z_t](a^{m,j}_{t}|x_t^{m,j})  \\
\pi_{t+1} &= \underline{F}(\underline{\pi_t},z_t,\theta_t^{l,m}[\underline{\pi_t},z_t],a^{l,m}_t)\\
z_{t+1} &= \phi(\underline{\pi_t},z_t,\theta^{l,m}_t[\underline{\pi_t},z_t])
}

\begin{theorem}
\label{Thm:Main}
A strategy profile $\tsigma$, as constructed through backward-forward recursion algorithm above is an SMFE of the game
\end{theorem}
\begin{IEEEproof}
We will prove this theorem in three parts. In Part 1 for the minor follower, we prove that $\tsigma^f \in BR^f(z,\tsigma^{l},\tsigma^{m})$ i.e. $\ \forall \ t\in[T]$, $\forall \sigma^f, z_{1:t},a_{1:t-1}^{l,m},x_{1:t}^f$
\eq{
 &\E^{\tsigma_{t:T}^{l,m},\tsigma_{t:T}^f,\underline{\pi_t}} \big\{ \sum_{n=t}^T \delta^{n-t}R_n^f(Z_n,X_n,A_n) |\underline{\pi_t},z_{1:t},a_{1:t-1}^{l,m},x_{1:t}^f\big\} \geq \nn\\
 &\E^{\tsigma_{t:T}^{l,m},\sigma_{t:T}^f,\underline{\pi_t}} \big\{ \sum_{n=t}^T \delta^{n-t}R_n^f(Z_n,X_n,A_n) |\underline{\pi_t},z_{1:t},a_{1:t-1}^{l,m},x_{1:t}^f\big\} \label{eq:Thm_f}.
}

In Part 2 for the major follower j, we prove that $\tsigma^{m,j} \in BR^f(z,\tsigma^{l},\tsigma^{m,-j},\tsigma_t^f)$ i.e. $\ \forall \ t\in[T]$, $\forall \sigma^{m,j}, z_{1:t},a_{1:t-1}^{l,m},x_{1:t}^{m,j}$
\eq{
 &\E^{\tsigma_{t:T}^{l,m},\tsigma_{t:T}^f,\underline{\pi_t}} \big\{ \sum_{n=t}^T \delta^{n-t}R_n^f(Z_n,X_n,A_n) |\underline{\pi_t},z_{1:t},a_{1:t-1}^{l,m},x_{1:t}^{m,j}\big\} \geq \nn\\
 &\E^{\tsigma_{t:T}^{l},\sigma_{t:T}^{m,j},\tsigma_{t:T}^{m,-j},\tsigma_{t:T}^f,\underline{\pi_t}} \big\{ \sum_{n=t}^T \delta^{n-t}R_n^f(Z_n,X_n,A_n) |\underline{\pi_t},z_{1:t},a_{1:t-1}^{l,m},x_{1:t}^{m,j}\big\} \label{eq:Thm_f}.
}

In Part 3 for the leader $i$, we show that $\forall z,t,\sigma^{l,i},a_{1:t-1}^{l,m},x_{1:t}^{l,i}$
\eq{
&\E^{\tsigma_{t:T}^{l},{\tsigma}_{t:T}^{m},\tsigma_{t:T}^f,\underline{\pi_t}} \big\{ \sum_{n=t}^T \delta^{n-t}R_n^{l,i}(Z_n,X_n^{l,m},A_n^{l,m}) |\underline{\pi_t},z_{1:t},a_{1:t-1}^{l,m},x_{1:t}^{l,i}\big\} \geq\nn\\
 &\E^{\sigma_{t:T}^{l,i},\tsigma^{l,-i}_{t:T},\hat{\sigma}_{t:T}^{m},\hat{\sigma}_{t:T}^f,\underline{\pi_t}} \big\{ \sum_{n=t}^T \delta^{n-t}R_n^{l,i}(Z_n,X_n^{l,m},A_n^{l,m}) |\underline{\pi_t},z_{1:t},a_{1:t-1}^{l,m},x_{1:t}^{l,i}\big\},\\
 &\text{where } \hat{\sigma}^f\in BR^f(z,\sigma_{t:T}^{l,i},\tsigma^{l,-i}_{t:T},\hat{\sigma}_{t:T}^{m},\hat{\sigma}_{t:T}^{f}),\hat{\sigma}^m\in BR^m(z,\sigma_{t:T}^{l,i},\tsigma^{l,-i}_{t:T},\hat{\sigma}_{t:T}^{m},\hat{\sigma}_{t:T}^{f})
}
where $\tsigma^f\in BR^f(z,\tsigma^{l},\tsigma^{m})$, as shown in Part 1.

Finally the process $z_{1:T}$ is consistent with $\tsigma^{l},\tsigma^{m},\tsigma^f$ such that $z=\Lambda(\tsigma^{l},\tsigma^{m},\tsigma^f)$. 
Combining the above parts prove the above result. The proof is presented in Appendix~C.
\end{IEEEproof}

In the following, we show that every Stackelberg mean field equilibrium can be found using the above backward recursion. This also enables us to comment on the existence of the solution of the fixed-point equation~\eqref{eq:SMFE_FP}.

\begin{theorem}
Suppose there exists an SMFE $(\tilde{\sigma}^{l},\tsigma^m,\tsigma^f,z)$ that is a solution of the fixed point equation defined in Definition~\ref{def:MFE}. Then there exists an equilibrium generating function $\theta$ that satisfies \eqref{eq:SMFE_FP} in backward recursion $\forall \underline{\pi_t},z_t$
	such that  $(\tilde{\sigma}^{l},\tsigma^m,\tsigma^f,z)$ is defined through forward recursion using $\theta$. This also implies that there exists a solution of~\eqref{eq:SMFE_FP} for each time $t$.
\end{theorem}
\begin{IEEEproof}
Suppose there exists an SMFE $(\tilde{\sigma}^{l,m},\tsigma^f,z)$ of the game. The proof in Appendix~\ref{app:Proof_Exist} show that all SMFE can be found using backward/forward recursion. This proves that there exists a solution of~\eqref{eq:SMFE_FP} for every $t$.
\end{IEEEproof}

\textbf{Remark:}
When leader $i$ is social welfare maximizing, her utility can be given by 
\eq{
R^{l,i}(z_t,x_t^{l,i},a_t^{l,i},\gamma^{m,f}_t) &= \sum_{\substack{x_t^{l,-i},x_t^{m},x^f,\\a_t^{l,-i},a_t^m,a_t^f}}z_t(x_t^f)\underline{\pi_t}(x^{l}_t,x^{m}_t)\gamma_t^f(a_t^f|x_t^f)R^f(z_t,x_t,a_t) + \nn\\
&\sum_{j=1}^M \pi_t^{m,j}(x_t^{m,j})\gamma_t^{m,j}(a_t^{m,j}|x_t^{m,j})R^{m,j}(z_t,x_t^{m,j},a_t^{m,j}).
}

\section{ Special case 1: When the number of leaders is infinite}

In this subsection, we consider the special case when the number of leaders is infinite and there are no major followers. All the leaders are homogeneous and affect each other and the followers through a mean field statistics of their own
Let $\xi_t$ be a mean field of the leaders where
\eq
{\xi_t(x)= \lim_{N\to\infty}\frac{1}{N}\sum_{i=1}^N 1(x_t^{l,i}=x)
}

Then, as was the case for the followers, the leaders' mean field can be updated as follows
\eq{\xi_{t+1}(\cdot) = \sum_{x_t,a_t}z_t(x_t^f)\xi_t(x^{l}_t) Q^l( \cdot|z_t,\xi_t,x_t^{l}, a_t^{l})\gamma^f_t(a^f_t|x_t^f)\gamma_t^{l,m}(a^{l,m}_t|x^{l,m}_t) \label{eq:eta_def_1}
}
We also call the above equation as $\xi_{t+1} = \eta(\xi_t,z_t,\gamma_t)$

We adapt the backward forward recursion described in previous section to this case as follows.

\subsection{Backward Recursion}

In this section, we define an equilibrium generating function $\theta=(\theta^{l}_t,\theta^{f}_t)_{t\in[T]}$, where $\theta^{l}_t :  \mathcal{P}(\cX^{l})\times \cP(\cX^f) \to \big\{\cX^{{l}} \to \mathcal{P}(\cA^{l}) \big\}$, $\theta^{f}_t :  \mathcal{P}(\cX^{l}))\times \mathcal{P}(\cX^f) \to \big\{\cX^{f} \to \mathcal{P}(\cA^{f}) \big\}$
and a sequence of functions $(V_t^{l},V_t^f)_{t\in \{ 1,2, \ldots T+1\}}$, where $V^{l}_t : \mathcal{P}(\cX^{l})\times \cP(\cX^{f})\times \cX^{l} \to \mathbb{R}$, $V^{f}_t :\mathcal{P}(\cX^{l})\times  \mathcal{P}(\cX^f)\times \cX^f \to \mathbb{R}$, in a backward recursive way, as follows. 
\begin{itemize}
\item[1.] Initialize $\forall \xi_{T+1}\in\mathcal{P}(\mathcal{X}^{l}), z_{T+1}\in \mathcal{P}(\cX^f), x_{T+1}^{l}\in \cX^{l}, x_{T+1}^{f}\in \cX^f,$
\eq{
V^{l}_{T+1}(\xi_{T+1},z_{T+1},x_{T+1}^{l}) &\defeq 0.   \label{eq:VT+1_1}\\
V^f_{T+1}(\xi_{T+1},z_{T+1},x_{T+1}^f) &\defeq 0
}
\item[2.] For $t = T,T-1, \ldots 1, \ \forall \xi_{t}\in\mathcal{P}(\mathcal{X}^{l}), z_{t}\in \mathcal{P}(\cX^f)$, 
Set $\tilde{\gamma}_t = \theta_t[\xi_t,z_t]$, where $\tilde{\gamma}_t =( \tgamma_t^{l},\tilde{\gamma}_t^f)$ is the solution of the following fixed-point equation. For a given $\xi_t,z_t,\gamma_t^{l}$, define $\bar{BR}_t^f(\xi_t,z_t,\gamma_t^{l})$ as follows, 
  \eq{
 &\bar{BR}_t^f(\xi_t,z_t,\gamma_t^{l}) :=\big\{ \tgamma_t^f: \forall x_t^f\in \cX^f, \tgamma_t^f(\cdot|x_t^f)\in  \arg\max_{\gamma^{m,j}_t(\cdot|x_t^{m,j})}\nn\\
 & \E^{\gamma^{m,j}_t(\cdot|x_t^{m,j}) {\gamma}^{l}_t,\,z_t,\xi_t} 
\big\{ R_t^f(z_t, X_t^{l,f}, A_t) +\delta V_{t+1}^f(\eta(\xi_t,z_t,\gamma_t),\phi(\xi_t, z_t,\gamma_t^{l},\tilde{\gamma}^f_t), X^f_{t+1}) \big\lvert \xi_t,z_t,x_t^f \big\}  \big\}, \label{eq:m_FP_1}
}
where expectation in (\ref{eq:m_FP}) is with respect to random variables $(X_t^{l},A_t,X^f_{t+1})$ through the measure\\
$\xi_t(x_t^{l})\gamma^f_t(a^f_t|x^f_t) {\gamma}^{l}_t(a^{l}_t|x_t^{l})$ $Q^f(x^f_{t+1}|\xi_t,z_t,x^f_t,a_t^{l},a_t^f)$ and $\phi$ is defined in \eqref{eq:phi_def} and $\eta$ is defined in \eqref{eq:eta_def_1}.

Similarly define $\forall \xi_{t}\in\mathcal{P}(\mathcal{X}^{l}), z_{t}\in \mathcal{P}(\cX^f)$, 
$\gamma_t^{l}$, define $\bar{BR}_t^f(\xi_t,z_t,\gamma_t^{l})$ as follows, where expectation in (\ref{eq:m_FP_2}) is with respect to random variables $(X_t^{l},A_t,X^f_{t+1})$ through the measure\\
$\xi_t(x_t^{l})\gamma^f_t(a^f_t|x^f_t) {\gamma}^{l}_t(a^{l}_t|x_t^{l})$ $Q^f(x^f_{t+1}|\xi_t,z_t,x_t,a_t)$ and $\phi$ is defined in \eqref{eq:phi_def} and $\eta$ is defined in \eqref{eq:eta_def_1}.

Then let for all $\xi_t,z_t$, $\theta[\xi_t,z_t] =(\tgamma_t^{l},\tgamma_t^f)$ is a solution of the following fixed-point equation (if it exists),
\seq{
\eq{
\tgamma_t^f &\in \bar{BR}_t^f(\xi_t,z_t,\tgamma_t^{l})  \label{eq:FP1_1}
}
and 
\eq{
\tgamma_t^{l} &\in \arg\max_{\gamma_t^{l}}\E^{ \hat{\gamma}_t^f{\gamma}^{l}_t,\,z_t} \big\{ R_t^{l}(z_t,X^{l}_t,A_t^{l}) +\nn\\
&\hspace{3cm}\delta V_{t+1}^{l,i}(\eta(\xi_t,z_t,\gamma^{l}_t),\phi(\xi_t,z_t,\gamma_t^{l},\hat{\gamma}_t^f),X_{t+1}^{l})|\xi_t,z_t,x_t^{l,i}\big\},  \label{eq:FP2_1}\\
&\text{where } \hat{\gamma}_t^f\in \bar{BR}_t^f(\xi_t, z_t,\gamma_t^{l}),
}
where the above expectation is defined with respect to random variables $(X_t^{l},X^f_t,A_t)$ through the measure $\xi_t(x_t^{l})z_t(x^f_t)\hat{\gamma}^f_t(a^f_t|x_t^f) {\gamma}^{l}_t(a^{l}_t|x_t^{l})Q^{l}(x_{t+1}^{l}|\xi_t,z_t,x_t^{l},x_t^f,a_t^{l})$, and $\hat{\gamma}^f_t\in BR_t^f(\xi_t,z_t,\gamma_t^{l})$.
\label{eq:SMFE_FP_1}
}
Let $(\tgamma_t^{l},\tgamma_t^f)$ be a pair of solution of the above operation. Then set $\forall x_t^f\in\cX^f$,
 \seq{
 \label{eq:Vdef_1}
  \eq{
  V^f_{t}(\xi_t,z_t,x_t^f) \defeq  &\;\E^{\tilde{\gamma}^{f}_t(\cdot|x_t) \tilde{\gamma}^{l}_t}\big\{ {R}_t^f (z_t,\xi_t,X^{f}_t,A_t) + \nn\\
  &\delta V_{t+1}^f (\eta(\xi_t,z_t,\tgamma^{l}_t,A_t),\phi(\xi_t,z_t,\tgamma_t^f, \tilde{\gamma}^{l}_t), X_{t+1}^f)\big\lvert \xi_t,z_t, x_t^f \big\}.
  \\
   V^{l}_{t}(\xi_t,z_t,x_t^{l}) \defeq  &\;\E^{\tilde{\gamma}^{f}_t \tilde{\gamma}^{l}_t}\big\{ {R}_t^{l} (z_t,X^{l}_t,A_t^{l}) + \nn\\
   &\delta V_{t+1}^{l} (\eta(\xi_t,z_t,\tgamma^{l}_t,A_t),\phi(\xi_t,z_t,\tgamma_t^{l}, \tilde{\gamma}^f_t),X_{t+1}^{l})\big\lvert \xi_t, z_t, x_t^{l} \big\}
   }
   }
   \end{itemize}

Based on $\theta$ defined in the backward recursion above, we now construct a set of strategies $\tsigma$ through forward induction as follows. 

For $t =1,2 \ldots T, \xi_t, z_{t}, x_{1:t}^f \in(\cX^f)^t,x_{1:t}^{l} \in(\cX^{l})^t,a_{1:t-1}^{l}\in(\cA^{l})^{t-1}$
\eq{
\pi_1(x_1^{l}) &:= Q^{l}(x_t^{l})\nn\\
\tsigma_{t}^{f}(a_{t}^{f}|z_{1:t},a_{1:t-1}^{l}, x_{1:t}^{f}) &:= \theta_{t}^{f}[\xi_t,z_t](a^{f}_{t}| x_{t}^{f})\\
\tsigma_{t}^{l}(a_{t}^{l}|z_{1:t},a_{1:t-1}^{l},x_{1:t}^{l}) &:= \theta_{t}^{l}[\xi_t,z_t](a^{l}_{t}|x_t^{l})  \\
\xi_{t+1} &= \eta(\xi_t,z_t,\theta^{l,f}_t[\xi_t,z_t])\\
z_{t+1} &= \phi(\xi_t,z_t,\theta^{l,f}_t[\xi_t,z_t])
}
\begin{theorem}
\label{Thm:Main_1}
A strategy profile and mean field beliefs $(\tsigma^l,\tsigma^f,z)$, as constructed through backward-forward recursion algorithm above is an SMFE of the game
\end{theorem}
\begin{IEEEproof}
The proofs are similar in spirit to the proofs of Theorem~1 where $\underline{\pi}_t$ is replaced by $\xi_t$ and are thus skipped.
\end{IEEEproof}

\section{Special Case 2: When the leaders and the major players don't have private states}
In this section, we consider the special case when the leaders and the major players don't have private states. The algorithm in previous section simplifies as follows.
\subsection{Algorithm for SMFE computation: Backward Recursion}
\label{s_sec:Result_2}

In this section, we define an equilibrium generating function $\theta=(\theta^{l,i}_t,\theta_t^{m,j},\theta^{f}_t)_{i\in\{1\ldots K\}, j\in\{1\ldots M, \},t\in[T]}$, where $\theta^{l,i}_t :   \cP(\cX^f)\to \big\{\cX^{{l,i}} \to \mathcal{P}(\cA^{l,i}) \big\}$, $\theta^{m,j}_t :   \cP(\cX^f)\times \cX^{m,j} \to \big\{\cX^{{m,j}} \to \mathcal{P}(\cA^{m,j}) \big\}$, $\theta^{f}_t :   \mathcal{P}(\cX^f)\times \cX^f \to \big\{\cX^{f} \to \mathcal{P}(\cA^{f}) \big\}$
and a sequence of functions $(V_t^{l,i}, V_t^{m,j},V_t^f)_{ \{i=1\ldots N,j=1\ldots, M, 1,2,t=1 \ldots T+1\}}$, where $V^{l,i}_t :   \cP(\cX^{f})\times \cX^{l,i} \to \mathbb{R}$, $V^{m,j}_t :  \cP(\cX^f)\times  \cX^{m,j} \to \mathbb{R}, V^{f}_t :   \mathcal{P}(\cX^f)\times \cX^f \to \mathbb{R} $, in a backward recursive way, as follows. 
\begin{itemize}
\item[1.] Initialize $\forall z_{T+1}\in \mathcal{P}(\cX^f), i=1\ldots K, j=1\ldots M,  x_{T+1}^{f}\in \cX^f,$
\eq{
V^{l,i}_{T+1}(z_{T+1}) &\defeq 0.   \label{eq:VT+1}\\
V^{m,j}_{T+1}(z_{T+1}) &\defeq 0.   \label{eq:VT+1}\\
V^f_{T+1}(z_{T+1},x_{T+1}^f) &\defeq 0
}
\item[2.] For $t = T,T-1, \ldots 1, \ \forall z_{t}\in \mathcal{P}(\cX^f)$, 
Set $\tilde{\gamma}_t = \theta_t[z_t]$, where $\tilde{\gamma}_t =( \tgamma_t^{l},\tgamma_t^{m},\tilde{\gamma}_t^f)$ is the solution of the following fixed-point equation. For a given $z_t,\gamma_t^{l},\gamma_t^{m}$, define $\bar{BR}_t^f(z_t,\gamma_t^{l},\gamma_t^{m})$ as follows, 
\seq{
  \eq{
 &\bar{BR}_t^f(z_t,\gamma_t^{l},\gamma_t^{m}) :=\big\{ \tgamma_t^f: \forall x_t^f\in \cX^f, \tgamma_t^f(\cdot|x_t^f)\in  \arg\max_{\gamma^{m,j}_t(\cdot|x_t^{m,j})}\nn\\
 & \E^{\gamma^{m,j}_t(\cdot|x_t^{m,j}) {\gamma}^{l}_t,\,z_t} 
\big\{ R_t^f(z_t, X^f_t,A_t) +\delta V_{t+1}^f(\phi( z_t,\gamma_t^{l},\gamma_t^{m},\tilde{\gamma}^f_t), X^f_{t+1}) \big\lvert z_t,x_t^f \big\}  \big\}, \label{eq:m_FP_2}
}
where expectation in (\ref{eq:m_FP_2}) is with respect to random variables $(A_t,X^f_{t+1})$ through the measure\\
$\gamma^f_t(a^f_t|x^f_t) {\gamma}^{l}_t(a^{l}_t)$ $Q^f(x^f_{t+1}|z_t,x^f_t,A^{l,m}_t,a_t^f)$ and $\phi$ is defined in \eqref{eq:phi_def}.

Similarly define $\forall z_{t}\in \mathcal{P}(\cX^f)$, 
$\gamma_t^{l},\gamma_t^{m,-j},\gamma_t^{f}$, define $\bar{BR}_t^f(z_t,\gamma_t^{l},\gamma_t^{m,-j},\gamma_t^f)$ as follows, 

  \eq{
 &\bar{BR}_t^{m,j}(z_t,\gamma_t^{l},\gamma_t^{m,-j},\gamma_t^f) :=\big\{ \tgamma_t^{m,j}: \forall \tgamma_t^{m,j}\in  \arg\max_{\gamma^{m,j}_t}\nn\\
 & \E^{\gamma^{m,j}_t {\gamma}^{l}_t,\gamma_t^{m,-j},\gamma_t^f,\,z_t} 
\big\{ R_t^{m,j}(z_t,A_t^{l,m}) +\delta V_{t+1}^{m,j}(\phi( z_t,\gamma_t^{l},\tgamma_t^{m,j},\gamma_t^{m,-j},{\gamma}^f_t)) \big\lvert z_t \big\}  \big\}, \label{eq:m_FP_2}
}
where expectation in (\ref{eq:m_FP_2}) is with respect to random variables $(A_t,X^f_{t+1})$ through the measure\\
$\gamma^f_t(a^f_t|x^f_t) {\gamma}^{l}_t(a^{l}_t)\gamma_t^m(a_t^m)$ $Q^f(x^f_{t+1}|z_t,x^f_t,a^{l,f}_t,a_t^f)$ and $\phi$ is defined in \eqref{eq:phi_def}.
}
Then let for all $\theta[z_t] =(\tgamma_t^{l},\tgamma_t^{m},\tgamma_t^f)$ is a solution of the following fixed-point equation (if it exists),
\seq{
\eq{
\tgamma_t^f &\in \bar{BR}_t^f(z_t,\tgamma_t^{l},\tgamma_t^{m})  \label{eq:FP1_2}\\
\tgamma_t^{m} &\in \bar{BR}_t^{m}(z_t,\tgamma_t^{l},\tgamma_t^{m},\tgamma_t^f) 
}
and 
\eq{
\tgamma_t^{l,i} &\in \arg\max_{\gamma_t^{l,i}}\E^{ {\gamma}^{l,i}_t,{\gamma}^{l,-i}_t, {\gamma}^{m}_t,\hat{\gamma}_t^f\,z_t} \big\{ R_t^{l,i}(z_t,A^{l,m}_t)+ \delta V_{t+1}^{l,i}(\phi(z_t,\gamma_t^{l,i},\tgamma_t^{l,-i},\hat{\gamma}_t^{m},\hat{\gamma}_t^f))|z_t\big\},  \label{eq:FP2_2}\\
&\text{where } \hat{\gamma}_t^f\in \bar{BR}_t^f( z_t,\gamma_t^{l,i},\tgamma_t^{l,-i},\hat{\gamma}_t^{m}), \hat{\gamma}_t^m \in \bar{BR}_t^{m}( z_t,\gamma_t^{l,i},\tgamma_t^{l,-i},\hat{\gamma}_t^{m},\hat{\gamma}_t^f)
}
where the above expectation is defined with respect to random variables $(X^f_t,A_t)$ through the measure $z_t(x^f_t)\hat{\gamma}^f_t(a^f_t|x_t^f) {\gamma}^{l}_t(a^{l}_t) \gamma_t^m(a_t^m)$, and $\hat{\gamma}^f_t\in BR_t^f(z_t,\gamma_t^{l},\gamma_t^{m})$.
\label{eq:SMFE_FP_2}
}
Let $(\tgamma_t^{l},\tgamma_t^{m},\tgamma_t^f)$ be a pair of solution of the above operation. Then set $\forall x_t^f\in\cX^f$,
 \seq{
 \label{eq:Vdef_2}
  \eq{
  V^f_{t}(z_t,x_t^f) \defeq  &\;\E^{\tilde{\gamma}^{f}_t(\cdot|x_t^f) \tilde{\gamma}^{l}_t}\big\{ {R}_t^f (z_t,X^f_t,A_t) + \delta V_{t+1}^f (\phi(z_t,\tgamma_t^f, \tilde{\gamma}^{l,m}_t), X_{t+1}^f)\big\lvert z_t, x_t^f \big\}. 
  \\
   V^{l,i}_{t}(z_t) \defeq  &\;\E^{\tilde{\gamma}^{f}_t \tilde{\gamma}^{l}_t}\big\{ {R}_t^{l,i} (z_t,A^{l,m}_t) + \delta V_{t+1}^{l,i} (\phi(z_t,\tgamma^{l}_t,A^{m}_t,A_t),z_t,\tgamma_t^{l},\tgamma_t^{m}, \tilde{\gamma}^f_t))\big\lvert z_t\big\}\\
   V^{m,j}_{t}(z_t) \defeq  &\;\E^{\tilde{\gamma}^{f}_t \tilde{\gamma}^{l}_t}\big\{ {R}_t^{m,j} (z_t,A^{l,m}_t) + \delta V_{t+1}^{m,j} (\phi(z_t,\tgamma_t^{l},\tgamma_t^{m}, \tilde{\gamma}^f_t))\big\lvert  z_t \big\}
   }
   }
   \end{itemize}

Based on $\theta$ defined in the backward recursion above, we now construct a set of strategies $\tsigma$ through forward induction as follows. 

For $t =1,2 \ldots T,  z_{t}, x_{1:t}^f \in(\cX^f)^t,a_{1:t-1}^{l,m}\in(\cA^{l,m})^{t-1}$
\eq{
\tsigma_{t}^{f}(a_{t}^{f}|z_{1:t},a_{1:t-1}^{l,m}, x_{1:t}^{f}) &:= \theta_{t}^{f}[z_t](a^{f}_{t}| x_{t}^{f})\\
\tsigma_{t}^{l,i}(a_{t}^{l,i}|z_{1:t},a_{1:t-1}^{l,m}) &:= \theta_{t}^{l,i}[z_t](a^{l,i}_{t})  \\
\tsigma_{t}^{m,j}(a_{t}^{m,j}|z_{1:t},a_{1:t-1}^{l,m}) &:= \theta_{t}^{m,j}[z_t](a^{m,j}_{t})  \\
z_{t+1} &= \phi(z_t,\theta^{l,m}_t,A^{l,m}_t[z_t])
}
\begin{corollary}
\label{Thm:Main2}
A strategy profile and mean field beliefs $\tsigma,z$, as constructed through backward-forward recursion algorithm above is an SMFE of the game
\end{corollary}
\begin{IEEEproof}
The result is implied by Theorem~1.
\end{IEEEproof}

\section{Special case 3: Infinite horizon case}
In this section we consider the case with infinite horizon. For this section we assume that the instantaneous rewards of the players $R^{l,i},R^{m,j},R^{f}$ are absolutely bounded and do not depend on time.

We design an algorithm to compute SMFE-ML of the infinite horizon game as follows.

\subsection{Backward Recursion}

In this section, we define an equilibrium generating function $\theta=(\theta^{l,i},\theta^{m,j},\theta^{f})_{i\in\{1\ldots N\}, j\in\{1\ldots M, \}}$, where $\theta^{l,i} :  (\prod_{i=1}^N\mathcal{P}(\cX^{l,i}))\times(\prod_{j=1}^M\cP(\cX^{m,j}))\times \cP(\cX^f) \to \big\{\cX^{{l,i}} \to \mathcal{P}(\cA^{l,i}) \big\}$, $\theta^{m,j} :  (\prod_{i=1}^K\mathcal{P}(\cX^{l,i}))\times(\prod_{j=1}^M\cP(\cX^{m,j}))\times \cP(\cX^f) \to \big\{\cX^{{m,j}} \to \mathcal{P}(\cA^{m,j}) \big\}$, $\theta^{f} :  (\prod_{i=1}^N\mathcal{P}(\cX^{l,i}))\times(\prod_{j=1}^M\cP(\cX^{m,j}))\times  \mathcal{P}(\cX^f) \to \big\{\cX^{f} \to \mathcal{P}(\cA^{f}) \big\}$
and a sequence of functions $(V^{l,i}, V^{m,j},V^f)$, where $V^{l,i} :  (\prod_{i=1}^N\mathcal{P}(\cX^{l,i}))\times(\prod_{j=1}^M\cP(\cX^{m,j}))\times \cP(\cX^{f})\times \cX^{l,i} \to \mathbb{R}$, $V^{m,j} :  (\prod_{i=1}^N\mathcal{P}(\cX^{l,i}))\times(\prod_{j=1}^M\cP(\cX^{m,j}))\times\cP(\cX^f)\times  \cX^{m,j} \to \mathbb{R}$,$V^{f} :  (\prod_{i=1}^N\mathcal{P}(\cX^{l,i}))\times(\prod_{j=1}^M\cP(\cX^{m,j}))\times  \mathcal{P}(\cX^f)\times \cX^f \to \mathbb{R}$, in a backward recursive way, as follows. 
\begin{itemize}
\item[1.] For $ \forall \underline{\pi}\in\prod_{i=1}^N\mathcal{P}(\mathcal{X}^{l,i})\times\prod_{j=1}^M\mathcal{P}(\mathcal{X}^{m,j}), z\in \mathcal{P}(\cX^f)$, 
For a given $\underline{\pi},z,\gamma^{l,m}$, define $\bar{BR}^f(\underline{\pi},z,\gamma^{l,m})$ as follows, 
\seq{
  \eq{
 &\bar{BR}^f(\underline{\pi},z,\gamma^{l,m}) :=\big\{ \tgamma^f: \forall x^f\in \cX^f, \tgamma^f(\cdot|x^f)\in  \arg\max_{\gamma^f(\cdot|x^f)}\E^{\gamma^f(\cdot|x^f) {\gamma}^{l,m},\,z,\underline{\pi}}\nn\\
 &  
\big\{ R^f(z_t, X_t,A_t) +\delta V^f(\underline{F}(\underline{\pi},z,\gamma^{l,m},A^{l,m}),\phi(\underline{\pi}, z,\gamma^{l,m},\tilde{\gamma}^f), X^{f,'}) \big\lvert \underline{\pi}, z,x^f \big\}  \big\}, \label{eq:m_FP3}
}
where expectation in (\ref{eq:m_FP3}) is with respect to random variables $(X^{l,m},A,X^{f,'})$ through the measure\\
$\underline{\pi_t}(x^{l,m}_t)\gamma^f_t(a^f_t|x^f_t) {\gamma}^{l,m}_t(a^{l,m}_t|x^{l,m}_t)$ $Q^f(x^{f,'}|z_t,x^{l,m}_t,x^f_t,a^{l,m}_t,a_t^f)$ and $\phi$ is defined in \eqref{eq:phi_def}.

Similarly define $\forall \underline{\pi}_{t}\in\prod_{i=1}^N\mathcal{P}(\mathcal{X}^{l,i})\times\prod_{j=1}^M\mathcal{P}(\mathcal{X}^{m,j}), z_{t}\in \mathcal{P}(\cX^f)$, 
$\gamma_t^{l},\gamma_t^{m,-j}\gamma_t^{f}$, define $\bar{BR}_t^{m,j}(\underline{\pi},z,\gamma^{l},\gamma^{m,-j},\gamma^f)$ as follows, 
\eq{
 &\bar{BR}^{m,j}(\underline{\pi},z,\gamma^{l},\gamma^{m,-j},\gamma^f) :=\big\{ \tgamma^{m,j}: \forall x^{m,j}\in \cX^{m,j}, \tgamma^{m,j}(\cdot|x^{m,j})\in  \arg\max_{\gamma^{m,j}(\cdot|x^{m,j})}\E^{\gamma^{m,j}(\cdot|x^{m,j}) {\gamma}^{l}\gamma_t^{m,-j}\gamma_t^f,\,z,\underline{\pi}} \nn\\
 & 
\big\{ R^{m,j}(z, X^{l,m},A^{l,m}) +\delta V^f(\underline{F}(\underline{\pi},z,\gamma^{l},\tgamma^{m,j},\gamma_t^{m,-j},{\gamma}^f,A^{l,m}),\phi(\underline{\pi}, z,\gamma^{l},\tgamma^{m,j},\gamma_t^{m,-j},{\gamma}^f), X^{m,j,'}) \big\lvert \underline{\pi}, z,x^{m,j} \big\}  \big\}, \label{eq:m_FP_mj3}
}
where  
the expectation in (\ref{eq:m_FP_mj3}) is with respect to random variables $(X^{l,m},A,X^{f,'})$ through the measure
$\underline{\pi}(x^{l,m})\gamma^f(a^f|x^f) {\gamma}^{l,m}(a^{l,m}|x^{l,m})$ $Q^f(x^{f,'}|z,x^{l,m,f},a^{l,m,f})$ and $\phi$ is defined in \eqref{eq:phi_def}.
}
Then let for all $\underline{\pi},z$, $\theta[\underline{\pi},z] =(\tgamma^{l},\tgamma^{m},\tgamma^f)$ is a solution of the following fixed-point equation (if it exists),
\seq{
\eq{
\tgamma^f &\in \bar{BR}^f(\underline{\pi},z,\tgamma^{l,m})  \label{eq:FP1}\\
\tgamma^{m} &\in \bar{BR}^{m}(\underline{\pi},z,\tgamma^{l},\tgamma^{m},\tgamma^f) \label{eq:FP1b}
}
and for all $i=1\ldots N$
\eq{
\tgamma_t^{l,i} &\in \arg\max_{\gamma^{l,i}}\E^{ {\gamma}^{l,i},\tgamma^{l,-j},\hat{\gamma}^m,\hat{\gamma}^f\,z} \big\{ R^{l,i}(z,X^{l,m},A^{l,m}) +\nn\\
&\hspace{3cm}\delta V^{l,i}(\underline{F}(\underline{\pi},z,\gamma^{l,i},\tgamma^{l,-i},\hat{\gamma}^{m},A^{l,m}),\phi(\underline{\pi},z,\gamma^{l,i},\tgamma^{l,-i},\hat{\gamma}^{m},\hat{\gamma}^f),X^{l,i,'})|\underline{\pi},z,x^{l,i}\big\},  \label{eq:FP2}\\
&\text{where } \hat{\gamma}^f\in \bar{BR}^f(\underline{\pi}, z,\gamma^{l,i},\tgamma^{l,-i},\hat{\gamma}^{m}), \hat{\gamma}^m \in \bar{BR}^{m}(\underline{\pi}, z,\gamma^{l,i},\tgamma^{l,-i},\hat{\gamma}^{m},\hat{\gamma}_t^f),
}
where the above expectation is defined with respect to random variables $(X^{l,m},A^{l,m}_t)$ through the measure\\ $\underline{\pi}(x^{l,m})z_t(x^f)\hat{\gamma}^f(a^f|x^f) {\gamma}^{l,i}(a^{l,i}|x^{l,i}){\tgamma}^{l,-i}_t(a^{l,-i}_t|x^{l,-i})\hat{\gamma}^{m}(a^{m}|x_t^{m})Q^{l,m}(x^{l,m,'}|z,x^{l,m},a^{l,m})$.

}
Let $(\tgamma^{l},\tgamma^{m},\tgamma^f)$ be a tuple of solution of the above operation. Then set $\forall i,j, x^f,x^{l,i},x^{m,j}$,
 \seq{
 \label{eq:Vdef}
  \eq{
  V^f(\underline{\pi},z,x^f) \defeq  &\;\E^{\tilde{\gamma}^{f}(\cdot|x) \tilde{\gamma}^{l,m}}\big\{ {R}^f (z,X,A) + \nn\\
  &\delta V^f (\underline{F}(\underline{\pi},z,\tgamma^{l,m},A^{l,m}),\phi(\underline{\pi},z,\tgamma), X^f)\big\lvert \underline{\pi},z, x^f \big\}. 
  \\
   V^{l,i}(\underline{\pi},z,x^{l,i}) \defeq  &\;\E^{\tilde{\gamma}^{f} \tilde{\gamma}^{l,m}}\big\{ {R}^{l,i} (z,X^{l,m},A^{l,m}) + \nn\\
   &\delta V^{l,i} (\underline{F}(\underline{\pi},z,\tgamma^{l,m},A^{l,m}),\phi(\underline{\pi},z,\tgamma),X^{l,i,'})\big\lvert \underline{\pi}, z, x^{l,i} \big\}\\
   V^{m,j}(\underline{\pi},z,x^{m,j}) \defeq  &\;\E^{\tilde{\gamma}^{f} \tilde{\gamma}^{l,m}}\big\{ {R}^{m,j} (z,X^{l,m},A^{l,m}) + \nn\\
   &\delta V^{m,j} (\underline{F}(\underline{\pi},z,\tgamma^{l,m},A^{l,m}),\phi(\underline{\pi},z,\tgamma),X^{m,j,'})\big\lvert \underline{\pi}, z, x^{m,j} \big\}
   }
   }
   \end{itemize}

Based on $\theta$ defined in the backward recursion above, we now construct a set of strategies $\tsigma$ through forward induction as follows. 

For $t =1,2 \ldots \infty,i,j, \underline{\pi_t}, z_{t}, x_{1:t}^f \in(\cX^f)^t,x_{1:t}^{l} \in(\cX^{l})^t,x_{1:t}^{m} \in(\cX^{m})^t,a_{1:t-1}^{l}\in(\cA^{l})^{t-1},a_{1:t-1}^{m}\in(\cA^{m})^{t-1}$
\eq{
\pi_1(x_1^{l,m}) &:= Q^{l,m}(x^{l}_1,x^{m}_1)\nn\\
\tsigma_{t}^{f}(a_{t}^{f}|z_{1:t},a_{1:t-1}^{l,m}, x_{1:t}^{f}) &:= \theta_{t}^{f}[\underline{\pi_t},z_t](a^{f}_{t}| x_{t}^{f})\\
\tsigma_{t}^{l,i}(a_{t}^{l,i}|z_{1:t},a_{1:t-1}^{l,i},x_{1:t}^{l,i}) &:= \theta_{t}^{l,i}[\underline{\pi_t},z_t](a^{l,i}_{t}|x_t^{l,i})  \\
\tsigma_{t}^{m,j}(a_{t}^{m,j}|z_{1:t},a_{1:t-1}^{m,j},x_{1:t}^{m,j}) &:= \theta_{t}^{m,j}[\underline{\pi_t},z_t](a^{m,j}_{t}|x_t^{m,j})  \\
\pi_{t+1} &= \underline{F}(\underline{\pi_t},z_t,\theta_t^{l,m}[\underline{\pi_t},z_t],a^{l,m}_t)\\
z_{t+1} &= \phi(\underline{\pi_t},z_t,\theta^{l,m}_t[\underline{\pi_t},z_t])
}

\begin{theorem}
\label{Thm:Main}
A strategy profile $\tsigma$, as constructed through backward-forward recursion algorithm above is an SMFE of the game
\end{theorem}
\begin{IEEEproof}
The proof is similar to the extension of finite horizon problems to infinite horizon problems in standard stochastic control problems and for now we omit the proof.
\end{IEEEproof}

\section{Conclusion}
\label{sec:Conclusion}
In this paper, we present the equivalent of Master equation for discrete time Stackelberg mean field games with multiple leaders and both major and minor followers, where the leaders and the followers observe Markovian states privately and publicly observe a mean field population state. The leader commits to a dynamic policy that the followers respond to optimally. The leader, knowing that the followers will do best response, commits to a policy that maximizes her total expected reward. We define Stackelberg Mean field equilibrium with multiple leaders (SMFE-ML) of the game which consists of solution of a fixed-point equation across time, which consists of best response of the leader, follower and the evolution of the mean field state. We propose an algorithm to compute all SMFE-ML of the game in a sequential manner. 
\appendices

\section{}
\label{app:0}
\begin{claim}
	For any policy profile $g$ and $\forall t$,
	\eq{
	\p^{\sigma}(x_{1:t}^{}|a_{1:t-1}) =  \p^{\sigma^{l}}(x_{1:t}^{l}|z_{1:t},a_{1:t-1})\p^{\sigma^m}(x_{1:t}^{m}|z_{1:t},a_{1:t-1})\p^{\sigma^f}(x_{1:t}^{f}|z_{1:t},a_{1:t-1})
	}
	\label{claim:CondInd}
	\end{claim}
	\begin{IEEEproof}
	\seq{
	\eq{
	&\p^{\sigma}(x_{1:t}|z_{1:t},a_{1:t-1})= \frac{\p^{\sigma}(x_{1:t},z_{1:t},a_{1:t-1})}{\sum_{\bar{x}_{1:t}} \p^{\sigma}(\bar{x}_{1:t},z_{1:t},a_{1:t-1})}
	}
	Here, we will take numerator and the denominator separately.
	\eq{
	&Nr =  \left(Q_1^{l}(x^{l}_1)\sigma^{l}_1(a_1^{l}|x_{1}^{l})\prod_{n=2}^t Q_n^{l}(x^{l}_{n}|z_{n-1},a_{n-1}^{l,m},x^{l}_{n-1})1(z_n=\phi(\pi_{n-1},z_{n-1},\gamma_t)) \sigma^{l}_n(a_n^{l}|z_{1:n},a^{l,m}_{1:n-1},x_{1:n}^{l}) \right)\\
	&\times \left(Q_1^{m}(x^{m}_1)\sigma^{m}_1(a_1^{m}|x_{1}^{m})\prod_{n=2}^t Q_n^{m}(x^{m}_{n}|z_{n-1},a_{n-1}^{l,m},x^{m}_{n-1}) \sigma^{m}_n(a_n^{m}|z_{1:n},a^{l,m}_{1:n-1},x_{1:n}^{m}) \right)
	\nn\\
	&\times \left(Q_1^f(x^f_1)\sigma^f_1(a_1^f|x_{1}^{f})\prod_{n=2}^t Q_n^f(x^f_{n}|z_{n-1},a^{l,m}_{n-1},x_{n-1}^f)
	\sigma^f_n(a_n^f|z_{1:n},a^{l,m}_{1:n-1},x_{1:n}^f) \right)\\
	%
	}
	and 
	\eq{
	Dr&=\sum_{x_{1:t}^l} =  \left(Q_1^{l}(x^{l}_1)\sigma^{l}_1(a_1^{l}|x_{1}^{l})\prod_{n=2}^t Q_n^{l}(x^{l}_{n}|z_{n-1},a_{n-1}^{l,m},x^{l}_{n-1})1(z_n=\phi(\pi_{n-1},z_{n-1},\gamma_t)) \sigma^{l}_n(a_n^{l}|z_{1:n},a^{l,m}_{1:n-1},x_{1:n}^{l}) \right)\\
	&\times\sum_{x_{1:t}^m} \left(Q_1^{m}(x^{m}_1)\sigma^{m}_1(a_1^{m}|x_{1}^{m})\prod_{n=2}^t Q_n^{m}(x^{m}_{n}|z_{n-1},a_{n-1}^{l,m},x^{m}_{n-1}) \sigma^{m}_n(a_n^{m}|z_{1:n},a^{l,m}_{1:n-1},x_{1:n}^{m}) \right)
	\nn\\
	&\times \sum_{x_{1:t}^f} \left(Q_1^f(x^f_1)\sigma^f_1(a_1^f|x_{1}^{f})\prod_{n=2}^t Q_n^f(x^f_{n}|z_{n-1},a^{l,m}_{n-1},x_{n-1}^f)
	\sigma^f_n(a_n^f|z_{1:n},a^{l,m}_{1:n-1},x_{1:n}^f) \right)
	%
	}
	
	Thus
	\eq{
	\p^{\sigma}(x_{1:t}^{l},x_{1:t}^{l},x_{1:t}^f|z_{1:t},a_{1:t-1}) =  \p^{\sigma^{l}}(x_{1:t}^{l}|z_{1:t},a_{1:t-1})\p^{\sigma^{m}}(x_{1:t}^{m}|z_{1:t},a_{1:t-1})\p^{\sigma^f}(x_{1:t}^{f}|z_{1:t},a_{1:t-1})
	}
	}
	\end{IEEEproof}
	\section{}
	\label{app:gsm}
	For any player $i$ (leader, major player, or minor player), we use the notation $g$ to denote a general policy of the form $A_t^i\sim g_t^i(\cdot|z_{1:t},a_{1:t-1}, x_{1:t}^{i})$, notation $s$ to denote a policy of the form $A_t^i \sim s_t^i(\cdot|z_{1:t},a_{1:t-1},x_t^i)$, and notation $m$ to denote a policy of the form $A_t^i \sim m_t^i(\cdot|\pi_t,z_t,x_t^i)$. It should be noted that since $\pi_t$ is a function of random variables $z_{1:t},a_{1:t-1}$, $m$ policy is a special type of $s$ policy, which in turn is a special type of $g$ policy.

Using the agent-by-agent approach~\cite{Ho80}, we show in Lemma~\ref{fact:G2S} that any expected reward profile of the players that can be achieved by any general strategy profile $g$ can also be achieved by a strategy profile $s$.
\begin{lemma}
Given a fixed strategy $g^{-i}$ of all players other than player $i$ and for any strategy $g^i$ of player $i$, there exists a strategy $s^i$ of player $i$ such that $\forall t \in \mathcal{T}, x_t\in \cX, a_t\in \cA,$
\eq{P^{s^i g^{-i}}(z_t,x_t, a_t) &= P^{g^ig^{-i}}(z_t,x_t, a_t) \;\;\;\;\;
}
which implies $ J^{i,s^ig^{-i}} = J^{i,g^ig^{-i}}$.\label{fact:G2S}
\end{lemma}
\begin{IEEEproof}
	The proof is on the similar lines as the proof of Lemma~1 in~\cite{VaSiAn19} 
\end{IEEEproof}
Since any $s^i$ policy is also a $g^i$ type policy, the above lemma can be iterated over all players which implies that for any $g$ policy profile there exists an $s$ policy profile that achieves the same reward profile i.e., $(J^{i,s})_{i\in \cN} = (J^{i,g})_{i\in \cN}$.
In the following lemma, we show that the space of profiles of type $s$ is outcome-equivalent to the space of profiles of type $m$.
\begin{lemma}
\label{fact:L1}
For any given strategy profile $s$ of all players, there exists a strategy profile $m$ such that
\eq{P^m(x_t, a_t) &= P^s(x_t, a_t) \;\;\;\;\forall t \in \mathcal{T}, x_t\in \cX, a_t\in \cA ,
}
which implies $ (J^{i,m})_{i\in \cN} = (J^{i,s})_{i\in \cN} $.
\label{fact:S2M}
\end{lemma}
\begin{IEEEproof}
	The proof is on the similar lines as the proof of Lemma~2 in~\cite{VaSiAn19}
\end{IEEEproof}

\section{Part 1: Minor Followers}
\label{app:P1}
\label{app:B}
\label{app:A}
\begin{IEEEproof}
We prove Theorem~\ref{Thm:Main} using induction and the results in Lemma~\ref{lemma:2}, and \ref{lemma:1} proved in \ref{app:B}. Let $\tsigma$ be the strategies computed by the methodology in Section~III.

\seq{
For the base case at $t=T$, $ z_{1:T},a_{1:T-1}^f,x_{1:T}^{f}, \sigma^{f}$
\eq{
\E^{\tsigma_T^{l},\tsigma_T^{m},\tsigma_{T}^{f},\underline{\pi_T}}\big\{  R^f_T(Z_T,X_T,A_T) \big\lvert \underline{\pi_T},z_{1:T},a_{1:T-1}^{l,m},x_{1:T}^f \big\}
&=
V^f_T(\underline{\pi_T},z_T, x_T^f)  \label{eq:T2a}\\
&\geq \E^{\tsigma^{l},\tsigma^{m},\sigma_{T}^{f},\underline{\pi_t}} \big\{ R^f_T(Z_T,X_T,A_T) \big\lvert\underline{\pi_t}, z_{1:T}, a_{1:T-1}^{l,m},x_{1:T}^f \big\},  \label{eq:T2}
}
}
where \eqref{eq:T2a} follows from Lemma~\ref{lemma:1} and \eqref{eq:T2} follows from Lemma~\ref{lemma:2} in Appendix~\ref{app:B}.

Let the induction hypothesis be that for $t+1$, $\forall , z_{1:t+1}, a_{1:t}^{l,m},x_{1:t+1}^f \in (\cX)^{t+1}, \sigma^f$,
\seq{
\eq{
 \E^{\tsigma_{t+1:T}^{l,m},\tsigma_{t+1:T}^{f},\pi_{t+1} } \big\{ \sum_{n=t+1}^T \delta^{n-t-1}R^f_n(Z_n,X_n,A_n) \big\lvert \pi_{t+1}, z_{1:t+1},a_{1:t}^{l,m}, x_{1:t+1}^f \big\} \\
 \geq
  \E^{\tsigma_{t+1:T}^{l,m},\sigma_{t+1:T}^{f},\pi_{t+1} } \big\{ \sum_{n=t+1}^T \delta^{n-t-1} R^f_n(Z_n,X_n,A_n) \big\lvert \pi_
  {t+1},z_{1:t+1}, a_{1:t}^{l,m}, x_{1:t+1}^f \big\}. \label{eq:PropIndHyp}
}
}
\seq{
Then $\forall z_{1:t},a_{1:t-1}^{l,m}, x_{1:t}^f, \sigma^f$, we have
\eq{
&\E^{\tsigma_{t:T}^{l,m},\tsigma_{t:T}^{f},\underline{\pi_t} } \big\{ \sum_{n=t}^T \delta^{n-t-1}R^f_n(Z_n,X_n,A_n) \big\lvert\underline{\pi_t}, z_{1:t},a_{1:t-1}^{l,m}, x_{1:t}^f \big\} \nonumber \\
&= V^f_t(\underline{\pi_t},z_{t}, x_t^f)\label{eq:T1}\\
&\geq \E^{\tsigma_t^{l},\tsigma_t^{m},\sigma_t^f,\underline{\pi_t}} \big\{ R^f_t(Z_t,X_t,A_t) + \delta V^f_{t+1} (\underline{F}(\underline{\pi_t},z_t,\tgamma^{l,m}_t,A^{l,m}_t),\phi(\underline{\pi_t},z_{t},\tgamma_t), X_{t+1}^f) \big\lvert\underline{\pi_t}, z_{1:t},a_{1:t-1}^{l,m}, x_{1:t}^f \big\}  \label{eq:T3}\\
&= \E^{\tsigma_t^{l},\tsigma_t^{m},\sigma_t^f,\underline{\pi_t} } \big\{ R^f_t(Z_t,X_t,A_t) + \delta \E^{\tsigma_{t+1:T}^{l,m},\tsigma_{t+1:T}^{f},\underline{F}(\underline{\pi_t},z_t,\tgamma_t^{l,m},A_t^{l,m})} \nn\\
&\big\{ \sum_{n=t+1}^T \delta^{n-t-1}R^f_n(Z_n,X_n,A_n) \big\lvert \underline{F}(\underline{\pi_t},z_t,\tgamma^{l,m}_t,A^{l,m}_t),z_{1:t},\phi(\underline{\pi_t},z_{t},\tgamma_t), x_{1:t}^f,X_{t+1}^f \big\}  \big\vert\underline{\pi_t}, z_{1:t},a_{1:t-1}^{l,m}, x_{1:t}^f \big\}  \label{eq:T3b}\\
&\geq \E^{\tsigma_t^{l,m},\sigma_t^f,\underline{\pi_t} } \big\{ R^f_t(Z_t,X_t,A_t) +  \delta\E^{\tsigma_{t+1:T}^{l,m},\sigma_{t+1:T}^{f},\underline{F}(\underline{\pi_t},z_t,\tgamma^{l,m}_t,A^{l,m}_t) } \nn\\
&\big\{ \sum_{n=t+1}^T \delta^{n-t-1}R^f_n(Z_n,X_n,A_n) \big\lvert \underline{F}(\underline{\pi_t},z_t,\tgamma^{l,m}_t,A^{l,m}_t), z_{1:t},\phi(\underline{\pi_t},z_{t},\tgamma_t), x_{1:t}^f,X_{t+1}^f\big\} \big\vert\underline{\pi_t}, z_{1:t},a_{1:t-1}^{l,m}, x_{1:t}^f \big\}  \label{eq:T4} \\
&= \E^{\tsigma_t^{l,m},\sigma_t^f,\underline{\pi_t} } \big\{ R^f_t(Z_t,X_t,A_t) +  \delta\E^{\tsigma_{t:T}^{l,m},\sigma_{t:T}^{f},\underline{\pi_t} } \nn\\
&  \big\{ \sum_{n=t+1}^T \delta^{n-t-1}R^f_n(Z_n,X_n,A_n) \big\lvert \underline{F}(\underline{\pi_t},z_t,\tgamma^{l,m}_t,A^{l,m}_t), z_{1:t}, \phi(\underline{\pi_t},z_{t},\tgamma_t), x_{1:t}^f,X_{t+1}^f\big\} \big\vert\underline{\pi_t}, z_{1:t},a_{1:t-1}^{l,m}, x_{1:t}^f \big\}  
\label{eq:T5}\\
&=\E^{\tsigma_{t:T}^{l,m},\sigma_{t:T}^{f},\underline{\pi_t} } \big\{ \sum_{n=t}^T \delta^{n-t}R^f_n(Z_n,X_n,A_n) \big\lvert\underline{\pi_t}, z_{1:t},a_{1:t-1}^{l,m},x_{1:t}^f \big\}  \label{eq:T6},
}
}
where \eqref{eq:T1} follows from Lemma~\ref{lemma:1}, \eqref{eq:T3} follows from Lemma~\ref{lemma:2}, \eqref{eq:T3b} follows from Lemma~\ref{lemma:1}, \eqref{eq:T4} follows from induction hypothesis in \eqref{eq:PropIndHyp} and \eqref{eq:T5} follows from the fact that the probability on $Z_{t+1:T},X_{t+1:T},A_{t+1:T}$ conditioned on $\tsigma^{l,m}_{t:T},\sigma^f_{t:T},\underline{\pi}_t$ only depends on $\tsigma_{t+1:T}^{l,m},\sigma_{t+1:T}^{f},\underline{F}(\underline{\pi_t},z_t,\tgamma_t^{l,m},A_t^{l,m})$ as the follower's strategy $\sigma_t^f$ doesn't affect either the update of the belief $\underline{\pi}_t$ or the update of the mean field $z_t$.
\end{IEEEproof}

\section{}
\label{app:B}
\label{app:lemmas}
\begin{lemma}
\label{lemma:2}
Let $\tsigma$ be the strategies computed by the methodology in Section~III. 
Then $\forall t\in [T],z_{1:t},a_{1:t-1}^{l,m},  x_{1:t}^f, \sigma^f_t$
\eq{
&V^f_t(\underline{\pi_t},z_t, x_t^f) \geq \nn\\
&\E^{\tsigma_t^{l},\tsigma_t^{m},\sigma_t^f,\underline{\pi_t}} \big\{ R^f_t(Z_t,X_t,A_t) + \delta V^f_{t+1} (\underline{F}(\underline{\pi_t},z_t,\tgamma^{l,m}_t,A^{l,m}_t),\phi(\underline{\pi_t},z_{t},\tgamma_t), X_{t+1}^f) \big\lvert \underline{\pi_t}, z_{1:t},a_{1:t-1}^{l,m}, x_{1:t}^f \big\}.\label{eq:lemma2}
}
\end{lemma}

\begin{IEEEproof}
We prove this lemma by contradiction.

 Suppose the claim is not true for time $t$. This implies $\exists \hat{\sigma}_t^f, \hat{z}_{1:t},\hat{a}_{1:t-1}^{l,m}, \hat{x}_{1:t}^f$ such that
\eq{
&\E^{\tsigma_t^{l},\tsigma_t^{m},\hat{\sigma}_t^f,\underline{\pi_t}} \big\{ R^f_t(Z_t,X_t,A_t) +  \delta V^f_{t+1} (\underline{F}(\underline{\pi_t},z_t,\tgamma^{l,m}_t,A_t^{l,m}),\phi(\underline{\pi_t},z_{t},\tgamma_t), X_{t+1}^f) \big\lvert \underline{\pi_t}, \hat{z}_{1:t},\hat{a}_{1:t-1}^{l,m},\hat{x}_{1:t}^f \big\}\nn\\ 
&> V^f_t(\underline{\pi_t},z_t, \hat{x}_{t}^f).\label{eq:E8}
}
We will show that this leads to a contradiction.
Construct 
\begin{equation}
\hat{\gamma}^f_t(a_t^f|x_t^f) = \lb{\hat{\sigma}_t^f(a_t^f|\hat{z}_{1:t},\hat{a}_{1:t-1}^{l,m},\hat{x}_{1:t}^f) \;\;\;\;\; x_t^f = \hat{x}_t^f \\ \text{arbitrary} \;\;\;\;\;\;\;\;\;\;\;\;\;\; \text{otherwise.}  }
\end{equation}

Then for $ \hat{z}_{1:t},\hat{a}_{1:t-1},\hat{x}_{1:t}^f$, we have
\seq{
\eq{
&V^f_t(\underline{\pi_t},z_t, \hat{x}_t^f)= \max_{\gamma_t^f(\cdot|\hat{x}_t^f)} \E^{\tsigma^{l},\tsigma^{m},\gamma^f_t(\cdot|\hat{x}_t^f),\underline{\pi_t}} \big\{ R^f_t(z_t,X^{l,m}_T,\hat{x}_t^f,A_t) +
\nn \\
& \delta V^f_{t+1} (\underline{F}(\underline{\pi_t},z_t,\tgamma^{l,m}_t,A^{l,m}_t),\phi(\underline{\pi_t},z_t, \tgamma_t), X_{t+1}^f) \big\lvert \underline{\pi_t},\hat{z}_{t}, \hat{x}_{t}^f \big\}, \label{eq:E11}\\
&\geq\E^{\tsigma^{l},\tsigma^{m},\hat{\gamma}_t^f(\cdot|\hat{x}_t^f),\underline{\pi_t}} \big\{ R^f_t(z_t,x^{l,m}_t,\hat{x}_t^f,a_t) + \delta V^f_{t+1} (\underline{F}(\underline{\pi_t},z_t,\tgamma^{l,m}_t,A_t^{l,m}),\phi(\underline{\pi_t},z_t, \tgamma_t), {X}_{t+1}^f) \big\lvert \underline{\pi_t}, \hat{z}_t,\hat{x}_{t}^f \big\}   
\\ 
&=\sum_{x^{l,m}_t,a_t^f,x_{t+1}^f}   \big\{ R^f_t(z_t,x^{l,m}_T,\hat{x}_t^f,a_t) + \delta V^f_{t+1} (\underline{F}(\underline{\pi_t},z_t,\tgamma^{l,m}_t,a_t^{l,m}),\phi(\underline{\pi_t},z_t, \tgamma_t), x_{t+1}^f)\big\}
\underline{\pi_t}(x^{l,m}_t)\nn\\
&\gamma_t^{l,m}(a_t^{l,m}|x_t^{l,m})\hat{\gamma}^f_t(a^f_t|\hat{x}_t^f)Q_t^f(x_{t+1}^f|\hat{z}_t,\hat{x}_t^{l,m},\hat{x}_t^f,a_t) 
\\ 
&= \sum_{x^{l,m}_t,a_t^f,x_{t+1}^f}  \big\{ R^f_t(z_t,x^{l,m}_t,\hat{x}^f_t,a_t) + \delta V^f_{t+1} (\underline{F}(\underline{\pi_t},z_t,\tgamma^{l,m}_t,a_t^{l,m}),\phi(\underline{\pi_t},z_t, \tgamma_t), x_{t+1}^f)\big\}
\underline{\pi_t}(x^{l,m}_t)\nn\\
&\sigma_t^{l,m}(a_t^{l,m}|\hat{z}_{1:t},\hat{a}_{1:t-1}^{l,m},x_{1:t}^{l,m})\hat{\sigma}^f_t(a_t^f|\hat{z}_{1:t},\hat{a}_{1:t-1}^{l,m},\hat{x}_{1:t}^f)Q_t^f(x_{t+1}^f|\hat{z}_t,\hat{x}_t^{l,m},\hat{x}_t^f,a_t) \label{eq:E9}\\
&= \E^{\tsigma^{l,m},\hat{\sigma}_t^f,\underline{\pi_t} } \big\{ R^f_t(z_t,x^{l,m}_t,\hat{x}^f_t,A_t)+ \delta V^f_{t+1} (\underline{F}(\underline{\pi_t},z_t,\tgamma^{l,m}_t,a_t^{l,m}),\phi(\underline{\pi_t},z_t, \tgamma_t), X_{t+1}^f) \big\lvert \underline{\pi_t}, \hat{z}_{1:t},\hat{a}_{1:t-1}^{l,m}, \hat{x}_{1:t}^f \big\}  \\
&> V^f_t(\underline{\pi_t},\hat{z}_t, \hat{x}_{t}^f), \label{eq:E10}
}
where \eqref{eq:E11} follows from definition of $V^f_t$ in \eqref{eq:Vdef}, \eqref{eq:E9} follows from definition of $\hat{\gamma}_t^f$ and \eqref{eq:E10} follows from \eqref{eq:E8}. However this leads to a contradiction.
}
\end{IEEEproof}

\begin{lemma}
\label{lemma:1}
Let $\tsigma$ be the strategies computed by the methodology in Section~III. 
Then $\forall t\in [T], z_{1:t},a_{1:t-1}^{l,m},x_{1:t}^f$,
\begin{gather}
V^f_t(\underline{\pi_t},z_{t}, x_t^f) =
\E^{\tsigma_{t:T}^{l,m},\tsigma_{t:T}^{f},\underline{\pi_t}} \big\{ \sum_{n=t}^T \delta^{n-t}R^f_n(Z_n,X_n,A_n) \big\lvert \underline{\pi_t},  z_{1:t},a_{1:t-1}^{l,m},x_{1:t}^f \big\} .
\end{gather} 
\end{lemma}

\begin{IEEEproof}
%
\seq{
We prove the lemma by induction. For $t=T$,
\eq{
 &\E^{\tsigma_t^{l},\tsigma_t^{m},\tsigma_{T}^{f},\underline{\pi_t} } \big\{  R(Z_T, X_T,A_T) \big\lvert\underline{\pi_t}, z_{1:T},a_{1:T-1}^{l,m},x_{1:T}^f \big\}\nn\\
 &= \sum_{a_T^f} R^f_T(z_T,x_T,a_T) \underline{\pi_t}(x^{l}_T,x^{m}_T)\tsigma_{T}^{f}(a_T^f|z_{T},x_{T}^f)\tsigma_{T}^{l}(a_T^l|z_{T},x_{T}^l)\tsigma_{T}^{m}(a_T^m|z_{T},x_{T}^m) \\
 &= V^f_T(\underline{\pi_t},z_{T}, x_T^f) \label{eq:C1},
}
}
where \eqref{eq:C1} follows from the definition of $V^f_t$ in \eqref{eq:Vdef}.
Suppose the claim is true for $t+1$, i.e., $\forall  t\in [T],  z_{1:t+1},a_{1:t}^{l,m},x_{1:t+1}^f$
\begin{gather}
V^f_{t+1}(\pi_{t+1},z_{t+1}, x_{t+1}^f) = \E^{\tsigma_{t+1:T}^{l,m},\tsigma_{t+1:T}^{f},\pi_{t+1}}
\big\{ \sum_{n=t+1}^T \delta^{n-t-1}R^f_n(Z_n,X_n,A_n) \big\lvert \pi_{t+1}, z_{1:t+1},a_{1:t}^{l,m}, x_{1:t+1}^f \big\} 
\label{eq:CIndHyp}.
\end{gather}
Then $\forall  t\in [T],z_{1:t},a_{1:t-1}^{l,m}, x_{1:t}^f$, we have
\seq{
\eq{
&\E^{\tsigma_{t:T}^{l,m},\tsigma_{t:T}^{f},\underline{\pi_t} } \big\{ \sum_{n=t}^T \delta^{n-t} R^f_n(Z_n,X_n,A_n) \big\lvert \underline{\pi_t}, z_{1:t}, a_{1:t-1}^{l,m},x_{1:t}^f \big\} 
\nonumber 
\\
&=  \E^{\tsigma_{t:T}^{l,m},\tsigma_{t:T}^{f},\underline{\pi_t}} \big\{R^f_t(Z_t,X_t,A_t) +\delta \E^{\tsigma_{t:T}^{l,m},\tsigma_{t:T}^{f},\underline{\pi_t} } 
\nonumber \\ 
& \big\{ \sum_{n=t+1}^T \delta^{n-t-1}R^f_n(Z_n,X_n,A_n)\big\lvert \underline{F}(\underline{\pi_t},z_t,\tgamma_t^{l,m},A^{l,m}_t), z_{1:t},\phi(\underline{\pi_t},z_t,\tgamma_t), x_{1:t}^f,X_{t+1}^f\big\} \big\lvert \underline{\pi_t}, z_{1:t}, a_{1:t-1}^{l,m}, x_{1:t}^f \big\} \label{eq:C2}
\\
&=  \E^{\tsigma_{t:T}^{l,m},\tsigma_{t:T}^{f},\underline{\pi_t}} \big\{R^f_t(Z_t,X_t,A_t) +\delta\E^{\tsigma_{t+1:T}^{l,m},\tsigma_{t+1:T}^{f},\underline{F}(\underline{\pi_t},z_t,\tgamma^{l,m}_t,A^{l,m}_t)}
\nonumber 
\\
&\big\{ \sum_{n=t+1}^T \delta^{n-t-1}R^f_n(Z_n,X_n,A_n)\big\lvert \underline{F}(\underline{\pi_t},z_t,\tgamma^{l,m}_t,A^{l,m}_t), z_{1:t},\phi(\underline{\pi_t},z_t,\tgamma_t), x_{1:t}^f,X_{t+1}^f\big\} \big\lvert \underline{\pi_t}, z_{1:t}, a_{1:t-1}^{l,m},x_{1:t}^f \big\} \label{eq:C3}
\\
&=  \E^{\tsigma_{t:T}^{l,m},\tsigma_{t:T}^{f},\underline{\pi_t}} \big\{R^f_t(Z_t,X_t,A_t) +  \delta V^f_{t+1}(\underline{F}(\underline{\pi_t},z_t,\tgamma^{l,m}_t,A^{l,m}_t),\phi(\underline{\pi_t},z_t,\tgamma_t), X_{t+1}^f) \big\lvert \underline{\pi_t},  z_{1:t},a_{1:t-1}^{l,m}, x_{1:t}^f \big\} 
\label{eq:C4}
\\
&=V^f_{t}(\underline{\pi_t},z_t, x_t^f) \label{eq:C6},
}
}
where ~\eqref{eq:C3} follows from the fact that the probability on $Z_{t+1:T},X_{t+1:T},A_{t+1:T}$ conditioned on $\tsigma^{l,m}_{t:T},\sigma^f_{t:T},\underline{\pi}_t$ only depends on $\tsigma_{t+1:T}^{l,m},\sigma_{t+1:T}^{f},\underline{F}(\underline{\pi_t},z_t,\tgamma_t^{l,m},A_t^{l,m})$ as the follower's strategy $\sigma_t^f$ doesn't affect either the update of the belief $\underline{\pi}_t$ or the update of the mean field $z_t$, 
\eqref{eq:C4} follows from the induction hypothesis in \eqref{eq:CIndHyp} and \eqref{eq:C6} follows from the definition of $V^f_t$ in \eqref{eq:Vdef}.
\end{IEEEproof}

\section{Part 2: Major Followers}
\label{b_app:P1}
\label{b_app:B}
\label{b_app:A}
\begin{IEEEproof}
We prove Theorem~\ref{Thm:Main} using induction and the results in Lemma~\ref{b_lemma:2}, and \ref{b_lemma:1} proved in Appendix~\ref{b_app:B}. Let $\tsigma$ be the strategies computed by the methodology in Section~III. 
\seq{
For the base case at $t=T$, $ z_{1:T},a_{1:T-1}^{l,m},x_{1:T}^{m,j},\sigma^{m,j}$
\eq{
&\E^{\tsigma_T^{l},\tsigma_{T}^{m,j},\tsigma_T^{m,-j},\tsigma_{T}^{f},\underline{\pi_t}}\big\{  R^{m,j}_T(Z_T,X^{l,m}_T,A^{l,m}_T) \big\lvert \underline{\pi_t},z_{1:T},a_{1:T-1}^{l,m},x_{1:T}^{m,j} \big\}
=
V^{m,j}_T(\underline{\pi_T},z_T, x_T^{m,j})  \label{b_eq:T2a}\\
&\geq \E^{\tsigma_T^{l},\sigma_{T}^{m,j},\tsigma_T^{m,-j},\tsigma_{T}^{f},\underline{\pi_T}} \big\{ R^{m,j}_T(Z_T,X^{l,m}_T,A^{l,m}_T) \big\lvert\underline{\pi_t}, z_{1:T}, a_{1:T-1}^{l,m},x_{1:T}^{m,j} \big\},  \label{b_eq:T2}
}
}
where \eqref{b_eq:T2a} follows from Lemma~\ref{b_lemma:1} and \eqref{b_eq:T2} follows from Lemma~\ref{b_lemma:2} in Appendix~\ref{b_app:B}.

Let the induction hypothesis be that for $t+1$. Then $\forall t, z_{1:t+1}, a_{1:t}^{l,m},x_{1:t+1}^{m,j} \in (\cX^{m,j})^{t+1}, \sigma^{m,j}$,
\seq{
\eq{
 \E^{\tsigma_{t+1:T}^{l,m},\tsigma_{t+1:T}^f,\underline{\pi}_{t+1} } \big\{ \sum_{n=t+1}^T \delta^{n-t-1}R^{m,j}_n(Z_n,X^{l,m}_n,A^{l,m}_n) \big\lvert \pi_{t+1}, z_{1:t+1},a_{1:t}^{l,m}, x_{1:t+1}^{m,j} \big\} \\
 \geq
  \E^{\tsigma_{t+1:T}^{l},\sigma_{t+1:T}^{m,j},\tsigma_{t+1:T}^{m,-j},\tsigma_{t+1:T}^{f},\underline{\pi}_{t+1} } \big\{ \sum_{n=t+1}^T \delta^{n-t-1} R^{m,j}_n(Z_n,X^{l,m}_n,A^{l,m}_n) \big\lvert \pi_
  {t+1},z_{1:t+1}, a_{1:t}^{l,m}, x_{1:t+1}^{m,j} \big\}. \label{b_eq:PropIndHyp}
}
}
\seq{
Then $\forall z_{1:t},a_{1:t-1}^{l,m}, x_{1:t}^{m,j}, \sigma^{m,j}$, we have
\eq{
&\E^{\tsigma_{t:T}^{l,m,f},\underline{\pi_t} } \big\{ \sum_{n=t}^T \delta^{n-t-1}R^{m,j}_n(Z_n,X^{l,m}_n,A^{l,m}_n) \big\lvert\underline{\pi_t}, z_{1:t},a_{1:t-1}^{l,m},x_{1:t}^{m,j} \big\} \nonumber \\
&= V^{m,j}_t(\underline{\pi_t},z_{t}, x_t^{m,j})\label{b_eq:T1}\\
&\geq \E^{\tsigma_t^{l},\sigma_t^{m,j},\tsigma_t^{m,-j},\tsigma_t^{f},\underline{\pi_t}} \big\{ R^{m,j}_t(Z_t,X^{l,m}_t,A^{l,m}_t) + \delta V^{m,j}_{t+1} (\underline{F}(\underline{\pi_t},z_t,\tgamma^{l,m}_t,A^{l,m}_t),\phi(\underline{\pi_t},z_{t},\tgamma_t),X_{t+1}^{m,j}) \big\lvert\underline{\pi_t}, z_{1:t},a_{1:t-1}^{l,m}, x_{1:t}^{m,j} \big\}  \label{b_eq:T3}\\
&= \E^{\tsigma_t^{l},\sigma_t^{m,j},\tsigma_t^{m,-j},\tsigma_t^{f},\underline{\pi_t} } \big\{ R^{m,j}_t(Z_t,X^{l,m}_t,A^{l,m}_t) + \delta \E^{\tsigma_{t+1:T}^{l,m},\tsigma_{t+1:T}^{m,j},\underline{F}(\underline{\pi_t},z_t,\tgamma_t^{l,m},A^{l,m}_t)} \nn\\
&\big\{ \sum_{n=t+1}^T \delta^{n-t-1}R^{m,j}_n(Z_n,X^{l,m}_n,A^{l,m}_n) \big\lvert \underline{F}(\underline{\pi_t},z_t,\tgamma^{l,m}_t,A^{l,m}_t),z_{1:t},Z_{t+1}, x_{1:t}^{m,j},X_{t+1}^{m,j} \big\}  \big\vert\underline{\pi_t}, z_{1:t},a_{1:t-1}^{l,m}, x_{1:t}^{m,j} \big\}  \label{b_eq:T3b}\\
&\geq \E^{\tsigma_t^{l},\sigma_t^{m,j},\tsigma_t^{m,-j},\tsigma_t^{f},\underline{\pi_t} } \big\{ R^{m,j}_t(Z_t,X^{l,m}_t,A^{l,m}_t) +  \delta\E^{\tsigma_{t+1:T}^{l,m},\sigma_{t+1:T}^{m,j},\underline{F}(\underline{\pi_t},z_t,\tgamma^{l,m}_t,A_t^{l,m}) } \nn\\
&\big\{ \sum_{n=t+1}^T \delta^{n-t-1}R^{m,j}_n(Z_n,X^{l,m}_n,A^{l,m}_n) \big\lvert \underline{F}(\underline{\pi_t},z_t,\tgamma^{l,m}_t,A^{l,m}_t), z_{1:t},Z_{t+1}, x_{1:t}^{m,j},X_{t+1}^{m,j}\big\} \big\vert\underline{\pi_t}, z_{1:t},a_{1:t-1}^{l,m}, x_{1:t}^{m,j} \big\}  \label{b_eq:T4} \\
&= \E^{\tsigma_t^{l},\tsigma_t^{m,j},\sigma_t^{m,-j},\tsigma_t^{f},\underline{\pi_t} } \big\{ R^{m,j}_t(Z_t,X^{l,m}_t,A^{l,m}_t) + \delta\E^{\tsigma_{t:T}^{l,m},\sigma_{t:T}^{m,j},\underline{\pi_t} }   \nn\\
& \big\{ \sum_{n=t+1}^T \delta^{n-t-1}R^{m,j}_n(Z_n,X^{l,m}_n,A^{l,m}_n) \big\lvert \underline{F}(\underline{\pi_t},z_t,\tgamma^{l,m}_t,A_t^{l,m}), z_{1:t}, Z_{t+1}, x_{1:t}^{m,j},X_{t+1}^{m,j}\big\} \big\vert\underline{\pi_t}, z_{1:t},a_{1:t-1}^{l,m}, x_{1:t}^{m,j} \big\}
\label{b_eq:T5}\\
&=\E^{\tsigma_{t:T}^{l,m},\sigma_{t:T}^{m,j},\underline{\pi_t} } \big\{ \sum_{n=t}^T \delta^{n-t}R^{m,j}_n(Z_n,X^{l,m}_n,A^{l,m}_n) \big\lvert\underline{\pi_t}, z_{1:t},a_{1:t-1}^{l,m},x_{1:t}^{m,j} \big\}  \label{b_eq:T6},
}
}
where \eqref{b_eq:T1} follows from Lemma~\ref{b_lemma:1}, \eqref{b_eq:T3} follows from Lemma~\ref{b_lemma:2}, \eqref{b_eq:T3b} follows from Lemma~\ref{b_lemma:1}, \eqref{b_eq:T4} follows from induction hypothesis in \eqref{b_eq:PropIndHyp} and \eqref{b_eq:T5} follows from Lemma~\ref{b_lemma:3}.
\end{IEEEproof}

\section{}
\label{b_app:B}
\label{b_app:lemmas}
\begin{lemma}
\label{b_lemma:2}
Let $\tsigma$ be the strategies computed by the methodology in Section~III. 
Then $\forall t\in [T],z_{1:t},a_{1:t-1}^{l,m}, x_{1:t}^{m,j}, \sigma^{m,j}_t$
\eq{
&V^{m,j}_t(\underline{\pi_t},z_t,x_t^{m,j}) \nn\\
&\geq \E^{\tsigma_t^{l},\sigma_t^{m,j},\tsigma_t^{m,-j},\tsigma_t^{f},\underline{\pi_t}} \big\{ R^{m,j}_t(Z_t,X^{l,m}_t,A^{l,m}_t) + \delta V^{m,j}_{t+1} (\underline{F}(\underline{\pi_t},z_t,\tgamma^{l,m}_t,A^{l,m}_t),\phi(\underline{\pi_t},z_{t},\tgamma_t), X_{t+1}^{m,j}) \big\lvert \underline{\pi_t}, z_{1:t},a_{1:t-1}^{l,m}, x_{1:t}^{m,j} \big\}.\label{b_eq:lemma2}
}
\end{lemma}

\begin{IEEEproof}
We prove this lemma by contradiction. Suppose the claim is not true for $t$. This implies $\exists i, \hat{\sigma}_t^{m,j}, \hat{z}_{1:t},\hat{a}_{1:t-1}^{l,m}$, $ \hat{x}_{1:t}^{m,j}$ such that
\eq{
&\E^{\tsigma_t^{l},\hat{\sigma}_t^{m,j},\tsigma_t^{m,-j},{\tsigma}_t^{f},\underline{\pi_t}} \big\{ R^{m,j}_t(Z_t,X^{l,m}_t,A^{l,m}_t) +  \delta V^{m,j}_{t+1} (\underline{F}(\underline{\pi_t},z_t,\tgamma^{l,m}_t,A_t^{l,m}),\phi(\underline{\pi_t},z_{t},\tgamma_t), X_{t+1}^{m,j}) \big\lvert \underline{\pi_t}, \hat{z}_{1:t},\hat{a}_{1:t-1}^{l,m},\hat{x}_{1:t}^{m,j} \big\} \nn\\
&> V^{m,j}_t(\underline{\pi_t},z_t, \hat{x}_{t}^{m,j}).\label{b_eq:E8}
}
We will show that this leads to a contradiction.
Construct 
\begin{equation}
\hat{\gamma}^{m,j}_t(a_t^{m,j}|x_t^{m,j}) = \lb{\hat{\sigma}_t^{m,j}(a_t^{m,j}|\hat{z}_{1:t},\hat{a}_{1:t-1}^{l,m},\hat{x}_{1:t}^{m,j}) \;\;\;\;\; x_t^{m,j} = \hat{x}_t^{m,j} \\ \text{arbitrary} \;\;\;\;\;\;\;\;\;\;\;\;\;\; \text{otherwise.}  }
\end{equation}

Then for $ \hat{z}_{1:t},\hat{a}_{1:t-1},\hat{x}_{1:t}^{m,j}$ and $\underline{\hat{\pi}_t}(x_t^{l,m}) = P^{\sigma}(x_t^{l,m}|\hat{z}_{1:t},\hat{a}_{1:t-1})$, we have
\seq{
\eq{
&V^{m,j}_t(\underline{\pi}_t,z_t, \hat{x}_t^{m,j})
 = \max_{\gamma_t^{m,j}(\cdot|\hat{x}_t^{m,j})} \E^{\tsigma^{l},\tsigma^{m},\gamma^{m,j}_t(\cdot|\hat{x}_t^{m,j}),\underline{\hat{\pi}_t} }\big\{ R^{m,j}_t(z_t,x^{l}_t,x_t^{m,-j},\hat{x}_t^{m,j},A_t^{l,m})\nn \\
&+ \delta V^{m,j}_{t+1} (\underline{F}(\underline{\hat{\pi}_t}  ,z_t,\tgamma^{l,m}_t,A^{l,m}_t),\phi(\underline{\hat{\pi}_t},z_t, \tgamma_t), X_{t+1}^{m,j}) \big\lvert \underline{\hat{\pi}_t},\hat{z}_{t}, \hat{x}_{t}^{m,j} \big\}, \label{b_eq:E11}\\
&\geq\E^{\tsigma^{l},\tsigma^{m},\hat{\gamma}_t^{m,j}(\cdot|\hat{x}_t^{m,j}),\underline{\hat{\pi}_t}} \big\{ R^{m,j}_t(z_t,x^{l}_t,x_t^{m,-j},\hat{x}_t^{m,j},A_t^{l,m}) + \delta V^{m,j}_{t+1} (\underline{F}(\underline{\hat{\pi}_t},z_t,\tgamma^{l,m}_t,A_t^{l,m}),\phi(\underline{\hat{\pi}_t},z_t, \tgamma_t), {X}_{t+1}^{m,j}) \big\lvert \underline{\hat{\pi}_t}, \hat{z}_t,\hat{x}_{t}^{m,j} \big\}   
\\ 
&=\sum_{x^{l}_t,x^{m,-j}_t,a_t^{m,j},x_{t+1}^{m,j}}   \big\{ R^{m,j}_t(z_t,x^{l}_t,x_t^{m,-j},\hat{x}_t^{m,j},a_t^{l,m}) + \delta V^{m,j}_{t+1} (\underline{F}(\underline{\hat{\pi}_t},z_t,\tgamma^{l,m}_t,a^{l,m}_t),\phi(\underline{\hat{\pi}_t},z_t, \tgamma_t), x_{t+1}^{m,j})\big\}
\nn\\
&\underline{\hat{\pi}_t}(x^{l,m}_t)\gamma_t^{l}(a_t^l|x_t^l)\gamma_t^{m,-j}(a_t^{m,-j}|x_t^{m,-j})\hat{\gamma}^{m,j}_t(a^{m,j}_t|\hat{x}_t^{m,j})Q_t^{m,j}(x_{t+1}^{m,j}|\hat{z}_t,{x}_t^{l},\hat{x}_t^{m,j},x_t^{m,-j},a_t^{l,m}) 
\\ 
&= \sum_{x^{l}_t,x_t^{m,-j},a_t^{m,j},x_{t+1}^{m,j}}  \big\{ R^{m,j}_t(z_t,x^{l}_t,x_t^{m,-j},\hat{x}_t^{m,j},a_t^{l,m}) + \delta V^{m,j}_{t+1} (\underline{F}(\underline{\hat{\pi}_t},z_t,\tgamma^{l,m}_t,a^{l,m}_t),\phi(\underline{\hat{\pi}_t},z_t, \tgamma_t), x_{t+1}^{m,j})\big\}
\nn\\
&\underline{\hat{\pi}_t}(x^{l,m}_t)\gamma_t^{l}(a_t^l|x_t^l)\gamma_t^{m,-j}(a_t^{m,-j}|x_t^{m,-j})\hat{\sigma}^{m,j}_t(a_t^{m,j}|\hat{z}_{1:t},\hat{a}_{1:t-1}^{l},\hat{x}_{1:t}^{m,j})Q_t^{m,j}(x_{t+1}^{m,j}|\hat{z}_t,{x}_t^{l},\hat{x}_t^{m,j},x_t^{m,-j},a^{l,m}_t) \label{b_eq:E9}\\
&= \E^{\tsigma^{l},\tsigma^{m},\hat{\sigma}_t^{m,j},\underline{\hat{\pi}_t} } \big\{ R^{m,j}_t(z_t,x^{l}_t,x_t^{m,-j},\hat{x}_t^{m,j},A_t^{l,m})+ \nn\\
&\delta V^{m,j}_{t+1} (\underline{F}(\underline{\hat{\pi}_t},z_t,\tgamma^{l,m}_t,a_t^{l,m}),\phi(\underline{\hat{\pi}_t},z_t, \tgamma_t), X_{t+1}^{m,j}) \big\lvert \underline{\hat{\pi}_t}, \hat{z}_{1:t},\hat{a}_{1:t-1}^{l,m}, \hat{x}_{1:t}^{m,j} \big\}  \\
&> V^{m,j}_t(\underline{\hat{\pi}_t},\hat{z}_t, \hat{x}_{t}^{m,j}), \label{b_eq:E10}
}
where \eqref{b_eq:E11} follows from definition of $V^{m,j}_t$ in \eqref{eq:Vdef}, \eqref{b_eq:E9} follows from definition of $\hat{\gamma}_t^{m,j}$ and \eqref{b_eq:E10} follows from \eqref{b_eq:E8}. However this leads to a contradiction.
}

\end{IEEEproof}

\begin{lemma}
\label{b_lemma:1}
Let $\tsigma$ be the strategies computed by the methodology in Section~III. 
Then $\forall t\in [T], z_{1:t},a_{1:t-1}^{l,m},x_{1:t}^{m,j}$,
\begin{gather}
V^{m,j}_t(\underline{\pi_t},z_{t}, x_t^{m,j}) =
\E^{\tsigma_{t:T}^{l,m},\tsigma_{t:T}^{f},\underline{\pi_t}} \big\{ \sum_{n=t}^T \delta^{n-t}R^{m,j}_n(z_t,X_t^{l,m},A_t^{l,m}) \big\lvert \underline{\pi_t},  z_{1:t},a_{1:t-1}^{l,m},x_{1:t}^{m,j} \big\} .
\end{gather} 
\end{lemma}

\begin{IEEEproof}
%
\seq{
We prove the lemma by induction. For $t=T$,
\eq{
 &\E^{\tsigma_t^{l},\tsigma_t^{m},\tsigma_{T}^{},\underline{\pi_t} } \big\{  R^{m,j}(Z_T,X_T^{l,m},A_T^{l,m}) \big\lvert\underline{\pi_t}, z_{1:T},a_{1:T-1}^{l,m},x_{1:T}^{m,j} \big\}\nn\\
 &= \sum_{a_T^{m,j}} R^{m,j}_T(z_T,x^{l,m}_T,a^{l,m}_T) \underline{\pi_t}(x^{l}_T,x^{m,-j}_T)\tsigma_{T}^{l,m}(a_T^{l,m}|z_{T},x_{T}^{l,m}) \\
 &= V^{m,j}_T(\underline{\pi_t},z_{T}, x_T^{m,j}) \label{b_eq:C1},
}
}
where \eqref{b_eq:C1} follows from the definition of $V^{m,j}_t$ in \eqref{eq:Vdef}.
Suppose the claim is true for $t+1$, i.e., $\forall  t\in [T],  z_{1:t+1},a_{1:t}^{l,m}$,  $x_{1:t+1}^{m,j}$
\begin{gather}
V^{m,j}_{t+1}(\pi_{t+1},z_{t+1}, x_{t+1}^{m,j}) = \E^{\tsigma_{t+1:T}^{l,m},\tsigma_{t+1:T}^{f},\pi_{t+1}}
\big\{ \sum_{n=t+1}^T \delta^{n-t-1}R^{m,j}_n(Z_n,X^{l,m}_n,A^{l,m}_n) \big\lvert \pi_{t+1}, z_{1:t+1},a_{1:t}^{l,m}, x_{1:t+1}^{m,j} \big\} 
\label{b_eq:CIndHyp}.
\end{gather}
Then $\forall  t\in [T],z_{1:t},a_{1:t-1}^{l,m}, x_{1:t}^{m,j}$, we have
\seq{
\eq{
&\E^{\tsigma_{t:T}^{l,m},\tsigma_{t:T}^{m,j},\underline{\pi_t} } \big\{ \sum_{n=t}^T \delta^{n-t} R^{m,j}_n(Z_n,X^{l,n}_n,A^{l,n}_n) \big\lvert \underline{\pi_t}, z_{1:t}, a_{1:t-1}^{l,m},x_{1:t}^{m,j} \big\} 
\nonumber 
\\
&=  \E^{\tsigma_{t:T}^{l,m},\tsigma_{t:T}^{m,j},\underline{\pi_t}} \big\{R^{m,j}_t(Z_t,X^{l,n}_t,A^{l,n}_t)+\delta \E^{\tsigma_{t:T}^{l,m},\tsigma_{t:T}^{m,j},\underline{\pi_t} } 
\nonumber \\ 
& \big\{ \sum_{n=t+1}^T \delta^{n-t-1}R^{m,j}_n(Z_n,X_n,A_n)\big\lvert \underline{F}(\underline{\pi_t},z_t,\gamma_t^{l,m},A^{l,m}_t), z_{1:t},Z_{t+1}, x_{1:t}^{m,j},X_{t+1}^{m,j}\big\} \big\lvert \underline{\pi_t}, z_{1:t}, a_{1:t-1}^{l,m}, x_{1:t}^{m,j} \big\} \label{b_eq:C2}
\\
&=  \E^{\tsigma_{t:T}^{l,m},\tsigma_{t:T}^{m,j},\underline{\pi_t}} \big\{R^{m,j}_t(Z_t,X_t,A_t) +\delta\E^{\tsigma_{t+1:T}^{l,m},\tsigma_{t+1:T}^{m,j},\underline{F}(\underline{\pi_t},z_t,\tgamma^{l,m}_t,A_t^{l,m})}
\nonumber 
\\
&\big\{ \sum_{n=t+1}^T \delta^{n-t-1}R^{m,j}_n(Z_n,X_n,A_n)\big\lvert \underline{F}(\underline{\pi_t},z_t,\gamma^{l,m}_t,A^{l,m}_t), z_{1:t},Z_{t+1}, x_{1:t}^{m,j},X_{t+1}^{m,j}\big\} \big\lvert \underline{\pi_t}, z_{1:t}, a_{1:t-1}^{l,m},x_{1:t}^{m,j} \big\} \label{b_eq:C3}
\\
&=  \E^{\tsigma_{t:T}^{l,m},\tsigma_{t:T}^{m,j},\underline{\pi_t}} \big\{R^{m,j}_t(Z_t,X_t,A_t) + \nn\\
&\delta V^{m,j}_{t+1}(\underline{F}(\underline{\pi_t},z_t,\tgamma^{l,m}_t,A^{l,m}_t),\phi(\underline{\pi_t},z_t,\tgamma_t), X_{t+1}^{m,j}) \big\lvert \underline{\pi_t},  z_{1:t},a_{1:t-1}^{l,m}, x_{1:t}^{m,j} \big\} 
\label{b_eq:C4}
\\
&=  \E^{\tsigma_t^{l},\tsigma_t^{m},\tsigma_{t}^{m,j},\underline{\pi_t}} \big\{R^{m,j}_t(Z_t,X_t,A_t) +  \delta V^{m,j}_{t+1}(\underline{F}(\underline{\pi_t},z_t,\tgamma^{l,m}_t,A^{l,m}_t),\phi(\underline{\pi_t},z_t,\tgamma_t), X_{t+1}^{m,j}) \big\lvert \underline{\pi_t},  z_{1:t},a_{1:t-1}^{l,m}, x_{1:t}^{m,j} \big\} 
\label{b_eq:C5}
\\
&=V^{m,j}_{t}(\underline{\pi_t},z_t, x_t^{m,j}) \label{b_eq:C6},
}
} where \eqref{b_eq:C3} follows from Lemma~\ref{b_lemma:3}, 
\eqref{b_eq:C4} follows from the induction hypothesis in \eqref{b_eq:CIndHyp} and \eqref{b_eq:C6} follows from the definition of $V^{m,j}_t$ in \eqref{eq:Vdef}.
\end{IEEEproof}

\begin{lemma}
\label{b_lemma:3}
$\forall  t\in \mathcal{T}, (z_{1:t+1},a_{1:t}^{l,m}, x_{1:t+1}^{m,j})$ and
$\sigma^{m,j}_{t}$ 
\eq{
&\E^{\tsigma_{t:T}^{l},\sigma_{t:T}^{m,j},\tsigma_{t:T}^{m,-j},\tsigma^{f}_{t:T},\,\underline{\pi_t}}  \big\{ \sum_{n=t+1}^T R_n^{m,j}(Z_n,X^{l,m}_n,A^{l,m}_n) \big\lvert \pi_{t}, z_{1:t+1},a_{1:t}^{l,m}, x_{1:t+1}^{m,j} \big\} =\nn\\
& \E^{\tsigma^{l}_{t+1:T},\sigma_{t+1:T}^{m,j},\tsigma_{t+1:T}^{m,-j},\tsigma^{f}_{t+1:T},\underline{F}(\underline{\pi_t},z_t,\tgamma^{l,m}_t,A^{l,m}_t)}  \big\{ \sum_{n=t+1}^T R_n^{m,j}(Z_n,X^{l,m}_n,A^{l,m}_n) \big\lvert \pi_{t+1}, z_{1:t+1},a_{1:t}^{l,m}, x_{1:t+1}^{m,j} \big\}. \label{eq:F1}
}

\end{lemma}
\begin{IEEEproof} 
Since the above expectations involve random variables $X_{t+1}^{l,\{m,-j\}}, Z_{t+1:T},A^{l,m}_{t+1:T}, X^{l,m}_{t+2:T}$, we consider the probability 
\seq{
\eq{
&\p^{\tsigma^{l}_{t:T},\sigma^{m,j}_{t:T},\tsigma_{t:T}^{m,-j},\sigma_{t:T}^{f}\,\underline{\pi_t}} (x_{t+1}^{l,\{m,-j\}}, z_{t+1:T},a^{l,m}_{t+1:T}, x^{l,m}_{t+2:T}\big\lvert \underline{\pi_t}, z_{1:t+1}, a_{1:t}^{l,m}, x_{1:t+1}^{m,j} ) = \frac{Nr}{Dr} \label{b_eq:F2}
}
\vspace{-0.4cm}
\eq{
&\text{where}\nn\\ 
&Nr 
=\sum_{x^{l,\{m,-j\}}_t,a_t^{l,m}}\p^{\tsigma^{l}_{t:T},\sigma^{m,j}_{t:T},\tsigma_{t:T}^{m,-j},\sigma_{t:T}^{f}\,\underline{\pi_t}} (x_t^{l,\{m,-j\}},a^{l,m}_t,z_{t+1}, x^{l,\{m,-j\}}_{t+1}, z_{t+2:T},a^{l,m}_{t+1:T},x^{l,m}_{t+2:T} \big\lvert\underline{\pi_t}, z_{1:t},a_{1:t-1}^{l,m}, x_{1:t}^{m,j} ) \\
&= \sum_{x_t^{l,\{m,-j\}},a_t^{l,m}}\p^{\tsigma^{l}_{t:T},\sigma^{m,j}_{t:T},\tsigma_{t:T}^{m,-j},\sigma_{t:T}^{f}\,\underline{\pi_t}} (x^{l,\{m,-j\}}_t \big\lvert \underline{\pi_t}, z_{1:t},a^{l,m}_{t+1:T}, x_{1:t}^{m,j} )1(z_{t+1}=\phi(\pi_{t},z_{t},\tgamma_t))
\nonumber 
\\
&\tsigma_t^{l,m}(a_t^{l,m}|\underline{\pi_t},z_{t},x_t^{l,m})Q^{l,\{m,-j\}}(x^{l,\{m,-j\}}_{t+1}|z_t,x^{l,m}_t, a^{l,m}_t)\p^{\tsigma^{l,m,f}_{t:T}, \,\underline{\pi_t}} (z_{t+2:T}, a^{l,m}_{t+1:T},x^{l,m}_{t+2:T}| z_{1:t},a_{1:t-1}^{l,m} ,x_{1:t-1}^{m,j}, x^{l,m}_{t:t+1}) 
\\
=&\sum_{x_t^{l,\{m,-j\}}}\underline{\pi_t}(x_t^{l,\{m,-j\}})1(z_{t+1}=\phi(\pi_{t},z_{t},\tgamma_t)) \tsigma_t^{l,m}(a_t^{l,m}|\underline{\pi_t},z_{t},x_t^{l,m}) Q^{l,\{m,-j\}}(x^{l,\{m,-j\}}_{t+1}|z_t,x^{l,m}_t, a^{l,m}_t)
\\
&\p^{\tsigma^{l}_{t+1:T},\sigma^f_{t+1:T} ,\, \pi_{t+1}} (z_{t+2:T}, a^{l,m}_{t+1:T},x^{l,m}_{t+2:T}| \underline{\pi_t},z_{1:t} ,a_{1:t-1}^{l,m},x_{1:t-1}^{m,j},x^{l,m}_{t:t+1}),\label{b_eq:Nr2}
}
where \eqref{b_eq:Nr2} follows from the conditional independence of types given common information, as shown in Claim~1 in~Appendix~\ref{app:0}, and the fact that probability on $(z_{t+1:T},a^{l,m}_{t+1:T},x^{l,m}_{2+t:T})$ given $z_{1:t}, x_{1:t}^{l,m},x^{l,m}_{t+1}, \pi_{t} $ depends on $z_{1:t}, a_{1:t-1}^{l,m},x_{1:t}^{l,m},x^{l,m}_{t+1}, \pi_{t+1} $ through ${\sigma_{t+1:T}^{f}, \tsigma_{t+1:T}^{l} }$. Similarly, the denominator in \eqref{b_eq:F2} is given by
\eq{
Dr &= \sum_{\tilde{x}_{t}^{l,\{m,-j\}},a_t^{l,m}} \p^{ \tsigma^{l}_{t:T},\sigma^{m,j}_{t:T},\tsigma_{t:T}^{m,-j},\sigma_{t:T}^{f}\,\underline{\pi_t}} (\tilde{x}_t^{l,\{m,-j\}}, a^{l,\{m,-j\}}_t, z_{t+1},x_{t+1}^{m,j}\big\lvert \underline{\pi_t}, z_{1:t-1},a_{1:t-1}^{l,m}, x_{1:t}^{m,j} )\\
%
%
=&\sum_{\tilde{x}_{t}^{l,m\{m,-j\}},a_t^{l,\{m,-j\}}} \underline{\pi_t}(\tilde{x}_t^{l,\{m,-j\}})1(z_{t+1}=\phi(\underline{\pi_t},z_t,\tgamma_t)) \sigma_t^{m,j}(a_t^{m,j}|z_{1:t}, a_{1:t-1}^{l,m},x_{1:t}^{l,})  \nonumber\\
&\tsigma_t^{l,\{m,-j\}}(a_t^{l,\{m,-j\}}|\underline{\pi_t},z_{t}, \tilde{x}_t^{l,\{m,-j\}}) Q^{l,\{m,j\}}(x^{\{m,j\}}_{t+1}|z_t,x^{l,m}_t, a^{l,m}_t)\label{b_eq:F4}
%
}

By canceling the terms $\phi(\cdot),\sigma_t^f(\cdot)$ and $Q^f(\cdot)$ in the numerator and the denominator, \eqref{b_eq:F2} is given by
\eq{
&\frac{\sum_{x_t^{l,\{m,-j\}}}\underline{\pi_t}(x_t^{l,\{m,-j\}}) \tsigma_t^{l,\{m,-j\}}(a_t^{l,\{m,-j\}}|\underline{\pi_t},z_{t}, x_t^{l,\{m,-j\}}) Q_{t+1}^{l,\{m,-j\}}(x^{l,m\{m,-j\}}_{t+1}|z_t,x^{l,m}_t, a^{l,m}_t)}{\sum_{\tilde{x}_{t}^{l,\{m,-j\}}} \underline{\pi_t}(\tilde{x}_t^{l,\{m,-j\}}) \tsigma_t^{l,\{m,-j\}}(a_t^{l,\{m,-j\}}|\underline{\pi_t},z_{t}, \tilde{x}_t^{l,m})}  \nonumber \\
&\times\p^{\tsigma^{l,m}_{t+1:T}, \sigma^f_{t+1:T} ,\, \pi_{t+1}} (z_{t+2:T}, a^{l,m}_{t+1:T},x^{l.m}_{t+2:T}|\underline{\pi_t}, z_{1:t},a^{l,m}_{1:t-1},x_{1:t}^{m,j}, x^{l,m}_{t+1})\\
=&\underline{\pi}_{t+1}^{l,\{m,-j\}}(x_{t+1}^{l,\{m,-i\}}) \p^{ \sigma_{t+1:T}^{m,j},\tsigma^{l,\{m,-j\}}_{t+1:T},\tsigma^f_{t+1:T}\, \underline{\pi}_{t+1}} (z_{t+2:T}, a^{l,m}_{t+1:T},x^{l,m}_{t+2:T}|\underline{\pi_t}, z_{1:t+1} ,a_{1:t-1}^{l,m},x_{1:t}^{m,j}, x^{l,m}_{t+1})\label{b_eq:F6}\\
=& \p^{\tsigma^{l}_{t+1:T},\sigma^{m,j}_{t+1:T},\tsigma_{t+1:T}^{m,-j},\tsigma_{t+1:T}^{f}\,\underline{\pi}_{t+1} } (x_{t+1}^{l,\{m,-j\}} ,z_{t+1:T},a_{1:t}^{l,m},x^{l,m}_{t+2:T} |\pi_{t}, z_{1:t+1},a_{1:t-1}^{l,m}, x_{1:t+1}^{m,j} ),
}
}
where \eqref{b_eq:F6} follows from using the definition of $\pi^{l,\{m,-j\}}_{t+1}(x_{t+1}^{l,\{m,-j\}})$ in \eqref{eq:piupdate}.

\end{IEEEproof}

\section{Part 3: Leaders}
\label{app:P2}
In the following, we will show that, $\forall t, i=1,\ldots N, z_{1:t},a_{1:t-1}^{l,m}, x_{1:t}^{l,i}, \sigma^{l,i}$
\eq{
&\E^{\tsigma^{l},\tsigma^{m},\tsigma^f,\underline{\pi_t}} \big\{ \sum_{n=t}^T \delta^{n-t}R_n^{l,i}(Z_n,X_n^{l,m},A_n^{l,m}) |\underline{\pi_t},z_{1:t},a_{1:t-1}^{l,m},x_{1:t}^{l,i}\big\} \nn\\
&\geq
 \E^{\sigma^{l,i}\tsigma^{l,-i},\hat{\sigma}^{m},\hat{\sigma}^f,\underline{\pi_t}} \big\{ \sum_{n=t}^T \delta^{n-t}R_n^{l,i}(Z_n,X_n^{l,m},A_n^{l,m}) |\underline{\pi_t},z_{1:t},a_{1:t-1}^{l,m},x_{1:t}^{l,i}\big\},
}
where $\tsigma^f\in BR^f(z,\tsigma^{l},\tsigma^{m}), \tsigma^m\in BR^f(z,\tsigma^{l},\tsigma^{m},\tsigma^f)$ as shown in Part 1 and $\hat{\sigma}^f\in BR^f(z,\sigma^{l,i},\tsigma^{l,-i},\hat{\sigma}^{m}),\hat{\sigma}^{m}\in BR^{m}(z,\sigma^{l,i},\tsigma^{l,-i},\hat{\sigma}^{m},\hat{\sigma}^f)$.

\begin{IEEEproof}
We prove the above result using induction and from results in Lemma~\ref{l_lemma:2} and \ref{l_lemma:1} proved in Appendix~\ref{l_app:lemmas}. 

For the base case at $t=T$, $\forall z_{1:T},a_{1:T-1}^{l,m}, x_{1:T}^{l,i}, \sigma^{l,i}$
\seq{
\eq{
&\hspace{-10pt}\E^{\tsigma_{T}^{l},\tsigma_T^m,\tsigma_{T}^{f}, \underline{\pi_t}}\big\{  R_T^{l,i}(Z_T,X^{l,m}_T,A^{l,m}_T) \big\lvert \underline{\pi_T}, z_{1:T},a_{1:T-1}^{l,m},x_{1:T}^{l,i}\big\}\nn \\
&\hspace{-10pt}=V^{l,i}_T(\underline{\pi_T},z_{T},x^{l,i}_T)  \label{l_eq:T2a}\\
&\hspace{-10pt}\geq \E^{ \sigma_{T}^{l,i},\tsigma_T^{l,-i},\hat{\sigma}_T^m,\hat{\sigma}_T^f,\underline{\pi}_T}\big\{ R_T^{l,i}(Z_T,X^{l,m}_T,A^{l,m}_T) \big\lvert \underline{\pi_T}, z_{1:T},a_{1:T-1}^{l,m},x_{1:T}^{l,i} \big\} \label{l_eq:T2},\nn\\
&\text{ where } \hat{\sigma}_T^f\in BR_T^f(\underline{\pi_T},z_{1:T},a_{1:T-1}^{l,m},\sigma_T^{l,i},\tsigma_T^{l,-i},\hat{\sigma}_T^{m}) \text{ and } \hat{\sigma}_T^{m}\in BR_T^{m}(\underline{\pi_T},z_{1:T},a_{1:T-1}^{l,m},\sigma_T^{l,i},\tsigma_T^{l,-i},\hat{\sigma}_T^{m},\hat{\sigma}_T^f)
}
}
where (\ref{l_eq:T2a}) follows from Lemma~\ref{l_lemma:1} and (\ref{l_eq:T2}) follows from Lemma~\ref{l_lemma:2} in Appendix~\ref{l_app:lemmas}. Let the induction hypothesis be that for $t+1$, $\forall z_{1:t+1}, a_{1:t}^{l,m},x_{1:t+1}^{l,m}, \sigma^{l,i}$,
\seq{
\eq{
 & \E^{ \tsigma_{t+1:T}^{l},\tsigma^m_{t:T},\tsigma_{t+1:T}^{f}\pi_{t+1}} \big\{ \sum_{n=t+1}^T R_n^{l,i}(Z_n,X_n^{l,m},A_n^{l,m}) \big\lvert \pi_{t+1}, z_{1:t+1},a_{1:t}^{l,m},x_{1:t+1}^{l,i} \big\} \nn\\
  &\geq \E^{ \sigma_{t+1:T}^{l,i},\tsigma_{t:T}^{l,-i},\hat{\sigma}_{t:T}^m,\hat{\sigma}_{t+1:T}^f,\pi_{t+1}} \big\{ \sum_{n=t+1}^T R_n^{l,i}(Z_n,X_n^{l,m},A_n^{l,m}) \big\lvert \pi_{t+1}, z_{1:t+1},a_{1:t}^{l,m},x_{1:t+1}^{l,m} \big\} \label{l_eq:PropIndHyp}\\
  &\text{where }  \hat{\sigma}^f_{t+1:T}\in BR_{t+1}^f(\pi_{t+1},z_{1:t+1},a_{1:t}^{l,m},\sigma_{t+1:T}^{l,i},\tsigma_{t+1:T}^{l,-i},\hat{\sigma}_{t+1:T}^{m})\nn\\
&  \hat{\sigma}^{m}_{t+1:T}\in BR_{t+1}^{m}(\pi_{t+1},z_{1:t+1},a_{1:t}^{l,m},\sigma_{t+1:T}^{l,i},\tsigma_{t+1:T}^{l,-i},\hat{\sigma}_{t+1:T}^{m},\hat{\sigma}_{t+1:T}^f)
}
Then $\forall z_{1:t},a_{1:t-1}^{l,m},x_{1:t}^{l,i}, \sigma^{l,i}$, we have
\eq{
&\E^{ \tsigma_{t:T}^{l,m},\tsigma_{t:T}^{f},\underline{\pi_t}} \big\{ \sum_{n=t}^T R_n^{l,i}(Z_n,X_n^{l,m},A_n^{l,m}) \big\lvert \pi_{t}, z_{1:t},a_{1:t-1}^{l,m},x_{1:t}^{l,i}\big\} \nn \\
&= V^{l,i}_t(\underline{\pi_t},z_{t},x^{l,i}_t)\label{l_eq:T1}\\
&\geq \E^{\gamma_t^{l,i},\tgamma_t^{l,-i},\hat{\gamma}_t^m,\hat{\gamma}^f_t,\underline{\pi_t}} \big\{ R_t^{l,i}(Z_t,X^{l,m}_t,A^{l,m}_t) + \nn\\
&V_{t+1}^{l,i} (\underline{F}(\underline{\pi_t},z_t,\gamma_t^{l,i},\tgamma_t^{l,-i},\hat{\gamma}_t^m,A_t^{l,m}),\phi(\underline{\pi_t},z_t,\gamma_t^{l,i},\tgamma_t^{l,-i},\hat{\gamma}_t^m,\hat{\gamma}^f_t),X_{t+1}^{l,i})\big\vert \pi_{t}, z_{1:t},a_{1:t-1}^{l,m},x_{1:t}^{l,i} \big\}  \label{l_eq:T3}\\
&= \E^{ \sigma_t^{l,i},\tsigma_t^{l,-i},\hat{\sigma}_t^m,\hat{\sigma}_t^f,\underline{\pi_t}} \big\{ R_t^{l,i}(Z_t,X^{l,m}_t,A^{l,m}_t) +\E^{\tsigma_{t+1:T}^{l,m,f},\underline{F}(\underline{\pi_t},z_t,\gamma_t^{l,i},\tgamma_t^{l,-i},\hat{\gamma}_t^m,A_t^{l,m})}  \nn\\
&\big\{ \sum_{n=t+1}^T R_n^{l,i}(Z_n,X^{l,m}_n,A^{l,m}_n)  \big\lvert z_{1:t},\phi(\underline{\pi_t},z_t,\gamma_t^{l,i},\tgamma_t^{l,-i},\hat{\gamma}_t^m,\hat{\gamma}_t^f),x_{1:t}^{l,i},X_{t+1}^{l,i}\big\}  \big\vert \pi_{t}, z_{1:t},a_{1:t-1}^{l,m},x_{1:t}^{l,i}\big\}  \label{l_eq:T3b}\\
&\geq \E^{\sigma_t^{l,i},\tsigma_t^{l,-i},\hat{\sigma}_t^m,\hat{\sigma}_t^f,\underline{\pi_t}} \big\{ R_t^{l,i}(Z_t,X^{l,m}_t,A^{l,m}_t)+\E^{\sigma_{t+1:T}^{l,i},\tsigma_{t+1:T}^{l,-i}\hat{\sigma}_{t+1:T}^{m},\hat{\sigma}_{t+1:T}^f \underline{F}(\underline{\pi_t},z_t,\gamma_t^{l,i},\tgamma_t^{l,-i},\hat{\gamma}_t^m,A_t^{l,m})}  \nn\\ &\big\{ \sum_{n=t+1}^T R_n^{l,i}(Z_n,X_n^{l,m},A_n^{l,m})  \big\lvert z_{1:t},\phi(\underline{\pi_t},z_t,\gamma_t^{l,i},\tgamma_t^{l,-i},\hat{\gamma}_t^m,\hat{\gamma}_t^f),x_{1:t}^{l,i},X_{t+1}^{l,i}\big\} \big\vert \pi_{t}, z_{1:t},a_{1:t-1}^{l,m},x_{1:t}^{l,i} \big\}  \label{l_eq:T4} \\
%
%
&= \E^{{\sigma}_t^{l,i},\tsigma_t^{l,-i},\hat{\sigma}_t^m,\hat{\sigma}_t^f,\underline{\pi_t}} \big\{ R_t^{l,i}(Z_t,X^{l,m}_t,A^{l,m}_t) + \E^{\sigma_{t:T}^{l,i},\tsigma_{t:T}^{l,-i},\hat{\sigma}_{t:T}^m,\hat{\sigma}_{t:T}^f,\underline{\pi_t}}\big\{ \sum_{n=t+1}^T R_n^{l,i}(Z_n,X_n^{l,m},A_n^{l,m})\nn\\
&   \big\lvert z_{1:t},\phi(\underline{\pi_t},z_t,\gamma_t^{l,i},\tgamma_t^{l,-i},\hat{\gamma}_t^m,\hat{\gamma}_t^f),x_{1:t}^{l,i},X_{t+1}^{l,i}\big\} \big\vert \pi_{t}, z_{1:t},a_{1:t-1}^{l,m},x_{1:t}^{l,i}\big\}  \label{l_eq:T5}\\
&=\E^{\sigma_{t:T}^{l,i},\tsigma_{t:T}^{l,-i},\hat{\sigma}_{t:T}^m,\hat{\sigma}_{t:T}^f,\underline{\pi_t}} \big\{ \sum_{n=t}^T R_n^{l,i}(Z_n,X_n^{l,m},A_n^{l,m}) \big\lvert \pi_{t}, z_{1:t},a_{1:t-1}^{l,m},x_{1:t}^{l,i}\big\}  \label{l_eq:T6},
}
}
where $\hat{\gamma}_t^f\in \bar{BR}_t^f(\underline{\pi_t}, z_t,\gamma_t^{l,i},\tgamma_t^{l,-i},\hat{\gamma}_t^m)$, $\hat{\sigma}_t^m \in BR_t^m(\underline{\pi_t},z_{1:t},a_{1:t-1}^{l,m},\sigma_t^{l,i},\tsigma_t^{l,-i},\hat{\sigma}_t^m,\hat{\sigma_t^f,}\tsigma_{t+1:T}^{l,m,f}), \hat{\gamma}_t^f = \hat{\sigma}_t^f(\cdot|z_{1:t},a_{1:t-1}^{l,m},\cdot)$,\\ $ \hat{\sigma}^f_{t+1:T}\in BR_{t+1}^f(\pi_{t+1},z_{1:t+1},a_{1:t}^{l,m},x_{1:t+1}^{l,m},\sigma_{t+1:T}^{l,m})$, (\ref{l_eq:T1}) follows from Lemma~\ref{l_lemma:1}, (\ref{l_eq:T3}) follows from Lemma~\ref{l_lemma:2}, (\ref{l_eq:T3b}) follows from Lemma~\ref{l_lemma:1} and 
(\ref{l_eq:T4}) follows from induction hypothesis in (\ref{l_eq:PropIndHyp}), (\ref{l_eq:T5}) follows from the fact that probability on $(z_{t+1:T},a^{l,m}_{t+1:T},x^{l,m}_{2+t:T})$ given $\underline{\pi_t},z_{1:t+1}, a_{1:t}^{l,m},x_{1:t+1}^{l,m}$ depends on $\pi_{t+1},z_{1:t+1}, a_{1:t}^{l,m},x_{1:t+1}^{l,m}$ through ${\hat{\sigma}_{t+1:T}^{f}, \tsigma_{t+1:T}^{l} }$. 
\end{IEEEproof}

\section{}
\label{l_app:lemmas}
\begin{lemma}
\label{l_lemma:2}
$\forall t\in [T], \pi_t,z_{1:t},a_{1:t-1}^{l,m},x_{1:t}^{l,i}, \forall \sigma^{l,i}$
\eq{
&V_t^{l,i}(\underline{\pi_t},z_{t},x^{l,i}_t) \geq \E^{\sigma_t^{l,i},\tsigma_t^{l,-i},\bar{\sigma}_t^m,\bar{\sigma}_t^f,\underline{\pi_t}}\nn\\
& \big\{ R_t^{l,i}(Z_t,X^{l,m}_t,A^{l,m}_t) + V_{t+1}^{l,i} (\underline{\pi}_{t+1},z_{t+1},X_{t+1}^{l,i}) \big\lvert \underline{\pi}_{t}, z_{1:t},a_{1:t-1}^{l,m},x_{1:t}^{l,i}\big\}\label{l_eq:lemma2}
}
where 
\eq{
\underline{\pi}_{t+1}&=P^{\tsigma_{1:t-1}^l,\sigma_t^{l,i},\tsigma_t^{l,-i},\tsigma_{t+1:T}^l,\hat{\sigma}^m,\hat{\sigma}^f}(\cdot|z_{1:t+1},a_{1:t})\\
z_{t+1}(\cdot) &= \sum_{x_t,a_t}z_{t}(x_{t}^f)P^{\tsigma_{1:t}^{l,m}}(x^{l,m}_{1:t}|z_{1:t},a_{1:t-1}) Q^f( \cdot|z_t,x_t^f, a_t^f,a^{l,m}_t)\bar{\sigma}^f_t(a^f_t|z_{1:t},a_{1:t-1}^{l,m},x_{1:t}^f)\\
&\sigma_t^{l,i}(a^{l,i}_t|z_{1:t},a_{1:t-1}^{l,m},x_{1:t}^{l,i})\tsigma_t^{l,-i}(a^{l,-i}_t|z_{1:t},a_{1:t-1}^{l,m},x_{1:t}^{l,-i})\bar{\sigma}_t^m(a^{m}_t|z_{1:t},a_{1:t-1}^{l,m},x_{1:t}^{m})\\
\bar{\sigma}_t^f&\in \bigcap_{x_{1:t}^{f}}BR_t^f(\underline{\pi_t},z_{1:t},a_{1:t-1}^{l,m},x_{1:t}^{f},\sigma_t^{l,i},\tsigma_t^{l,-i},\bar{\sigma}_{t}^{m},\tsigma_{t+1:T}^{l,m}),\\
\forall j, \bar{\sigma}_t^{m,j}&\in \bigcap_{x_{1:t}^{m,j}}BR^{m,j}_t(\underline{\pi_t},z_{1:t},a_{1:t-1}^{l,m},x_{1:t}^{m,j},\sigma_t^{l,i},\tsigma_t^{l,-i},\bar{\sigma}_{t}^{m},\bar{\sigma}^f_{t},\tsigma_{t+1:T}^{l,m,f})\\ 
{\tgamma}^{m}_t &= {\tsigma}_t^{m}(\cdot|{z}_{1:t},a_{1:t-1}^{l,m},\cdot)\\
{\tgamma}^{l,-i}_t &= {\tsigma}_t^{l,-i}(\cdot|{z}_{1:t},a_{1:t-1}^{l,m},\cdot)\\
\gamma_t^i &= \sigma_t^i(\cdot|z_{1:t},a^{l,m}_{1:t-1},x^{l,i}_{1:t-1},\cdot)\\
\bar{\gamma}_t^f &\in \bar{BR}_t^f(\underline{\pi_t},z_t,\gamma_t^{l,i},{\tgamma}_t^{l,-i},\bar{\gamma}_t^m)\\
\bar{\gamma}_t^m &\in \bar{BR}_t^m(\underline{\pi_t},z_t,\gamma_t^{l,i},\bar{\gamma}_t^{l,-i},\bar{\gamma}_t^m,\bar{\gamma}_t^f)
}
where we assume that $\bar{\sigma}^f,\bar{\sigma}^m$ are of type $m$ (Since if they are not, one can find an equivalent policies of type $m$ that achieve same reward profile, as shown in Appendix~\ref{app:gsm}).

\end{lemma}

\begin{IEEEproof}
We first note that from Lemma~\ref{lemma:pi_t}
\seq{
\label{eq:piz_update}
\eq{
\underline{\pi}_{t+1}&=P^{\tsigma_{1:t-1}^l,\sigma_t^{l,i},\tsigma_t^{l,-i},\tsigma_{t+1:T}^l,\hat{\sigma}^m}(\cdot|z_{1:t+1},a_{1:t}^{l,m})\nn\\
&=\underline{F}(\underline{\pi_t},z_t,\gamma_t^i,\tgamma_t^{l,-i}, \tgamma_t^{m},a_t^{l,m})
}
where 
\eq{
\underline{\pi}_{t}(\cdot)&=P^{\tsigma_{1:t-1}^l,\hat{\sigma}_{1:t-1}^m}(\cdot|z_{1:t+1},a_{1:t}^{l,m})
}
and 
\eq{
z_{t+1}(\cdot) &= \sum_{x_t^f,x_{1:t}^{l,m},a_t}z_{t}(x_{t}^f)P^{\tsigma_{1:t}^{l,m}}(x^{l,m}_{1:t}|z_{1:t},a_{1:t-1}) Q^f( \cdot|z_t,x_t^f, a_t^f,a^{l,m}_t)\bar{\sigma}^f_t(a^f_t|z_{1:t},a_{1:t-1}^{l,m},x_{1:t}^f)\\
&\sigma_t^{l,i}(a^{l,i}_t|z_{1:t},a_{1:t-1}^{l,m},x_{1:t}^{l,i})\tsigma_t^{l,-i}(a^{l,-i}_t|z_{1:t},a_{1:t-1}^{l,m},x_{1:t}^{l,-i})\bar{\sigma}_t^m(a^{m}_t|z_{1:t},a_{1:t-1}^{l,m},x_{1:t}^{m})\\
&=\phi(\underline{\pi_t},z_{t},\gamma_t^i,\tgamma_t^{l,-i},\bar{\gamma}_t^m,\bar{\gamma}_t^f)
}
where 
\eq{
\underline{\pi}_{t}(\cdot)&=P^{\tsigma_{1:t-1}^{l,m},}(\cdot|z_{1:t+1},a_{1:t}^{l,m})
}
}
Thus~\eqref{l_eq:lemma2} reduces to
\eq{
&V_t^{l,i}(\underline{\pi_t},z_{t},x^{l,i}_t) \geq \E^{\sigma_t^{l,i},\tsigma_t^{l,-i},\bar{\sigma}_t^m,\bar{\sigma}_t^f,\underline{\pi_t}}\nn\\
& \big\{ R_t^{l,i}(Z_t,X^{l,m}_t,A^{l,m}_t) + V_{t+1}^{l,i} (\underline{F}(\underline{\pi_t},z_t,\gamma_t^i,\tgamma_t^{l,-i}, \hat{\gamma}_t^{m},A_t^{l,m}),\phi(\underline{\pi_t},z_{t},\gamma_t^i,\tgamma_t^{l,-i},\bar{\gamma}_t^m,\bar{\gamma}_t^f),X_{t+1}^{l,i}) \big\lvert \hat{\underline{\pi}}_{t}, z_{1:t},a_{1:t-1}^{l,m},x_{1:t}^{l,i}\big\}
}

We prove this by contradiction. Suppose the claim is not true for $t$. This implies $\exists\  \breve{\sigma}_t^{l,i},\hat{z}_{1:t},\hat{a}_{1:t-1}^{l,m}$, $\hat{x}_{1:t}^{l,i}$ such that $\hat{\pi}_t = P^{\tsigma^{l,m}_{1:t}}(\cdot|\hat{z}_{1:t},\hat{a}_{1:t-1}^{l,m},\hat{x}_{1:t-1}^{l,i},\cdot)$ and,
\eq{
&\E^{ \breve{\sigma}_t^{l,i},\tsigma^{l,-i},\hat{\sigma}_t^m,\hat{\sigma}_t^f,\underline{\pi_t}} \big\{ R_t^{l,i}(Z_t,X^{l,m}_t,A^{l,m}_t) + \nn\\
&V_{t+1}^{l,i} (\underline{F}(\underline{\hat{\pi}_t},z_t,\breve{\gamma}_t^{l,i},\tgamma_t^{l,-i},\hat{\gamma}_t^m, A^{l,m}_t),\phi(\underline{\hat{\pi}_t},\hat{z}_t,\breve{\gamma}_t^{l,i},\tgamma_t^{\{l,-i\}},\hat{\gamma}_t^{m,f}),X_{t+1}^{l,i}) \big\lvert \hat{\underline{\pi}}_{t}, \hat{z}_{1:t},\hat{a}_{1:t-1}^{l,m},\hat{x}_{1:t}^{l,i}\big\} >\nn\\ &V_t^{l,i}({\hat{\underline{\pi}}_t},\hat{z}_{t},\hat{x}_t^{l,i}),\label{l_eq:E8}
}
where $\breve{\gamma}^{l,i}_t = \breve{\sigma}^{l,i}_t(\cdot|\hat{z}_{1:t},\hat{a}_{1:t-1}^{l,m},x^{l,i}_{1:t-1},\cdot)$,  $\hat{\sigma}_t^f\in BR_t^f(\underline{\hat{\pi}_t},\hat{z}_{1:t},\hat{a}_{1:t-1}^{l,m}, \breve{\sigma}_t^{l,i},\tsigma^{l,-i},\hat{\sigma}_t^m,\hat{\sigma}_t^f,{\tsigma}_{t+1:T}^{l,m})$, \\$\hat{\sigma}_t^m\in BR_t^m(\underline{\hat{\pi}_t},\hat{z}_{1:t},\hat{a}_{1:t-1}^{l,m},\breve{\sigma}_t^{l,i},\tsigma^{l,-i},\hat{\sigma}_t^m,\hat{\sigma}_t^f,{\tsigma}_{t+1:T}^{l,m,f})$ and $\hat{\gamma}_t^{m}$ satisfies
\eq{
\hat{\gamma}^{m}_t &= \hat{\sigma}_t^{m}(\cdot|\hat{z}_{1:t},\hat{a}_{1:t-1}^{l,m},\hat{x}_{1:t-1}^{m},\cdot)
}


Then for $\hat{z}_{1:t},\hat{a}_{1:t-1}^{l,m},\hat{x}_{1:t}^{l,i}$, we have
\seq{
\eq{
&V_t^{l,i}(\underline{\pi_t},\hat{z}_{t},\hat{x}_t^{l,i})= \max_{\gamma^{l,i}_t} \E^{\gamma^{l,i}_t,\tgamma_t^{-i},\bar{\gamma}^{m}_t,\bar{\gamma}_t^f,\underline{\pi_t} } \big\{ R_t^{l,i}(Z_t,X^{l,m}_t,A^{l,m}_t) +\\
& V_{t+1}^{l,i} (\underline{F}(\underline{\pi_t},z_t,\gamma^{l,i}_t,\tgamma_t^{-i},\bar{\gamma}^{m}_t,A^{l,m}_t),\phi(\underline{\pi_t},\hat{z}_t,\gamma^{l,i}_t,\tgamma_t^{-i},\bar{\gamma}^{m}_t,\bar{\gamma}_t^f),X_{t+1}^{l,i}) \big\lvert \hat{\pi}_{t}, \hat{z}_{1:t},\hat{a}_{1:t-1}^{l,m},\hat{x}_{1:t}^{l,i}\big\} \label{l_eq:E11}\\
&\geq\E^{\breve{\gamma}^{l,i}_t,\tgamma_t^{-i},\hat{\gamma}^{m}_t,\hat{\gamma}_t^f,\underline{\pi_t}} \big\{ R_t^{l,i}(Z_t,X^{l,m}_t,A^{l,m}_t)+\nn\\
&V_{t+1}^{l,i} (\underline{F}(\underline{\pi_t},z_t,\breve{\gamma}^{l,i}_t,\tgamma_t^{-i},\hat{\gamma}^{m}_t,A^{l,m}_t),\phi(\underline{\pi_t},\hat{z}_{t},\breve{\gamma}^{l,i}_t,\tgamma_t^{-i},\hat{\gamma}^{m}_t,\hat{\gamma}_t^f),X_{t+1}^{l,i}) \big\lvert \hat{\pi}_{t}, \hat{z}_{1:t},\hat{a}_{1:t-1}^{l,m},\hat{x}_{1:t}^{l,i} \big\}   \\
&= \E^{\breve{\sigma}^{l,i}_t,\tsigma_t^{-i},\hat{\sigma}^{m}_t,\hat{\sigma}_t^f,\underline{\pi_t}} \big\{ R_t^{l,i}(Z_t,X^{l,m}_t,A^{l,m}_t)\nn\\
&+  V_{t+1}^{l,i} (\underline{F}(\underline{\hat{\pi}_t},z_t,\breve{\gamma}^{l,i}_t,\tgamma_t^{-i},\hat{\gamma}^{m}_t,A^{l,m}_t),\phi(\underline{\hat{\pi}_t},\hat{z}_{t},\breve{\gamma}^{l,i}_t,\tgamma_t^{-i},\hat{\gamma}^{m}_t,\hat{\gamma}_t^f),X_{t+1}^{l,i}) \big\lvert \hat{\underline{\pi}}_{t}, \hat{z}_{1:t},\hat{a}_{1:t-1}^{l,m},\hat{x}_{1:t}^{l,i}\big\}  \label{l_eq:E9}\\
&> V_t^{l,i}(\underline{\hat{\pi}_t},\hat{z}_{t},\hat{x}_t^{l,i}) \label{l_eq:E10} 
}
where (\ref{l_eq:E11}) follows from the definition of $V_t^{l,i}$ in (\ref{eq:Vdef}) and~\eqref{eq:piz_update}, (\ref{l_eq:E9}) follows from definition of $breve{\gamma}_t^{l,i},\hat{\gamma}^{m}_t,\hat{\gamma}_t^f$ and (\ref{l_eq:E10}) follows from (\ref{l_eq:E8}). However this leads to a contradiction. 
}
\end{IEEEproof}

\begin{lemma}
\label{l_lemma:1}
$\forall t\in [T],z_{1:t},a_{1:t-1}^{l,m},x_{1:t}^{l,i}$
\eq{
V^{l,i}_t(\underline{\pi_t},z_{t},x^{l,i}_t)&= \E^{\tsigma_{t:T}^{l,m,f} ,\underline{\pi_t}} \big\{ \sum_{n=t}^T R_n^{l,i}(Z_n,X_n^{l,m},A_n^{l,m}) \big\lvert \underline{\pi}_{t}, z_{1:t},a_{1:t-1}^{l,m},x_{1:t}^{l,i}\big\} .
}
\end{lemma}
\begin{IEEEproof}
\seq{
We prove the lemma by induction. For $t=T$, 
\eq{
 &\E^{\tsigma_{T}^{l,m},\tsigma_{T}^{f},\underline{\pi_T}} \big\{  R_T^{l,i}(Z_T,X^{l,m}_T,A^{l,m}_T) \big\lvert \pi_{T}, z_{1:T},a_{1:T-1}^{l,m},x_{1:T}^{l,i}\big\}\nn \\
 &= \sum_{x_T^{\{l,-i\},m}, a^{l,m}_T} \underline{\pi}_T(x_T^{\{l,-i\},m})R_T^{l,i}(z_T,x^{l,m}_T,a^{l,m}_T) \tsigma_{T}^{l,m}(a^{l,m}_T|\underline{\pi}_T,z_{T},x_T^{l,m})\\ 
 &=V^{l,i}_T(\underline{\pi_T},z_{T},x^{l,i}_T) \label{l_eq:C1}
}
}
where (\ref{l_eq:C1}) follows from the definition of $V_t^{l,i}$ in (\ref{eq:Vdef}).

Suppose the claim is true for $t+1$, i.e., $\forall  t\in [T], z_{1:t+1},a_{1:t}^{l,m},x_{1:t+1}^{l,i}$
\eq{
&V^{l,i}_{t+1}(\pi_{t+1},z_{t+1},x_{t+1}^{l,i}) = \E^{ \tsigma_{t+1:T}^{l,m},\tsigma_{t+1:T}^{f},\pi_{t+1}} \big\{ \sum_{n=t+1}^T R_n^{l,i}(Z_n,X_n^{l,m},A_n^{l,m}) \big\lvert\pi_{t+1}, z_{1:t+1},a_{1:t}^{l,m},x_{1:t+1}^{l,i}\big\} \label{l_eq:CIndHyp}.
}
Then $\forall  t\in [T], z_{1:t},a_{1:t-1}^{l,m},x_{1:t}^{l,i}$, we have
	\seq{
\eq{
&\E^{ \tsigma_{t:T}^{l,m}, \tsigma_{t:T}^{f},\underline{\pi_t}} \big\{ \sum_{n=t}^T R_n^{l,i}(Z_n,X_n^{l,m},A_n^{l,m}) \big\lvert \underline{\pi_t}, z_{1:t},a_{1:t-1}^{l,m},x_{1:t}^{l,i} \big\} \nn \\
&=  \E^{ \tsigma_{t:T}^{l,m},\tsigma_{t:T}^{f}, \underline{\pi_t} } \big\{R_t^{l,i}(Z_t,X^{l,m}_T,A^{l,m}_t) +\E^{ \tsigma_{t:T}^{l,m},\tsigma_{t:T}^{f},\underline{\pi_t} } \big\{ \sum_{n=t+1}^T R_n^{l,i}(Z_n,X_n^{l,m},A_n^{l,m})\nn\\
& \big\lvert \underline{F}(\underline{\pi_t},z_t,\tgamma^{l,m}_t,A^{l,m}_t),z_{1:t},  \phi(\underline{\pi_t},z_t,\tgamma_t),a_{1:t-1}^{l,m},A^{l,m}_t,x_{1:t}^{l,i},X_{t+1}^{l,i}\big\} \big\lvert \underline{\pi_t}, z_{1:t},a_{1:t-1}^{l,m},x_{1:t}^{l,i}\big\} \label{l_eq:C2}
\\
&=  \E^{\tsigma_{t:T}^{l,m},\tsigma_{t:T}^{f},\underline{\pi_t}} \big\{R_t^{l,i}(Z_t,X^{l,m}_T,A^{l,m}_t)+\E^{\tsigma_{t+1:T}^{l,m},\tsigma_{t+1:T}^{f} ,\underline{F}(\underline{\pi_t},z_t,\tgamma^{l,m}_t,A^{l,m}_t)}\big\{ \sum_{n=t+1}^T R_n^{l,i}(Z_n,X_n^{l,m},A_n^{l,m}) \big\lvert\nn\\
&  \underline{F}(\underline{\pi_t},z_t,\tgamma^{l,m}_t,A^{l,m}_t),z_{1:t},\phi(\underline{\pi_t},z_t,\tgamma_t),a_{1:t-1}^{l,m},A^{l,m}_t,x_{1:t}^{l,i},X_{t+1}^{l,i}\big\} \big\lvert \underline{\pi_t}, z_{1:t},a_{1:t-1}^{l,m},x_{1:t}^{l,i}\big\} \label{l_eq:C3}\\
&=  \E^{\tsigma_{t}^{l,m},\tsigma_{t}^{f},\underline{\pi_t}} \big\{R_t^{l,i}(Z_t,X^{l,m}_T,A^{l,m}_t) + V^{l,i}_{t+1}(\underline{F}(\underline{\pi_t},z_t,\tgamma^{l,m}_t,A^{l,m}_t),\phi(\underline{\pi_t},z_t,\tgamma_t),X_{t+1}^{l,i}) \big\lvert \underline{\pi_t}, z_{1:t},a_{1:t-1}^{l,m},x_{1:t}^{l,i}\big\} \label{l_eq:C5}\\
&=V^{l,i}_{t}(\underline{\pi_t},z_t,x^{l,i}_t) \label{l_eq:C6},
}
}
where (\ref{l_eq:C3}) follows from~Lemma~\ref{l_lemma:3}, (\ref{l_eq:C5}) follows from the induction hypothesis in (\ref{l_eq:CIndHyp}), 
and (\ref{l_eq:C6}) follows from the definition of $V_t^{l,i}$ in (\ref{eq:Vdef}).
\end{IEEEproof}

\begin{lemma}
\label{l_lemma:3}
$\forall t\in \mathcal{T}, \sigma_t^{l,i}, (z_{1:t+1},a_{1:t}^{l,m},x_{1:t+1}^{l,i})$ 
\eq{
&\E^{\sigma_{t:T}^{l,i},\tsigma_{t:T}^{\{l,-i\},m,f},\underline{\pi_t}}  \big\{ \sum_{n=t+1}^T R_n^{l,i}(Z_n,X^{l,m}_n,A^{l,m}_n) \big\lvert \underline{\pi_t}, z_{1:t+1},a_{1:t}^{l,m}, x_{1:t+1}^{l,i} \big\} =\nn\\
& \E^{\sigma_{t+1:T}^{l,i},\tsigma_{t+1:T}^{\{l,-i\},m,f},\underline{F}(\underline{\pi_t},z_t,\tgamma_t^{l,m},a_t^{l,m})}  \big\{ \sum_{n=t+1}^T R_n^{l,i}(Z_n,X^{l,m}_n,A^{l,m}_n) \big\lvert \underline{\pi_t}, z_{1:t+1},a_{1:t}^{l,m}, x_{1:t+1}^{l,i} \big\}, \label{l_eq:F1}
}
where ${\tgamma}_t^{l,m}=\tsigma^{l,m}_t(\cdot|\underline{\pi}_t,z_t,\cdot)$.
\end{lemma}
\begin{IEEEproof} 
Since the above expectations involve random variables $Z_{t+1:T},X^{\{l,-i\},m}_{t+1:T},A^{l,m}_t $, we consider the probability 
\seq{
\eq{
&\p^{\sigma_{t:T}^{l,i},\tsigma_{t:T}^{l,-i},\tsigma_{t:T}^{m,f},\underline{\pi_t}} (z_{t+1:T},x_{t+1}^{\{l,-i\},m},x_{t+2:T}^{l,m},a_{t+1:T}^{l,m}\big\lvert \underline{\pi_t}, z_{1:t+1}, a_{1:t}^{l,m}, x_{1:t+1}^{l,i} ) = \frac{Nr}{Dr} \label{l_eq:F2}
}
\vspace{-0.4cm}
\eq{
&\text{where}\nn\\ 
&Nr 
=\sum_{x_t^{\{l,-i\},m},a_t^{l,m}}\p^{\sigma_{t:T}^{l,i},\tsigma_{t:T}^{\{l,-i\},m,f},\underline{\pi_t}} (x_t^{\{l,-i\},m},a^{l,m}_t,z_{t+1}, x^{\{l,-i\},m}_{t+1}, z_{t+2:T},a^{l,m}_{t+1:T},x^{l,m}_{t+2:T} \big\lvert\underline{\pi_t}, z_{1:t},a_{1:t-1}^{l,m}, x_{1:t}^{l,i} ) \\
%
=&\sum_{x_t^{\{l,-i\},m},a_t^{l,m}}\underline{\pi}_t(x_t^{\{l,-i\},m})1(z_{t+1}=\phi(\pi_{t},z_{t},\tgamma_t))\sigma_t^{l,m}(a_t^{l,m}|z_{1:t},a_{1:t-1}^{l,m}, x_{1:t}^{l,m})   Q^{l,m}(x^{\{l,-i\},m}_{t+1}|z_t,x^{l,m}_t, a_t^{l,m})
\\
&\p^{\sigma_{t+1:T}^{l,i},\tsigma_{t+1:T}^{\{l,-i\},m,f},\underline{\pi}_{t}} (z_{t+2:T},x^{l,m}_{t+2:T}, a^{l,m}_{t+1:T}| \underline{\pi_t},z_{1:t+1} ,a_{1:t}^{l,m},x_{1:t}^{l,i},x_{t+1}^{l,m}),\label{l_eq:Nr2}
}
where \eqref{l_eq:Nr2} follows from the fact that probability on $(z_{t+1:T},x^{l,m}_{t+2:T},a^{l,m}_{t+1:T})$ given $\underline{\pi_t},z_{1:t+1} ,a_{1:t}^{l,m},x_{1:t}^{l,i},x_{t+1}^{l,m}$ depends on $\underline{\pi_t},z_{1:t+1} ,a_{1:t}^{l,m},x_{1:t}^{l,i},x_{t+1}^{l,m}$ through $\sigma_{t+1:T}^{l,i},\tsigma_{t+1:T}^{l,-i},\tsigma_{t+1:T}^{m,f}$. Similarly, the denominator in \eqref{l_eq:F2} is given by
\eq{
Dr &=  \sum_{x_t^{\{l,-i\},m}}\p^{ \sigma_{t:T}^{l,i},\tsigma_{t:T}^{l,-i},\tsigma_{t:T}^{m,f},\, \underline{\pi_t}} ( x_t^{\{l,-i\},m},a^{l,m}_t, z_{t+1},x_{t+1}^{l,i}\big\lvert \underline{\pi_t}, z_{1:t},a_{1:t-1}^{l,m}, x_{1:t}^{l,i} )\\
&=\sum_{x_t^{\{l,-i\},m}}\underline{\pi}_t(x_t^{\{l,-i\},m})1(z_{t+1}=\phi(\underline{\pi_t},z_t,\tgamma_t)) \tsigma_t^{l,m}(a_t^{l,m}|\underline{\pi_t},z_{t},x_t^{l,m})Q^{l,i}(x^{l,i}_{t+1}|z_t,x^{l,m}_t, a^{l,m}_t)\label{l_eq:F3}
%
%
}

By canceling the terms $1(z_{t+1} = \phi(\underline{\pi_t},z_t,\tgamma_t)),\sigma_t^{l,i}(\cdot)$ and $Q^{l,i}(\cdot)$ in the numerator and the denominator, \eqref{l_eq:F2} is given by
\eq{
&\frac{\sum_{x_t^{l}}\underline{\pi}_t(x_t^{\{l,-i\},m}) \tsigma_t^{\{l,-i\},m}(a_t^{\{l,-i\},m}|\underline{\pi_t},z_{t},x_t^{\{l,-i\},m}) Q_{t+1}^{\{l,-i\},m}(x^{\{l,-i\},m}_{t+1}|z_t,x^{l,m}_t,a^{l,m}_t)}{\sum_{\tilde{x}_{t}^{\{l,-i\},m}} \underline{\pi}_t(\tilde{x}_t^{\{l,-i\},m}) \tsigma_t^{\{l,-i\},m}(a_t^{\{l,-i\},m}|\underline{\pi_t},z_{t},\tilde{x}_t^{\{l,-i\},m})}  \nonumber \\
&\times\p^{\tsigma^{l}_{t+1:T}, \sigma^f_{t+1:T} ,\, \pi_{t+1}} (z_{t+1:T}, a_{t+1:T}^{l,m},x^{l,m}_{t+2:T}| z_{1:t},a_{1:t-1}^{l,m},x_{1:t}^f, x_{t+1})\\
=&\pi_{t+1}^{\{l,-i\},m}(x_{t+1}^{\{l,-i\},m}) \p^{\sigma_{t+1:T}^{l,i},\tsigma_{t+1:T}^{l,-i},\tsigma_{t+1:T}^{m,f} ,\, \pi_{t+1}} (z_{t+1:T}, a_{t+1:T}^{l,m},x^{l,m}_{t+2:T}| z_{1:t},a_{1:t-1}^{l,m},x_{1:t}^f, x_{t+1})\label{l_eq:F6}\\
=& \p^{\sigma_{t+1:T}^{l,i},\tsigma_{t+1:T}^{l,-i},\tsigma_{t+1:T}^{m,f}\, \pi_{t+1} } (z_{t+1:T},x_{t+1}^{\{l,-i\},m},x_{t+2:T}^{l,m},a_{t+1:T}^{l,m} | z_{1:t},a_{1:t-1}^{l,m}, x_{1:t+1}^f ),
}

where \eqref{l_eq:F6} follows from using the definition of $\underline{\pi}_{t+1}(x_{t+1}^{l,m})$ in \eqref{eq:piupdate}.
}
\end{IEEEproof}

\section{Extra Lemmas}
\begin{lemma}
\label{lemma:BR}
Let $\sigma_t^i$ be any strategy of player $i$ and fix $z_{1:T}$. Let
\eq{
\hat{\sigma}^f &\in BR^f(z_{1:T},\tsigma_{1:t-1}^l,\sigma_t^{l,i},\tsigma_t^{l,-i},\tsigma_{t+1:T}^l,\hat{\sigma}^m)\label{eq:EL1}\\
\hat{\sigma}^m &\in BR^f(z_{1:T},\tsigma_{1:t-1}^l,\sigma_t^{l,i},\tsigma_t^{l,-i},\tsigma_{t+1:T}^l,\hat{\sigma}^m,\hat{\sigma}^f)\label{eq:EL2}}
where we assume that $\hat{\sigma}^f,\hat{\sigma}^m$ are of type $m$ (Since if they are not, one can find an equivalent policies of type $m$ that achieve same reward profile, as shown in Appendix~\ref{app:gsm}).
Let
\eq{
\hat{\gamma}_t^f &\in \bar{BR}_t^f(\pi_t,z_t,\gamma_t^{l,i},\tgamma_t^{l,-i},\hat{\gamma}_t^m)\label{eq:EL3}\\
\hat{\gamma}_t^m &\in \bar{BR}_t^m(\pi_t,z_t,\gamma_t^{l,i},\tgamma_t^{l,-i},\hat{\gamma}_t^m,\hat{\gamma}_t^f)\label{eq:EL4}
}
where $\gamma_t^{l,i} = \sigma_t^{l,i}(\cdot|a_{1:t-1}^{l,m},z_{1:t},x_{1:t-1}^{l,i},\cdot)$. Then $\forall a_{1:t-1},x_{1:t-1}^f,x_{1:t-1}^{m,j},x_{1:t-1}^{l,i}$, and for every $\hat{\gamma}_t^f,\hat{\gamma}_t^m$ that satisfy~\eqref{eq:EL3}, \eqref{eq:EL4} $\exists (\hat{\sigma}_t^f,\hat{\sigma}_t^m)$ that satisfy~\eqref{eq:EL1}~\eqref{eq:EL2} such that
\eq{
\hat{\gamma}_t^f &= \hat{\sigma}_t^f(\cdot|z_{1:t},a_{1:t-1},\cdot)\\
\hat{\gamma}_t^{m,j} &= \hat{\sigma}_t^{m,j}(\cdot|z_{1:t},a^{l,m}_{1:t-1},\cdot)
}where,

\eq{
&BR^f(z_{1:T},\sigma^{l},\sigma^{m}) :=\bigcap_{t}\bigcap_{a_{1:t-1}^{l,m}}\bigcap_{x_{1:t}^{f}}\arg\max_{\sigma^f} \E^{\sigma_{t:T}^{l,m},{\sigma}_{t:T}^f,\underline{\pi_t}}[\sum_{n=t}^T \delta^{n-t} R^f(Z_n,X_n,A_n)|z_{1:t},a_{1:t-1}^{l,m},x_{1:t}^{f} ]\\
&BR_t^{m,j}(\underline{\pi_t},z_{1:t},a_{1:t-1}^{l,m},x_{1:t}^{m,j},\sigma_{t:T}^{l},\sigma_{t:T}^{m,-j},\sigma_{t:T}^f) \nn\\
&:=\arg\max_{ \sigma^{m,j}} \E^{\sigma_{t:T}^{l}\sigma_{t:T}^{m,j}\sigma_{t:T}^{m,-j},\sigma_{t:T}^f,\underline{\pi_t}} \big\{ \sum_{n=t}^T \delta^{n-t}R_n^{m,j}(Z_n,X_n^{l,m},A_n^{l,m}) |\underline{\pi_t},z_{1:t},a_{1:t-1}^{l,m},x_{1:t}^{m,j}\big\}\\
&BR^{m,j}(z,\sigma^l,\sigma^{m,-j},\sigma^f) :=\bigcap_t \bigcap_{a_{1:t-1}^{m,j}}\bigcap_{x_{1:t}^{m,j}}  BR_t^{m,j}(\underline{\pi_t},z_{1:t},a_{1:t-1}^{l,m},x_{1:t}^{m,j},\sigma_{t:T}^{l},\sigma_{t:T}^{m,-j},\sigma_{t:T}^f).
}

where
\eq{
\pi_t(\cdot)=P^{\tsigma_{1:t-1}^{l,m,f}}(\cdot|z_{1:t},a_{1:t-1}). \label{eq:pi_def}
}
 \eq{
 &\bar{BR}_t^f(\underline{\pi_t},z_t,\gamma_t^{l,m}) :=\big\{ \tgamma_t^f: \forall x_t^f\in \cX^f, \tgamma_t^f(\cdot|x_t^f)\in  \arg\max_{\gamma^f_t(\cdot|x_t^f)}\E^{\gamma^f_t(\cdot|x_t^f) {\gamma}^{l,m}_t,\,z_t,\underline{\pi_t}}\nn\\
 &  
\big\{ R_t^f(z_t, X_t,A_t) +\delta V_{t+1}^f(\underline{F}(\underline{\pi_t},z_t,\gamma_t^{l,m},A^{l,m}_t),\phi(\underline{\pi_t}, z_t,\gamma_t^{l,m},\tilde{\gamma}^f_t), X^f_{t+1}) \big\lvert \underline{\pi_t}, z_t,x_t^f \big\}  \big\}, \label{eq:m_FP}
}
\eq{
 &\bar{BR}_t^{m,j}(\underline{\pi_t},z_t,\gamma_t^{l},\gamma_t^{m,-j},\gamma_t^f) :=\big\{ \tgamma_t^{m,j}: \forall x_t^{m,j}\in \cX^{m,j}, \tgamma_t^{m,j}(\cdot|x_t^{m,j})\in  \arg\max_{\gamma^{m,j}_t(\cdot|x_t^{m,j})}\E^{\gamma^{m,j}_t(\cdot|x_t^{m,j}) {\gamma}^{l}_t,\gamma_t^{m,-j},\gamma_t^f,\,z_t,\underline{\pi_t}} \nn\\
 & 
\big\{ R_t^f(z_t, X^{l,m,f}_t,A_t) +\delta V_{t+1}^f(\underline{F}(\underline{\pi_t},z_t,\gamma_t^{l},\tgamma_t^{m,j},\gamma_t^{m,-j},A^{l,m}_t),\phi(\underline{\pi_t}, z_t,\gamma_t^{l},\tgamma_t^{m,j},\gamma_t^{m,-j},{\gamma}^f_t), X^f_{t+1}) \big\lvert \underline{\pi_t}, z_t,x_t^f \big\}  \big\}, \label{eq:m_FP_mj}
}

and 
\eq{
\pi_t(\cdot)=P^{\tsigma_{1:t-1}^l,\hat{\sigma}_{1:t-1}^m,\hat{\sigma}_{1:t-1}^f}(\cdot|z_{1:t},a_{1:t-1}). \label{eq:pi_def}
}

\end{lemma}
\begin{IEEEproof}
Part 1: Minor Follower
\seq{
\eq{
& \E^{\gamma_t^{l,i},\tgamma_t^{l,-i},\hat{\gamma}_t^m,\hat{\gamma}^f_t(\cdot|x_t^f),\,\pi_t} \big\{ R_t^f(Z_t, X_t,A_t) +\nn\\
&\delta V_{t+1}^f(\underline{F}(\underline{\pi_t},z_t,\gamma_t^{l,i},\tgamma_t^{l,-i},\hat{\gamma}_t^m,A^{l,m}_t),\phi(\underline{\pi_t}, z_t,\gamma_t^{l,i},\tgamma_t^{l,-i},\hat{\gamma}_t^m,\hat{\gamma}^f_t), X^f_{t+1}) \big\lvert \underline{\pi_t}, z_t,x_t^f \big\}  \big\}\nn\\
&=\max_{\gamma^f_t(\cdot|x_t^f)} \E^{\gamma_t^{l,i},\tgamma_t^{l,-i},\hat{\gamma}_t^m,{\gamma}^f_t(\cdot|x_t^f), \,\pi_t} \big\{ R_t^f(Z_t, X_t,A_t) +\nn\\
&\delta V_{t+1}^f(\underline{F}(\underline{\pi_t},z_t,\gamma_t^{l,i},\tgamma_t^{l,-i},\hat{\gamma}_t^m,A^{l,m}_t),\phi(\underline{\pi_t}, z_t,\gamma_t^{l,i},\tgamma_t^{l,-i},\hat{\gamma}_t^m,\hat{\gamma}^f_t), X^f_{t+1}) \big\lvert \underline{\pi_t}, z_t,x_t^f \big\}  \big\}\label{eq:P0}\\\\
%
%
 %
&= \max_{\gamma^f_t(\cdot|x_t^f)} \E^{ \gamma_t^{l,i},\tgamma_t^{l,-i},\hat{\gamma}_t^m,{\gamma}^f_t(\cdot|x_t^f) \,\pi_t} \big\{ R_t^f(Z_t,X_t,A_t)+ \E^{\tsigma_{t+1:T}^{l,m,f},\underline{F}(\underline{\pi}_t,z_t, \gamma_t^{l,i},\tgamma_t^{l,-i},\hat{\gamma}_t^m, A^{l,m}_t)}\nn\\
& \big\{\sum_{n=t+1}^T \delta^{n-t}R_n^f(Z_n,X_n,A_n) |z_{1:t},\phi(\underline{\pi_t}, z_t,\gamma_t^{l,i},\tgamma_t^{l,-i},\hat{\gamma}_t^m,\hat{\gamma}^f_t),a_{1:t-1},A_t,x_{1:t}^f,X_{t+1}^f\big\}\big\lvert \pi_t,z_t,x_t^f\big\}\label{eq:P1}\\
&= \max_{\gamma^f_t(\cdot|x_t^f)} \E^{ \gamma_t^{l,i},\tgamma_t^{l,-i},\hat{\gamma}_t^m,{\gamma}^f_t(\cdot|x_t^f) \,\pi_t} \big\{ R_t^f(Z_t,X_t,A_t)+ \max_{\sigma_{t+1:T}^f}\E^{\tsigma_{t+1:T}^{l,m},\sigma_{t+1:T}^f,\underline{F}(\underline{\pi}_t, \gamma_t^{l,i},\tgamma_t^{l,-i},\hat{\gamma}_t^m, A^{l,m}_t)}\nn\\
& \big\{\sum_{n=t+1}^T \delta^{n-t}R_n^f(Z_n,X_n,A_n) |z_{1:t},\phi(\underline{\pi_t}, z_t,\gamma_t^{l,i},\tgamma_t^{l,-i},\hat{\gamma}_t^m,\hat{\gamma}^f_t),a_{1:t-1},A_t,x_{1:t}^f,X_{t+1}^f\big\}\big\lvert \pi_t,z_t,x_t^f\big\}\label{eq:P2}\\
%
%
%
&=\max_{\gamma^f_t(\cdot|x_t^f)} \E^{\gamma_t^{l,i},\tgamma_t^{l,-i},\hat{\gamma}_t^m,{\gamma}^f_t(\cdot|x_t^f) ,\,\pi_t} \big\{ R_t^f(Z_t,X_t,A_t) + \max_{\sigma_{t+1:T}^f}\E^{\sigma_t^{l,i},\tsigma_t^{l,-i},\hat{\sigma}_t^m,\hat{\sigma}_t^f,\tsigma_{t+1:T}^{l,m}\sigma_{t+1:T}^f,\underline{\pi}_{t}}  \nn\\
&
\big\{ \sum_{n=t+1}^T R_n^f(Z_n,X_n,A_n) \big\lvert z_{1:t},\phi(\underline{\pi_t}, z_t,\gamma_t^{l,i},\tgamma_t^{l,-i},\hat{\gamma}_t^m,\hat{\gamma}^f_t),a_{1:t-1},A_t,x_{1:t}^f,X_{t+1}^f\big\} \big\vert z_{1:t},a_{1:t-1}, x_{1:t}^f \big\}\label{eq:P3} \\
&=\max_{\sigma_t^f} \E^{\sigma_t^{l,i},\tsigma_t^{l,-i},\hat{\sigma}_t^m,{\sigma}_t^f,\underline{\pi}_{t}} \big\{ R_t^f(Z_t,X_t,A_t) + \max_{\sigma_{t+1:T}^f}\E^{\sigma_t^{l,i},\tsigma_t^{l,-i},\hat{\sigma}_t^m,\hat{\sigma}_t^f,\tsigma_{t+1:T}^{l,m}\sigma_{t+1:T}^f,\underline{\pi}_{t}} \nn\\
&\big\{ \sum_{n=t+1}^T R_n^f(Z_n,X_n,A_n) \big\lvert z_{1:t},\phi(\underline{\pi_t}, z_t,\gamma_t^{l,i},\tgamma_t^{l,-i},\hat{\gamma}_t^m,\hat{\gamma}^f_t),a_{1:t-1},A_t,x_{1:t}^f,X_{t+1}^f\big\} \big\vert z_{1:t},a_{1:t-1}, x_{1:t}^f \big\}\label{eq:P4} \\
&=\max_{\sigma_{t:T}^f} \E^{\sigma_t^{l,i},\tsigma_t^{l,-i},\hat{\sigma}_t^m,{\sigma}_{t:T}^f, \tsigma_{t+1:T}^{l,m},\underline{\pi}_{t}}\big\{ \sum_{n=t}^T \delta^{t-n} R_n^f(Z_n,X_n,A_n) \big\vert z_{1:t},a_{1:t-1}, x_{1:t}^f \big\}\label{eq:P5}
%
%
%
%
%
}
}
where \eqref{eq:P0} follows from definition of $\hat{\gamma}_t^f$ in~\eqref{eq:EL2}, \eqref{eq:P1} follows from Lemma~\ref{lemma:1} in Appendix~\ref{app:lemmas}, \eqref{eq:P3} follows from the fact that the probability on $Z_{t+1:T},X_{t+1:T},A_{t+1:T}$ conditioned on $\tsigma^{l,m}_{t:T},\sigma^f_{t:T},\underline{\pi}_t$ only depends on $\tsigma_{t+1:T}^{l,m},\sigma_{t+1:T}^{f},\underline{F}(\underline{\pi_t},z_t,\tgamma_t^{l,m},A_t^{l,m})$ as the follower's strategy $\sigma_t^f$ doesn't affect either the update of the belief $\underline{\pi}_t$ or the update of the mean field $z_t$, \eqref{eq:P4} follows from the fact that $\hat{\sigma}^{f,m}_t$ are of type $m$ and definition of $\hat{\gamma}_t^f,\hat{\gamma}_t^m$.

Part 2: Major Follower (The arguments are similar but now for major player $j$)
\seq{
\eq{
& \E^{\gamma_t^{l,i},\tgamma_t^{l,-i},\hat{\gamma}_t^{m},\hat{\gamma}^f_t,\,\pi_t} \big\{ R_t^{m,j}(Z_t,X^{l,m}_t,A^{l,m}_t) +\nn\\
&\delta V_{t+1}^{m,j}(\underline{F}(\underline{\pi_t},z_t,\gamma_t^{l,i},\tgamma_t^{l,-i},\hat{\gamma}_t^m,A^{l,m}_t),\phi(\underline{\pi_t}, z_t,\gamma_t^{l,i},\tgamma_t^{l,-i},\hat{\gamma}_t^m,\hat{\gamma}^{f}_t), X^{m,j}_{t+1}) \big\lvert \underline{\pi_t}, z_t,x_t^{m,j} \big\}  \big\}\nn\\
&=\max_{\gamma^{m,j}_t(\cdot|x_t^{m,j})} \E^{\gamma_t^{l,i},\tgamma_t^{l,-i},\gamma_t^{m,j}(\cdot|x_t^{m,j}),\hat{\gamma}_t^{m,-j}\hat{\gamma}^f_t, \,\pi_t} \big\{ R_t^{m,j}(Z_t,X^{l,m}_t,A^{l,m}_t) +\nn\\
&\delta V_{t+1}^{m,j}(\underline{F}(\underline{\pi_t},z_t,\gamma_t^{l,i},\tgamma_t^{l,-i},\hat{\gamma}_t^m,A^{l,m}_t),\phi(\underline{\pi_t}, z_t,\gamma_t^{l,i},\tgamma_t^{l,-i},\hat{\gamma}_t^m,\hat{\gamma}^f_t), X^{m,j}_{t+1}) \big\lvert \underline{\pi_t}, z_t,x_t^{m,j} \big\}  \big\}\nn\\
%
%
 %
&= \max_{\gamma^{m,j}_t(\cdot|x_t^{m,j})} \E^{ \gamma_t^{l,i},\tgamma_t^{l,-i},\gamma_t^{m,j}(\cdot|x_t^{m,j})\hat{\gamma}_t^{m,-j},{\gamma}^f_t \,\pi_t} \big\{ R_t^{m,j}(Z_t,X^{l,m}_t,A^{l,m}_t)+ \E^{\tsigma_{t+1:T}^{l,m,f},\underline{F}(\underline{\pi}_t,z_t, \gamma_t^{l,i},\tgamma_t^{l,-i},\hat{\gamma}_t^m, A^{l,m}_t)}\nn\\
& \big\{\sum_{n=t+1}^T \delta^{n-t}R_n^{m,j}(Z_n,X^{l,m}_n,A^{l,m}_n) |z_{1:t},\phi(\underline{\pi_t}, z_t,\gamma_t^{l,i},\tgamma_t^{l,-i},\hat{\gamma}_t^m,\hat{\gamma}^f_t),a^{l,m}_{1:t-1},A^{l,m}_t,x_{1:t}^{m,j},X_{t+1}^{m,j}\big\}\big\lvert \pi_t,z_t,x_t^{m,j}\big\}\label{b_eq:P1}\\
&= \max_{\gamma^{m,j}_t(\cdot|x_t^{m,j})} \E^{ \gamma_t^{l,i},\tgamma_t^{l,-i},\gamma_t^{m,j}(\cdot|x_t^{m,j})\hat{\gamma}_t^{m,-j},{\gamma}^f_t) \,\pi_t} \big\{ R_t^{m,j}(Z_t,X^{l,m}_t,A^{l,m}_t)+ \max_{\sigma_{t+1:T}^{m,j}}\E^{\tsigma_{t+1:T}^{l},\sigma_{t+1:T}^{m,j},\tsigma_{t+1:T}^{m,-j},\tsigma_{t+1:T}^{f},\underline{F}(\underline{\pi}_t,z_t, \gamma_t^{l,i},\tgamma_t^{l,-i},\hat{\gamma}_t^m, A^{l,m}_t)}\nn\\
& \big\{\sum_{n=t+1}^T \delta^{n-t}R_n^{m,j}(Z_n,X^{l,m}_n,A^{l,m}_n) |z_{1:t},\phi(\underline{\pi_t}, z_t,\gamma_t^{l,i},\tgamma_t^{l,-i},\hat{\gamma}_t^m,\hat{\gamma}^f_t),a^{l,m}_{1:t-1},A^{l,m}_t,x_{1:t}^{m,j},X_{t+1}^{m,j}\big\}\big\lvert \pi_t,z_t,x_t^{m,j}\big\}\label{b_eq:P2}\\
%
%
%
&=\max_{\gamma^{m,j}_t(\cdot|x_t^{m,j})} \E^{\gamma_t^{l,i},\tgamma_t^{l,-i},\gamma_t^{m,j}(\cdot|x_t^{m,j})\hat{\gamma}_t^{m,-j},{\gamma}^f_t ,\,\pi_t} \big\{ R_t^{m,j}(Z_t,X^{l,m}_t,A^{l,m}_t) + \max_{\sigma_{t+1:T}^{m,j}}\E^{\sigma_t^{l,i},\tsigma_t^{l,-i},\hat{\sigma}_t^m,\hat{\sigma}_t^f,\tsigma_{t+1:T}^{l}\sigma_{t+1:T}^{m,j},\sigma_{t+1:T}^{m,-j},\tsigma_{t+1:T}^f,\underline{\pi}_{t}}  \nn\\
&
\big\{ \sum_{n=t+1}^T R_n^{m,j}(Z_n,X^{l,m}_n,A^{l,m}_n) \big\lvert z_{1:t},\phi(\underline{\pi_t}, z_t,\gamma_t^{l,i},\tgamma_t^{l,-i},\hat{\gamma}_t^m,\hat{\gamma}^f_t),a^{l,m}_{1:t-1},A^{l,m}_t,x_{1:t}^{m,j},X_{t+1}^{m,j}\big\} \big\vert \pi_t,z_{1:t},a^{l,m}_{1:t-1}, x_{1:t}^{m,j} \big\}\label{b_eq:P3} \\
&=\max_{\sigma_t^{m,j}} \E^{\sigma_t^{l,i},\tsigma_t^{l,-i},\sigma_t^{m,j}\hat{\sigma}_t^{m,-j},\hat{\sigma}^f_t,\underline{\pi}_{t}} \big\{ R_t^{m,j}(Z_t,X^{l,m}_t,A^{l,m}_t) + \max_{\sigma_{t+1:T}^{m,j}}\E^{\sigma_t^{l,i},\tsigma_t^{l,-i},\hat{\sigma}_t^m,\hat{\sigma}_t^f,\tsigma_{t+1:T}^{l}\sigma_{t+1:T}^{m,j},\sigma_{t+1:T}^{m,-j},\tsigma_{t+1:T}^f,\underline{\pi}_{t}} \nn\\
&\big\{ \sum_{n=t+1}^T R_n^{m,j}(Z_n,X^{l,m}_n,A^{l,m}_n) \big\lvert z_{1:t},\phi(\underline{\pi_t}, z_t,\gamma_t^{l,i},\tgamma_t^{l,-i},\hat{\gamma}_t^m,\hat{\gamma}^f_t),a^{l,m}_{1:t-1},A^{l,m}_t,x_{1:t}^{m,j},X_{t+1}^{m,j}\big\} \big\vert\pi_t, z_{1:t},a^{l,m}_{1:t-1}, x_{1:t}^{m,j} \big\}\label{b_eq:P4} \\
&=\max_{\sigma_{t:T}^{m,j}} \E^{\sigma_t^{l,i},\tsigma_t^{l,-i},\sigma_t^{m,j}\hat{\sigma}_t^{m,-j},\hat{\sigma}_t^f,\tsigma_{t+1:T}^{l}\sigma_{t+1:T}^{m,j},\tsigma_{t+1:T}^{m,-j},\tsigma_{t+1:T}^f,\underline{\pi}_{t}}\big\{ \sum_{n=t}^T \delta^{t-n} R_n^{m,j}(Z_n,X^{l,m}_n,A^{l,m}_n) \big\vert \pi_t, z_{1:t},a^{l,m}_{1:t-1}, x_{1:t}^{m,j} \big\}\label{b_eq:P5}
%
%
%
%
%
}
}
where \eqref{eq:P1} follows from Lemma~\ref{lemma:1} in Appendix~\ref{app:lemmas}, \eqref{eq:P3} follows from Lemma~\ref{b_lemma:3}, \eqref{eq:P4} follows from the fact that $\hat{\sigma}^{f,m}_t$ are of type $m$ and definition of $\hat{\gamma}_t^f,\hat{\gamma}_t^m$.
  This proves the theorem.

\end{IEEEproof}

\begin{lemma}
Let $(\tsigma^{l,m,f},z)$ be an SMFE-ML for the game considered and let $\forall x_t^{l,m}$
\eq{
\pi_t(x_t^{l,m}) = P^{\tsigma_{1:t-1}^{l,m,f}}(x^{l,m}_{t}|z_{1:t},a^{l,m}_{1:t-1})
}. 
Then for any given $(\sigma_{t}^{l,i}, a_{1:t-1},x_{1:t}^l)$, let 
\eq{
{\gamma}^{l,i}_t &= {\sigma}_t^{l,i}(\cdot|z_{1:t},{a}^{l,m}_{1:t-1},\cdot)
}
Then
\eq{
&P^{\tsigma_{1:t-1}^l,\sigma_t^{l,i},\tsigma_t^{l,-i},\tsigma_{t+1:T}^l,\hat{\sigma}^m}(x^{l,m}_{t+1}|z_{1:t+1},a_{1:t})= F(\pi_t,\gamma_t^{l,i},\tgamma_t^{l,-i},\hat{\gamma}_t^m, a_t^{l,m})(x^{l,m}_{t+1})
}
\label{lemma:pi_t}
where $\hat{\sigma}^f \in BR^f(\tsigma_{1:t-1}^l,\sigma_t^{l,i},\tsigma_t^{l,-i},\tsigma_{t+1:T}^l,\hat{\sigma}^m)$,
$\hat{\sigma}^m \in BR^m(\tsigma_{1:t-1}^l,\sigma_t^{l,i},\tsigma_t^{l,-i},\tsigma_{t+1:T}^l,\hat{\sigma}^f,\hat{\sigma}^m)$
where  $\hat{\sigma}^f,\hat{\sigma}^m$ are of type $m$.
Also $\hat{\gamma}_t^m \in \bar{BR}_t^m(\pi_t,z_t,\gamma_t^{l,i},\tgamma_t^{l,-i},\hat{\gamma}_t^m)$
\end{lemma}

\begin{IEEEproof}
Let $\hat{\sigma}^f \in BR^f(\tsigma_{1:t-1}^l,\sigma_t^{l,i},\tsigma_t^{l,-i},\tsigma_{t+1:T}^l,\hat{\sigma}^m)$ and $\gamma_t^f\in \bar{BR}^f(\pi_t,\gamma_t^{l,i},\tgamma_t^{l,-i},\hat{\gamma}_t^m)$ be of type $m$. 
 Then $\forall a_{1:t-1},x_{1:t-1}^f,x_{1:t-1}^{m,j},x_{1:t-1}^{l,i}$, and for every $\hat{\gamma}_t^f,\hat{\gamma}_t^m$ that satisfy~\eqref{eq:EL3}, \eqref{eq:EL4} $\exists (\hat{\sigma}_t^m)$ that satisfy~\eqref{eq:EL1}~\eqref{eq:EL2} such that
\eq{
\hat{\gamma}_t^{m,j} &= \hat{\sigma}_t^{m,j}(\cdot|z_{1:t},a^{l,m}_{1:t-1},\cdot)
}where,

Thus
\eq{
&P^{\tsigma_{1:t-1}^l,\sigma_t^{l,i},\tsigma_t^{l,-i},\tsigma_{t+1:T}^l,\hat{\sigma}^m,\hat{\sigma}^f}(x^{l,m}_{t+1}|z_{1:t+1},a_{1:t})\nn\\
&=\frac{\sum_{x^{l,m}_{1:t}}P^{\tsigma_{1:t-1}^l,\sigma_t^{l,i},\tsigma_t^{l,-i},\tsigma_{t+1:T}^l,\hat{\sigma}^m,\hat{\sigma}^f}(x^{l,m}_{1:t+1},z_{t+1},a^{l,m}_t|z_{1:t},a^{l,m}_{1:t-1})}{\sum_{x^{l,m}_{1:t}}P^{\tsigma_{1:t-1}^l,\sigma_t^{l,i},\tsigma_t^{l,-i},\tsigma_{t+1:T}^l,\hat{\sigma}^m,\hat{\sigma}^f}(x_{1:t}^{l,m}|z_{1:t},a^{l,m}_{1:t-1})1(z_{t+1} = \phi(z_t,\tgamma_t))\sigma_t^{l,m}(a^{l,m}_t|z_{1:t},a^{l,m}_{1:t-1},x_{1:t}^{l,m}))}
}
\eq{
&=\frac{\sum_{x^{l,m}_{1:t}}P^{\tsigma_{1:t-1}^l,\sigma_t^{l,i},\tsigma_t^{l,-i},\tsigma_{t+1:T}^l,\hat{\sigma}^m,\hat{\sigma}^f}(x^{l,m}_{1:t}|z_{1:t+1},a^{l,m}_{1:t})\sigma_t^{l,i}(a_t^{l,i}|a_{1:t-1},x_{1:t}^{l,i}),\tsigma_t^{l,-i}(a_t^{l,-i}|a_{1:t-1},x_{1:t}^{l,-i}),\hat{\sigma}^m(a_t^{m}|a_{1:t-1},x_{1:t}^{m})Q(x^{l,m}_{t+1}|x^{l,m}_t,a^{l,m}_t)}{\sum_{x^{l,m}_{1:t}}P^{\tsigma_{1:t-1}^l,\sigma_t^{l,i},\tsigma_t^{l,-i},\tsigma_{t+1:T}^l,\hat{\sigma}^m}(x^{l,m}_{1:t}|z_{1:t},a^{l,m}_{1:t-1})\sigma_t^{l,i}(a_t^{l,i}|a_{1:t-1},x_{1:t}^{l,i}),\tsigma_t^{l,-i}(a_t^{l,-i}|a_{1:t-1},x_{1:t}^{l,-i}),\hat{\sigma}^m(a_t^{m}|a_{1:t-1},x_{1:t}^{m})}\\
&=\frac{\sum_{x_{t}}\pi_t(x^{l,m}_{t})\gamma_t^{l,i}(a^{l,i}_t|x_{t}^{l,i}),\tgamma_t^{l,-i}(a^{l,-i}_t|x_{t}^{l,-i}),\hat{\gamma}_t^m(a^{m}_t|x_{t}^{m})Q(x^{l,m}_{t+1}|x^{l,m}_t,a^{l,m}_t)}{\sum_{x^{l,m}_{t}}\pi_t(x_{t}^{l,m})\gamma_t^{l,i}(a^{l,i}_t|x_{t}^{l,i}),\tgamma_t^{l,-i}(a^{l,-i}_t|x_{t}^{l,-i}),\hat{\gamma}_t^m(a^{m}_t|x_{t}^{m})}\\
%
%
&=F(\pi_t,\gamma_t^{l,i},\tgamma_t^{l,-i},\hat{\gamma}_t^m,a^{l,m}_t)(x^{l,m}_{t+1})
}
Moreover,
\eq{
z_{t+1}(\cdot) &= \sum_{x_t^f,x_{t}^{l,m},a_t}z_{t}(x_{t}^f)P^{\tsigma_{t}^{l,m}}(x^{l,m}_{t}|z_{1:t},a_{1:t-1}) Q^f( \cdot|z_t,x_t^f, a_t^f,a^{l,m}_t){\sigma}^f_t(a^f_t|z_{1:t},a_{1:t-1}^{l,m},x_{t}^f)\\
&\sigma_t^{l}(a^{l}_t|z_{1:t},a_{1:t-1}^{l,m},x_{1:t}^{l}){\sigma}_t^m(a^{m}_t|z_{1:t},a_{1:t-1}^{l,m},x_{t}^{m})\\
&=\phi(\underline{\pi_t},z_{t},\gamma_t^{l,m,f})
}
where 
\eq{
\gamma_t^l =\sigma^l_t(\cdot|z_{1:t},a_{1:t-1}^{l,m},\cdot) \\
\gamma_t^m = \sigma^m_t(\cdot|z_{1:t},a_{1:t-1}^{l,m},\cdot)\\
\gamma_t^f= \sigma^f_t(\cdot|z_{1:t},a_{1:t-1}^{l,m},\cdot)
}
\eq{
\underline{\pi}_{t}(\cdot)&=P^{\tsigma_{1:t-1}^{l,m},}(\cdot|z_{1:t+1},a_{1:t}^{l,m})
}
\end{IEEEproof}
\section{Lemmas for converse}
\label{app:Proof_Exist}

\begin{IEEEproof}
We prove this by contradiction. This implies there exists $\underline{\pi_t},z_t$ such that either (a)~\eqref{eq:FP1}
 doesn't have a solution,  (b)~\eqref{eq:FP1b} doesn't have a solution, or (c)~\eqref{eq:FP2} doesn't have a solution. 

\bit{
\item[(a)] Suppose~\eqref{eq:FP1}
 doesn't have a solution (concerning the minor follower) i.e. for any equilibrium generating function $\theta$ that generates $(\tsigma^{l,m,f},z)$ through forward recursion, there exists
\seq{
 $t\in\cT, z_{1:t}, a_{1:t-1}^{l,m}$ such that for $\underline{\pi_t}(\cdot)= P^{\tsigma^{l},\tsigma^{m},\tsigma^f}(\cdot|z_{1:t},a_{1:t-1}^{l,m})$, \eqref{eq:FP1} is not satisfied for $\theta_t$
i.e. for $\tgamma^f_t = \theta_t^f[\underline{\pi_t},z_t] = \tsigma_t^f(\cdot|\underline{\pi_t},z_t,\cdot), \tgamma^{l}_t = \theta^{l}[\underline{\pi_t},z_t] = \tsigma_t^{l}(\cdot|\underline{\pi_t},z_t,\cdot)$, $\tgamma^{m}_t = \theta^{m}[\underline{\pi_t},z_t] = \tsigma_t^{m}(\cdot|\underline{\pi_t},z_t,\cdot)$, $\exists x_t^f$ such that
\eq{
\tgamma_t^f(\cdot|x_t^f)\notin  \arg\max_{\gamma^{m,j}_t(\cdot|x_t^{m,j})}\E^{ {\tgamma}^{l}_t,\tgamma_t^m,\gamma^{m,j}_t(\cdot|x_t^{m,j}),\,z_t,\underline{\pi_t}}\nn\\
 &\hspace{-5cm}  
\big\{ R_t^f(z_t,X_t,A_t) +\delta V_{t+1}^{m,j}(\underline{F}(\underline{\pi_t},z_t,\tgamma_t^{l,m},A_t^{l,m}),\phi(\underline{\pi_t}, z_t,\tgamma_t), X^f_{t+1}) \big\lvert \underline{\pi_t}, z_t,x_t^f \big\}  
  }
  Let $t$ be the first instance in the backward recursion when this happens. This implies $\exists\ \hat{\gamma}_t^f$ such that
  \eq{
  &\E^{ {\tgamma}^{l,m}_t,\hat{\gamma}^f_t(\cdot|x_t^f),\,z_t,\underline{\pi_t}} 
\big\{ R_t^f(z_t, X_t,A_t) +\delta V_{t+1}^{m,j}(\underline{F}(\underline{\pi_t},z_t,\tgamma_t^{l,m},A^{l,m}_t),\phi(\underline{\pi_t}, z_t,\tgamma_t), X^f_{t+1}) \big\lvert \underline{\pi_t}, z_t,x_t^f \big\}  
  \nn\\
  &>  \E^{ {\tgamma}^{l,m}_t,\tgamma^f_t(\cdot|x_t^f),\,z_t,\underline{\pi_t}} 
\big\{ R_t^f(z_t, X_t,A_t) +
\delta V_{t+1}^{m,j}(\underline{F}(\underline{\pi_t},z_t,\tgamma_t^{l,m},A^{l,m}_t),\phi(\underline{\pi_t}, z_t,\tgamma_t), X^f_{t+1}) \big\lvert \underline{\pi_t}, z_t,x_t^f \big\}   \label{a_eq:E1}
  }
  This implies for $\hat{\sigma}^f_t(\cdot|z_{1:t},a_{1:t-1}^{l,m},x_{1:t-1}^f,\cdot) = \hat{\gamma}_t^f$,
  \eq{
  &\E^{ \tsigma_{t:T}^{l,m},\tsigma_{t:T}^{f},\underline{\pi_t}} \big\{ \sum_{n=t}^T R_n^f(Z_n,X_n,A_n) \big\lvert \underline{\pi_t}, z_{1:t},a_{1:t-1}^{l,m}, x_{1:t}^f \big\}
  \nn\\
  &= \E^{ \tsigma_t^{l,m},\tsigma_t^{f}, \underline{\pi_t}} \big\{ R_t^f(Z_t,X_t,A_t) + \E^{\tsigma_{t:T}^{f} \tsigma_{t:T}^{l},\underline{\pi_t}}\nn\\
  &\big\{ \sum_{n=t+1}^T R_n^f(Z_n,X_n,A_n) \big\lvert \underline{\pi_t},z_{1:t},\phi(\underline{\pi_t},z_t,\tgamma_t),a_{1:t-1},A^{l,m}_t, x_{1:t}^f,X_{t+1}^f \big\}  \big\vert \underline{\pi_t}, z_{1:t}, a_{1:t-1}^{l,m}, x_{1:t}^f \big\}
\\
  &= \E^{\tsigma_t^{l,m},\tsigma_t^{f}, \,\underline{\pi_t}} \big\{ R_t^f(Z_t,X_t,A_t) + \E^{\tsigma_{t+1:T}^{f} \tsigma_{t+1:T}^{l},\underline{F}(\underline{\pi_t},z_t,\tgamma_t^{l,m},A^{l,m}_t)}\nn\\
  &\big\{ \sum_{n=t+1}^T R_n^f(Z_n,X_n,A_n) \big\lvert z_{1:t}, \phi(\underline{\pi_t},z_t,\tgamma_t), a_{1:t-1}^{l,m},A^{l,m}_t, x_{1:t}^f,X_{t+1}^f \big\}  \big\vert \underline{\pi_t}, z_{1:t}, a_{1:t-1}^{l,m},x_{1:t}^f \big\} \label{a_eq:E2}
  \\
  &=\E^{ \tilde{\gamma}^{l}_t, \tilde{\gamma}^{m}_t,\tgamma^f_t, \, \underline{\pi_t}} \big\{ R_t^f(Z_t,X_t,A_t) + V_{t+1}^{m,j} (\underline{F}(\underline{\pi_t}, z_t,\tilde{\gamma}^{l,m}_t, A^{l,m}_t),\phi(\underline{\pi_t},z_t,\tgamma_t), X_{t+1}^f) \big\lvert\underline{\pi_t},x_t^f \big\} \label{a_eq:E3}
  \\
  &< \E^{\tilde{\gamma}^{l,m}_t,\hat{\gamma}^f_t \, \underline{\pi_t}} \big\{ R_t^f(Z_t,X_t,A_t) + V_{t+1}^{m,j} (\underline{F}(\underline{\pi}_t,z_t, \tilde{\gamma}^{l,m}_t, A^{l,m}_t), \phi(\underline{\pi_t},z_t,\tgamma_t),X_{t+1}^f) \big\lvert \underline{\pi_t}, x_t^f \big\}\label{a_eq:E4}
  \\
  &= \E^{ \tsigma_t^{l,m},\hat{\sigma}_t^f, \underline{\pi_t}} \big\{ R_t^f(Z_t,X_t,A_t) +  \E^{ \tsigma_{t+1:T}^{l,m},\tsigma_{t+1:T}^{f}, \underline{F}(\underline{\pi}_t,z_t, \tilde{\gamma}^{l,m}_t, A^{l,m}_t)}\nn\\
  &\big\{ \sum_{n=t+1}^T R_n^f(Z_n,X_n,A_n) \big\lvert z_{1:t},\phi(\underline{\pi_t},z_t,\tgamma_t), a_{1:t-1},A_t, x_{1:t}^f,X_{t+1}^f\big\} \big\vert \underline{\pi_t},z_{1:t},a_{1:t-1}^{l,m}, x_{1:t}^f \big\}\label{a_eq:E5}
  \\
  &=\E^{\tsigma_{t:T}^{l,m},\hat{\sigma}_t^f,\tsigma_{t+1:T}^{f},\underline{\pi_t}} \big\{ \sum_{n=t}^T R_n^f(Z_n,X_n,A_n) \big\lvert \underline{\pi_t},z_{1:t}, a_{1:t-1}^{l,m}, x_{1:t}^f \big\},\label{a_eq:E6}
  }
  where \eqref{a_eq:E2} follows from the fact that the probability on $Z_{t+1:T},X_{t+1:T},A_{t+1:T}$ conditioned on $\tsigma^{l,m}_{t:T},\sigma^f_{t:T},\underline{\pi}_t$ only depends on $\tsigma_{t+1:T}^{l,m},\sigma_{t+1:T}^{f},\underline{F}(\underline{\pi_t},z_t,\tgamma_t^{l,m},A_t^{l,m})$ as the follower's strategy $\sigma_t^f$ doesn't affect either the update of the belief $\underline{\pi}_t$ or the update of the mean field $z_t$, \eqref{a_eq:E3} follows from the definitions of $\tgamma_t^f$ and $\underline{\pi_t}$ and Lemma~\ref{lemma:1}, \eqref{a_eq:E4} follows from \eqref{a_eq:E1} and the definition of $\hat{\sigma}_t^f$, \eqref{a_eq:E5} follows from Lemma~\ref{lemma:2}, \eqref{a_eq:E6} again from the fact that the probability on $Z_{t+1:T},X_{t+1:T},A_{t+1:T}$ conditioned on $\tsigma^{l,m}_{t:T},\sigma^f_{t:T},\underline{\pi}_t$ only depends on $\tsigma_{t+1:T}^{l,m},\sigma_{t+1:T}^{f},\underline{F}(\underline{\pi_t},z_t,\tgamma_t^{l,m},A_t^{l,m})$ as the follower's strategy $\sigma_t^f$ doesn't affect either the update of the belief $\underline{\pi}_t$ or the update of the mean field $z_t$.. However, this leads to a contradiction since $(\tsigma^{l,m,f},z)$ is a SMFE-ML of the game.
}

\item[(b)]If~\eqref{eq:FP1b} doesn't have a solution (concerning major follower):
Suppose for any equilibrium generating function $\theta$ that generates $(\tsigma^{l},\tsigma^{m},\tsigma^f,z)$ through forward recursion, there exists $t\in\cT, z_{1:t}, a_{1:t-1}^{l,m}$ such that for $\underline{\pi_t}(\cdot)= P^{\tsigma^{l,m,f}}(\cdot|z_{1:t},a_{1:t-1}^{l,m})$, \eqref{eq:FP1b} is not satisfied for $\theta$
i.e. for $\tgamma^f_t = \theta^f[\underline{\pi_t},z_t] = \tsigma_t^f(\cdot|\underline{\pi_t},z_t,\cdot), \tgamma^{l}_t = \theta^{l}[\underline{\pi_t},z_t] = \tsigma_t^{l}(\cdot|\underline{\pi_t},z_t,\cdot), \tgamma^{m}_t = \theta^{m}[\underline{\pi_t},z_t] = \tsigma_t^{m}(\cdot|\underline{\pi_t},z_t,\cdot)$, $\exists j, x_t^{m,j}$ such that
\seq{
\eq{
\tgamma_t^{m,j}(\cdot|x_t^{m,j})\notin  \arg\max_{\gamma^{m,j}_t(\cdot|x_t^{m,j})}\E^{ {\tgamma}^{l}_t,\gamma^{m,j}_t(\cdot|x_t^{m,j})\tgamma_t^{m,-j},\gamma^f_t\,z_t,\underline{\pi_t}}\nn\\
 &\hspace{-7cm}  
\big\{ R_t^{m,j}(z_t, X_t^{l,m},A^{l,m}_t) +\delta V_{t+1}^{m,j}(\underline{F}(\underline{\pi_t},z_t,\tgamma_t^{l,m},A_t^{l,m}),\phi(\underline{\pi_t}, z_t,\tgamma_t), X^{m,j}_{t+1}) \big\lvert \underline{\pi_t}, z_t,x_t^{m,j} \big\}  
  }
  Let $t$ be the first instance in the backward recursion when this happens. This implies $\exists\ \hat{\gamma}_t^{m,j}$ such that
  \eq{
  &\E^{{\tgamma}^{l}_t,\hat{\gamma}_t^{m,j}(\cdot|x_t^{m,j}),\tgamma_t^{m,-j},{\tgamma}^f_t\,z_t,\underline{\pi_t}} 
\big\{ R_t^{m,j}(z_t, X^{l,m}_t,A_t^{l,m}) +\nn\\
&\delta V_{t+1}^{m,j}(\underline{F}(\underline{\pi_t},z_t,\tgamma_t^{l,m},A^{l,m}_t),\phi(\underline{\pi_t}, z_t,\tgamma_t), X^{m,j}_{t+1}) \big\lvert \underline{\pi_t}, z_t,x_t^{m,j} \big\}  
  \nn\\
  &>  \E^{ {\tgamma}^{l}_t,\tgamma^{m,j}_t(\cdot|x_t^{m,j}), \tgamma^{m,-j}_t,\tgamma^{f}_t\,z_t,\underline{\pi_t}} 
\big\{ R_t^{m,j}(z_t, X^{l,m}_t,A^{l,m}_t) +\nn\\
&\delta V_{t+1}^{m,j}(\underline{F}(\underline{\pi_t},z_t,\tgamma_t^{l,m},A^{l,m}_t),\phi(\underline{\pi_t}, z_t,\tgamma_t), X^{m,j}_{t+1}) \big\lvert \underline{\pi_t}, z_t,x_t^{m,j} \big\}   \label{b_eq:E1}
  }
  This implies for $\hat{\sigma}^{m,j}_t(\cdot|z_{1:t},a_{1:t-1}^{l,m},x_{1:t-1}^{m,j},\cdot) = \hat{\gamma}_t^{m,j}$,
  \eq{
  &\E^{\tsigma_{t:T}^{l,m,f},\underline{\pi_t}} \big\{ \sum_{n=t}^T R_n^{m,j}(Z_n,X^{l,m}_n,A^{l,m}_n) \big\lvert \underline{\pi_t}, z_{1:t},a^{l,m}_{1:t-1}, x_{1:t}^{m,j} \big\}
  \nn\\
  &= \E^{ \tsigma_t^{l,m,f}, \underline{\pi_t}} \big\{ R_n^{m,j}(Z_t,X^{l,m}_t,A^{l,m}_t) + \E^{ \tsigma_{t:T}^{l,m,f},\underline{\pi_t}}\nn\\
  &\big\{ \sum_{n=t+1}^T R_n^{m,j}(Z_n,X^{l,m}_n,A^{l,m}_n) \big\lvert \underline{\pi_t},z_{1:t},\phi(\underline{\pi_t},z_t,\tgamma_t),a^{l,m}_{1:t-1},A^{l,m}_t, x_{1:t}^{m,j},X_{t+1}^{m,j} \big\}  \big\vert \underline{\pi_t}, z_{1:t}, a_{1:t-1}^{l,m}, x_{1:t}^{m,j} \big\}
\\
  &= \E^{ \tsigma_t^{l,m,f}, \,\underline{\pi_t}} \big\{ R_n^{m,j}(Z_t,X^{l,m}_t,A^{l,m}_t) + \E^{\tsigma_{t+1:T}^{f} \tsigma_{t+1:T}^{l},\underline{F}(\underline{\pi_t},z_t,\tgamma_t^{l,m},A^{l,m}_t)}\nn\\
  &\big\{ \sum_{n=t+1}^T R_n^{m,j}(Z_n,X^{l,m}_n,A^{l,m}_n) \big\lvert z_{1:t}, \phi(\underline{\pi_t},z_t,\tgamma_t), a_{1:t-1}^{l,m},A^{l,m}_t, x_{1:t}^{m,j},X_{t+1}^{m,j} \big\}  \big\vert \underline{\pi_t}, z_{1:t}, a_{1:t-1}^{l,m},x_{1:t}^{m,j} \big\} \label{b_eq:E2}
  \\
  &=\E^{\tilde{\gamma}^{l}_t,\tgamma^{m,j}_t(\cdot|x_t^{m,j}),\tgamma^{m,-j}_t,\tgamma^{f}_t   \, \underline{\pi_t}} \big\{ R_n^{m,j}(Z_t,X^{l,m}_t,A^{l,m}_t) + V_{t+1}^{m,j} (\underline{F}(\underline{\pi_t}, z_t,\tilde{\gamma}^{l,m}_t, A^{l,m}_t),\phi(\underline{\pi_t},z_t,\tgamma_t), X_{t+1}^{m,j}) \big\lvert\underline{\pi_t},x_t^{m,j} \big\} \label{b_eq:E3}
  \\
  &< \E^{\tsigma_t^l,\hat{\gamma}^{m,j}_t(\cdot|x_t^{m,j}),{\tsigma}_t^{m,-j} \tilde{\gamma}^{f}_t, \, \underline{\pi_t}} \big\{ R_n^{m,j}(Z_t,X^{l,m}_t,A^{l,m}_t) + V_{t+1}^{m,j} (F({\pi}_t, \tilde{\gamma}_t, A_t), \phi(\underline{\pi_t},z_t,\tgamma_t),X_{t+1}^{m,j}) \big\lvert \underline{\pi_t}, x_t^{m,j} \big\}\label{b_eq:E4}
  \\
  &= \E^{ \tsigma_t^{l},\hat{\sigma}_t^{m,j},{\tsigma}_t^{m,-j},{\tsigma}_t^{f}, \underline{\pi_t}} \big\{ R_n^{m,j}(Z_t,X^{l,m}_t,A^{l,m}_t) +  \E^{\tsigma_{t+1:T}^{f} \tsigma_{t+1:T}^{l} \pi_{t+1}}\nn\\
  &\big\{ \sum_{n=t+1}^T R_n^{m,j}(Z_n,X^{l,m}_n,A^{l,m}_n) \big\lvert z_{1:t},\phi(\underline{\pi_t},z_t,\tgamma_t), a^{l,m}_{1:t-1},A^{l,m}_t, x_{1:t}^{m,j},X_{t+1}^{m,j}\big\} \big\vert \underline{\pi_t},z_{1:t},a_{1:t-1}^{l,m}, x_{1:t}^{m,j} \big\}\label{b_eq:E5}
  \\
  &=\E^{ \tsigma_{t:T}^{l},\hat{\sigma}_t^{m,j},\tsigma_t^{m,-j},\tsigma_{t+1:T}^{f},\underline{\pi_t}} \big\{ \sum_{n=t}^T R_n^{m,j}(Z_n,X^{l,m}_n,A^{l,m}_n) \big\lvert \underline{\pi_t},z_{1:t}, a_{1:t-1}^{l,m}, x_{1:t}^{m,j} \big\},\label{b_eq:E6}
  }
  where \eqref{b_eq:E2} follows from Lemma~\ref{b_lemma:3}, \eqref{b_eq:E3} follows from the definitions of $\tgamma_t^{m,j}$ and $\underline{\pi_t}$ and Lemma~\ref{lemma:1}, \eqref{b_eq:E4} follows from \eqref{b_eq:E1} and the definition of $\hat{\sigma}_t^{m,j}$, \eqref{b_eq:E5} follows from Lemma~\ref{lemma:2}, \eqref{b_eq:E6} follows from Lemma~\ref{b_lemma:3}. However, this leads to a contradiction since $(\tsigma^{l,m,f},z)$ is a SMFE-ML of the game.
}

\seq{
\item[(c)] ~\eqref{eq:FP2}
 doesn't have a solution (concerning leader $i$)

Suppose for any equilibrium generating function $\theta$ that generates $(\tsigma^{l,m,f},z)$ through forward recursion, there exists $t\in\cT, z_{1:t}, a_{1:t-1}^{l,m}$ such that for $\underline{\pi_t}(\cdot)= P^{\tsigma^{l,m,f}}(\cdot|z_{1:t},a_{1:t-1}^{l,m})$, \eqref{eq:FP2} is not satisfied for $\theta$
i.e. for $\tgamma^f_t = \theta^f[\underline{\pi_t},z_t] = \tsigma_t^f(\cdot|\underline{\pi_t},z_t,\cdot), \tgamma^{l,m}_t = \theta^{l,m}[\underline{\pi_t},z_t] = \tsigma_t^{l,m}(\cdot|\underline{\pi_t},z_t,\cdot)$, $\exists i, x^{l,i}_t$ such that
\eq{
\tgamma_t^{l,i} &\notin \arg\max_{\gamma_t^{l,i}}\E^{ {\gamma}^{l,i}_t,\tgamma_t^{l,-i},\bar{\gamma}_t^m,\bar{\gamma}_t^f,\,z_t} \big\{ R_t^{l,i}(z_t,X^{l,m}_t,A^{l,m}_t)\nn\\
&+\delta V_{t+1}^{l,i}(\underline{F}(\underline{\pi_t},z_t,{\gamma}^{l,i}_t,\tgamma_t^{l,-i},\bar{\gamma}_t^m,A^{l,m}_t),\phi(\underline{\pi_t},z_t,{\gamma}^{l,i}_t,\tgamma_t^{l,-i},\bar{\gamma}_t^m,\bar{\gamma}_t^f),X_{t+1}^{l,i})|\underline{\pi_t},z_t,x^{l,i}_t\big\},  \label{c_eq:FP2}\\
&\text{where } \bar{\gamma}_t^m\in \bar{BR}_t^m(\underline{\pi_t}, z_t,\gamma_t^{l,i},{\tgamma}_t^{l,-i},\bar{\gamma}_t^m,\bar{\gamma}_t^f),\;\;\; \bar{\gamma}_t^f\in \bar{BR}_t^f(\underline{\pi_t}, z_t,\gamma_t^{l,i},{\tgamma}_t^{l,-i},\bar{\gamma}_t^m)
} 
  Let $t$ be the first instance in the backward recursion when this happens. This implies $\exists i\ \breve{\gamma}_t^{l,i}$ such that
  \eq{
  &\E^{ \breve{\gamma}^{l,i}_t,{\tgamma}_t^{l,-i},\hat{\gamma}_t^m, \hat{\gamma}_t^f,\,\underline{\pi}_t,z_t} \big\{ R_t^{l,i}(z_t,X^{l,m}_t,A^{l,m}_t)\nn\\
  &+\delta V_{t+1}^{l,i}(\underline{F}(\underline{\pi_t},z_t,\breve{\gamma}_t^{l,i},{\tgamma}^{l,-i}_t,\hat{\gamma}_t^{m,f},A^{l,m}_t),\phi(\underline{\pi_t},z_t,\breve{\gamma}_t^{l,i},{\tgamma}_t^{l,-i},\hat{\gamma}_t^{m,f}),X_{t+1}^{l,i})|\underline{\pi_t},z_t,x^{l,i}_t\big\}  
  \nn\\
  &>  \E^{ {\tgamma}^{l,m,f}_t,\,\underline{\pi}_t,z_t} \big\{ R_t^{l,i}(z_t,X^{l,m}_t,A^{l,m}_t) \nn\\
  &+\delta V_{t+1}^{l,i}(\underline{F}(\underline{\pi_t},z_t,\tgamma^{l,m}_t,A^{l,m}_t),\phi(\underline{\pi_t},z_t,\tgamma_t),X_{t+1}^{l,i})|\underline{\pi_t},z_t,x^{l,i}_t\big\}   \label{c_eq:E1}
  }
  \eq{
 &\text{where }
 \hat{\gamma}_t^f\in \bar{BR}_t^f(\hat{\gamma}^{l,i}_t,{\tgamma}_t^{l,-i},\hat{\gamma}_t^m),
 \hat{\gamma}_t^m\in \bar{BR}_t^f(\underline{\pi_t}, z_t,\breve{\gamma}^{l,i}_t,{\tgamma}_t^{l,-i},\hat{\gamma}_t^m, \hat{\gamma}_t^f)
 }
  This implies for $\hat{\sigma}^{l,i}_t(\cdot|z_{1:t},a_{1:t-1}^{l,m},x_{1:t-1}^{l,i},\cdot) = \hat{\gamma}_t^{l,i}$,
  \eq{
  &\E^{ \tsigma_{t:T}^{l,m,f},\underline{\pi_t}} \big\{ \sum_{n=t}^T R_n^{l,i}(Z_n,X^{l,m}_n,A^{l,m}_n) \big\lvert \underline{\pi_t}, z_{1:t},a_{1:t-1}^{l,m}, x_{1:t}^{l,i} \big\}
  \nn\\
  &= \E^{ \tsigma_t^{l,m,f},\underline{\pi_t}} \big\{ R_t^{l,i}(Z_t,X^{l,m}_t,A^{l,m}_t) + \E^{\tsigma_{t:T}^{f} \tsigma_{t:T}^{l},\underline{\pi_t},z_t}\nn\\
  &\big\{ \sum_{n=t+1}^T R_n^{l,i}(Z_n,X^{l,m}_n,A^{l,m}_n) \big\lvert \underline{\pi}_t,z_{1:t},\phi(\underline{\pi_t},z_t,\tgamma_t),a^{l,m}_{1:t-1},A^{l,m}_t, x_{1:t}^{l,i},X_{t+1}^i \big\}  \big\vert \underline{\pi}_t, z_{1:t}, a_{1:t-1}^{l,m}, x_{1:t}^{l,i} \big\}
\\
  &= \E^{\tsigma_t^{l,m,f}, \,\underline{\pi_t}} \big\{ R_t^{l,i}(Z_t,X^{l,m}_t,A^{l,m}_t) + \E^{ \tsigma_{t+1:T}^{l,m,f},\underline{F}(\underline{\pi_t},z_t,\tgamma_t^{l,f},A^{l,f}_t)}\nn\\
  &\big\{ \sum_{n=t+1}^T R_n^{l,i}(Z_n,X^{l,m}_n,A^{l,m}_n) \big\lvert \underline{\pi_t}, z_{1:t}, \phi(\underline{\pi_t},z_t,\tgamma_t), a^{l,m}_{1:t-1},A^{l,m}_t, x_{1:t}^{l,i},X_{t+1}^i \big\}  \big\vert \underline{\pi_t}, z_{1:t}, a_{1:t-1}^{l,m},x_{1:t}^{l,i} \big\} \label{c_eq:E2}
  \\
  &=\E^{ \tilde{\gamma}^{l,m}_t,\tgamma^f_t(\cdot|x_t^f), \, \underline{\pi_t}} \big\{ R_t^{l,i}(Z_t,X^{l,m}_t,A^{l,m}_t) \nn\\
  &+ V_{t+1}^{l,i} (\underline{F}(\underline{\pi_t},z_t, \tilde{\gamma}^{l,m}_t, A^{l,m}_t),\phi(\underline{\pi_t},z_t,\tgamma_t), X_{t+1}^{l,i}) \big\lvert\underline{\pi_t},x^{l,i}_t \big\} \label{c_eq:E3}
  \\
  &< \E^{ \breve{\gamma}^{l,i}_t,{\tsigma}^{l,-i}_t,\hat{\sigma}^m_t,\hat{\sigma}^f_t, \, \underline{\pi_t}} \big\{ R_t^{l,i}(Z_t,X^{l,m}_t,A^{l,m}_t) +\nn\\
  &V_{t+1}^{l,i} (F({\pi}_t,z_t, \breve{\gamma}^{l,i}_t,{\tsigma}^{l,-i}_t,\hat{\sigma}^m_t,\hat{\sigma}^f_t, A^{l,m}_t), \phi(\underline{\pi_t},z_t,\breve{\gamma}^{l,i}_t,{\tsigma}^{l,-i}_t,\hat{\sigma}^m_t,\hat{\sigma}^f_t),X_{t+1}^{l,i}) \big\lvert \underline{\pi_t},z_t, x^{l,i}_t \big\}\label{c_eq:E4}
  \\
  &= \E^{ \breve{\gamma}^{l,i}_t,{\tsigma}^{l,-i}_t,\hat{\sigma}^m_t,\hat{\sigma}^f_t, \underline{\pi_t}} \big\{ R_t^{l,i}(Z_t,X^{l,m}_t,A^{l,m}_t) +  \E^{ \tsigma_{t+1:T}^{l,m,f},F({\pi}_t,z_t, \breve{\gamma}^{l,i}_t,{\tsigma}^{l,-i}_t,\hat{\sigma}^m_t,\hat{\sigma}^f_t, A^{l,m}_t)}\nn\\
  &\big\{ \sum_{n=t+1}^T R_n^{l,i}(Z_n,X^{l,m}_n,A^{l,m}_n) \big\lvert z_{1:t},\phi(\underline{\pi_t},z_t,\breve{\gamma}^{l,i}_t,{\tsigma}^{l,-i}_t,\hat{\sigma}^m_t,\hat{\sigma}^f_t), a^{l,m}_{1:t-1},A^{l,m}_t, x_{1:t}^{l,i},X_{t+1}^{l,i}\big\} \big\vert \underline{\pi}_t,z_{1:t},a_{1:t-1}^{l,m}, x_{1:t}^{l,i} \big\}\label{c_eq:E5}
  \\
  &=\E^{\breve{\sigma}^{l,i}_t,{\tsigma}^{l,-i}_t,\tsigma_{t+1:T}^{l},\hat{\sigma}^m_t,\tsigma_{t+1:T}^{m}\hat{\sigma}^f_t,\tsigma_{t+1:T}^{f}\underline{\pi_t}} \big\{ \sum_{n=t}^T R_n^{l,i}(Z_n,X^{l,m}_n,A^{l,m}_n) \big\lvert \underline{\pi_t},z_{1:t}, a_{1:t-1}^{l,m}, x_{1:t}^{l,i} \big\},\label{c_eq:E6}
  }
  where \eqref{c_eq:E2} follows from~Lemma~\ref{l_lemma:3}, \eqref{c_eq:E3} follows Lemma~\ref{l_lemma:1}, \eqref{c_eq:E4} follows from \eqref{c_eq:E1} and the definition of $\hat{\sigma}_t^f$, \eqref{c_eq:E5} follows from Lemma~\ref{l_lemma:2}, \eqref{c_eq:E6} again follows from~Lemma~\ref{l_lemma:3}. However, this leads to a contradiction since $(\tsigma^{l,m,f},z)$ is an SMFE-ML of the game.
  }
  }
\end{IEEEproof}

\bibliographystyle{IEEEtran}

\end{document}